\documentstyle[preprint,tighten,aps,epsfig]{revtex} 

\newcommand{\vc}[1]{{\mathbf #1}}

\begin{document}

\title{Field Theory of Mesoscopic Fluctuations in Superconductor/Normal-Metal 
Systems}

\author{Alexander Altland$^{\dagger,\S}$, B. D. Simons$^{\S}$ 
and D. Taras-Semchuk$^{\S}$} 
\address{$^{\dagger}$ Institut f\"ur Theoretische Physik, Universit\"at zu
K\"oln, Z\"ulpicher Strasse 77, 50937 K\"oln, Germany \\
$^{\S}$ Cavendish Laboratory, Madingley Road, Cambridge CB3 OHE, UK}

\date{\today}
\maketitle 
 
\begin{abstract} 
  Thermodynamic and transport properties of mesoscopic conductors are
  strongly influenced by the proximity of a superconductor: An
  interplay between the large scale quantum coherent wave functions in
  the normal mesoscopic and the superconducting region, respectively,
  leads to unusual mechanisms of quantum interference. These manifest
  themselves in both the mean and the mesoscopic fluctuation behaviour
  of superconductor/normal-metal (SN) hybrid systems being strikingly
  different from those of conventional mesoscopic systems. After
  reviewing some established theories of SN-quantum interference
  phenomena, we introduce a new approach to the analysis of
  SN-mesoscopic physics. Essentially, our formalism
  represents a unification of the quasi-classical formalism for
  describing {\it mean} properties of SN-systems on the one hand, with
  more recent field theories of mesoscopic {\it fluctuations} on the
  other hand. Thus, by its very construction, the new approach is
  capable of exploring both averaged and fluctuation properties of
  SN-systems on the same microscopic footing. As an example, the
  method is applied to the study of various characteristics of the
  single particle spectrum of SNS-structures.
\end{abstract} 

\tableofcontents

\section{Introduction}
Physical properties of both superconductors and mesoscopic normal
metals are governed by mechanisms of macroscopic quantum coherence.
Their interplay in SN-systems, i.e. hybrid systems comprised of a
superconductor adjacent to a mesoscopic normal metal, gives rise to
qualitatively new phenomena (see Ref.~\cite{Beenakker97} for a review):
Aspects of the superconducting characteristics are imparted on the
behaviour of electrons in the normal region. This phenomenon, known as
the ``proximity effect'', leads to both the {\it mean} (disorder
averaged) properties of SN-systems being substantially different from
those of normal metals and various types of mesoscopic {\it
  fluctuations}.
 
Although these two classes of phenomena are rooted in the same fundamental
physical mechanism -- a tendency towards the formation of Cooper pairs in
the {\it normal} metal region of a SN-system -- there are also major
differences. Even so, it is notable that more than two decades passed
between the first analyses of the manifestations of the proximity
effect in the mean properties of SN-systems and the mere {\it
  observation} that the same effect may also be exhibited in
mesoscopic fluctuations. The intimate connection between mean and
fluctuation manifestations of the proximity effect will be discussed
in some detail below. At this stage we simply itemize some basic
proximity effect induced phenomena -- of both mean and fluctuation
type -- and briefly comment on the theoretical approaches that have
been applied to their analysis.

{\it Mean properties of SN-systems:} Superconductors strongly modify
the physical properties of adjacent normal metals.  For example, the
proximity of a superconducting condensate tends to induce singular
behaviour in the normal metal density of states (DoS). The properties
of these singularities depend on both the coupling to the
superconductor and purely intrinsic characteristics of the normal
mesoscopic component.  This complex behaviour indicates that we are
confronted with an {\it interplay} between mechanisms of quantum
coherence in the superconductor and in the mesoscopic N-region.  DoS
singularities are only one of many more examples of manifestations of
the proximity effect.  For example, supercurrents may flow through
normal metal regions and the conductance of the normal metal may vary
with the phase of the superconducting order parameter
\cite{Spivak,A+Spivak,S+Nazarov}.  Indeed the conductance may, quite
counterintuitively, {\it increase} as a function of the impurity
concentration, through the phenomenon known as reflectionless
tunneling \cite{Beenakker97,Marmorkos93}.

All these phenomena share the common drawback that they are
exceedingly difficult to describe within conventional perturbative
techniques of condensed matter physics. Broadly speaking, the reason
for these problems is that the conventional 'reference point' of
perturbative approaches to dirty metals, i.e. a filled Fermi sea of
electrons with an essentially structureless dispersion relation,
represents a poor starting point to the description of SN-hybrids. In
fact, the proximity of a superconductor leads to a strong modification
of the states in the vicinity of the Fermi-surface, which implies that
it is difficult to perturbatively interpolate between the conventional
weakly disordered metal limit and the true state of the N-component of
a SN-system.

In the late sixties, Eilenberger \cite{Eilenberger} 
also introduced a novel approach to the
description of {\it bulk} superconductors which subsequently turned
out to be extremely successful in the analysis of SN-systems.
Essentially this so-called quasi-classical approach provided a
controlled coarse-graining procedure by which the Gorkov equation for
the microscopic Green function of superconductors could be drastically
simplified. Since on the one hand the Green function contains all the
information that is needed to describe SN-phenomena, whilst on the
other hand -- for the reasons indicated above -- its computation in
proximity effect influenced environments is in general tremendously
difficult, it is clear that Eilenberger's method represented a
breakthrough. Extending Eilenberger's work, Usadel~\cite{Usadel} later
derived a non-linear, diffusion-type equation for the quasi-classical
Green function of {\it dirty} metals (for the precise definition of
the term 'dirty', see below), the so-called Usadel equation. Based
largely on the pioneering work of Eilenberger~\cite{Eilenberger} and
Usadel~\cite{Usadel}, a powerful array of quasi-classical methods
has since been developed to study the mean properties of SN-systems.

In view of what has been said above about the difficulties encountered
in perturbative approaches, it is instructive to re-interpret the
solution of the quasi-classical equations in terms of the language of
conventional diagrammatic perturbation theory. Referring for details
to later sections, we here merely notice that the quasi-classical
Green functions actually represent high-order summations of quantum
interference processes caused by multiple impurity scattering. More
precisely, the solution of the Usadel equation generally sums up
infinitely many so-called diffuson diagrams, the fundamental building
blocks of the perturbative approach to dirty metals. Whereas in normal
metals high order interference contributions to physical observables
are usually small in powers of the parameter $g^{-1}$ ($g \gg 1$ is
the dimensionless conductance of a weakly disordered metal), here they
represent the leading order contribution even to a single Green
function. In passing we note that, very much as the conductance of
normal metals may be expanded in terms of the weak localization
corrections, disorder averaged properties of SN-systems can be
systematically expanded beyond the leading quasi-classical
approximation in powers of $g^{-1}$.  We will come back to this issue
below.  Having made these observations one may anticipate that the
tendency to strong quantum interference in SN-systems not only affects
their mean properties but also leads to the appearance of unusual
mesoscopic fluctuation behaviour.

{\it Mesoscopic fluctuations:} It has been shown both experimentally
\cite{Hartog,Hecker} 
and theoretically
\cite{Spivak,A+Spivak,A+Zirn,Takane,Beenakker93,Hui,Lambert93,Brouwer95,Brouwer96,Frahm}
that mesoscopic fluctuations in SN-systems not only tend to be larger
than in the pure N-case, but also can be of qualitatively different
physical origin. After what was said above it should be no surprise
that the pronounced tendency to exhibit fluctuations again finds its
origin in an interplay between standard mechanisms of mesoscopic
quantum coherence and the proximity effect.

The list of examples of such SN-specific mesoscopic fluctuation
phenomena includes
\begin{itemize}
\item As in N-systems, sample-specific fluctuations of the conductance
  are universal. However, due to proximity effect induced coherence
  mechanisms, the fluctuations are generally larger than in normal
  systems \cite{Hartog,Hecker}.
\item Besides the normal current, the critical Josephson current,
  $I_c$, through
  SNS-junctions exhibits fluctuations \cite{A+Spivak}, which become
  universal in the limit of a short ($L \ll \xi$) junction
  \cite{Beenakker91}: 
\begin{eqnarray*}
\langle \delta I_c^2\rangle \sim \left\{\begin{array}{lc}
(e E_c)^2, & L \gg \xi, \\
(e\Delta)^2, & L \ll \xi.\end{array}\right.
\end{eqnarray*}
Here $\xi = (D/\Delta)^{1/2}$ is the coherence length of
dirty superconductors, $\Delta$ the order parameter, $D$ the diffusion
constant, $E_c = D/L^2$ the Thouless energy and $L$ the system size
(note that we set $\hbar=1$ throughout). 

\item Such fluctuations of the supercurrent are relatively robust
  \cite{A+Spivak}: for instance, a relatively strong magnetic field
  will exponentially suppress the average supercurrent, but reduces
  the variance of the supercurrent fluctuations by only a factor of 2.
  
\item The single particle spectrum in the vicinity of the DoS anomaly
  exhibits characteristic types of statistics \cite{A+Zirn,Frahm}.
\item Novel types of universal spectral fluctuations appear \cite{A+Zirn}.
\end{itemize}
As compared to the mean properties of SN-systems, the physics of
fluctuation phenomena is less well understood. Firstly, the
quasiclassical approach is not tailored to an analysis of
fluctuations. Although, to compute fluctuations, one
needs to average {\it products} of Green functions over disorder, 
the quasiclassical equations are derived for single disorder averaged
Green functions and can, to the best of our knowledge, not be extended
to the computation of higher order cumulants\footnote{Instead of
  deriving equations for the Green functions themselves one may
  attempt to set up a quasiclassical approximation for their {\it
    generating functional} (U.Eckern, private communication). Whether
  or not such an approach has been realized and/or made working in the
  concrete analysis of fluctuation phenomena is unknown to us.}.
Secondly, diagrammatic methods, for reasons similar to those outlined
above, are ruled out in cases where the proximity effect is fully
established.

Important progress has been made by extending the scattering
formulation of transport in N-mesoscopic systems to the
SN-case~\cite{BTK,Takane,Lambert91,Beenakker92}.  
This approach made possible an efficient
calculation of both fluctuation and weak localization contributions to
various global transport properties of SN-systems \cite{Beenakker97}. 
Unlike the
quasiclassical formalism, however, the transfer matrix approach is
not microscopic. Instead, the different components of an SN-system are
treated as black boxes which are described in terms of
phenomenological stochastic scattering matrices.  This approach,
whilst extremely powerful in the analysis of global transport features
(the conductance say), cannot address problems that
necessitate a local and truly microscopic description. For example, it is not
suitable for the calculation of spectral fluctuations, both global and
local, the analysis of local currents, and so on.

The purpose of this paper is to introduce a theoretical approach to
the study of SN-systems which essentially represents a unification of
the above quasiclassical concepts with more recent field theoretical
methods developed to study N-mesoscopic fluctuations. As a result we
will obtain a modelling of SN-systems that treats mean and fluctuation
manifestations of the proximity effect on the same footing, thereby
revealing their common physical origin. This work represents the
development of ideas that we have originally presented in a short letter
\cite{AST}. 

Our starting point will be a
connection, recently identified, between quasiclassical equations for
Green functions on the one hand and supersymmetric nonlinear
$\sigma$-models on the other. As was shown by Muzykhantskii and
Khmelnitskii~\cite{Muz2}, the former can be regarded as the classical
equations of motions of the latter. In other words, the $\sigma$-model
formulation has been shown to provide a variational principle
associated to quasiclassics. So far, these connections have not been
exploited within their natural context, superconductivity.  To fill
this gap, we will demonstrate here that by embedding concepts of
quasiclassics into a field theoretical framework, one obtains a
flexible and fairly general theoretical tool to the analysis of
SN-systems. In particular, it will be straightforward to extend the
quasiclassical equations so as to account for the consequences of
time-reversal symmetry, the connections to perturbative diagrammatic
approaches will become clear, and -- most importantly -- the effective
action approach may be straightforwardly extended to the computation
of mesoscopic fluctuations. In doing so, it will become clear in which
way both the characteristic features of the Usadel Green function and
SN-mesoscopic fluctuations originate in the same basic mechanisms of
quantum interference.

In this paper the emphasis will be on the {\it construction} of the
approach, that is, most of its applications will be deferred to
forthcoming publications. However, in order to demonstrate the
practical use of the formalism we will consider at least one important
representative of mesoscopic fluctuation phenomena, namely {\it
  fluctuations in the quasi-particle spectrum}, in some detail: The
DoS of N-mesoscopic systems exhibits quantum
fluctuations around its disorder averaged mean value which may be
described in terms of various types of universal statistics. The
analogous question for SN-systems -- What types of statistics govern
the disorder induced fluctuation behaviour of the {\it proximity
  effect influenced} DoS? -- has not been answered so far.  Below we
will show the emergence of some kind of modified Wigner Dyson
statistics\cite{Mehta}, within the newly constructed formalism. A
concise presentation of both the field theory and its application to
SNS-spectral statistics is contained in Ref.~\cite{AST}.

The organization of the paper is as follows. In section
\ref{sec:andreev}, we review the basic microscopic
mechanism responsible for SN-quantum interference phenomena.  
In section \ref{sec:quasi}, we discuss the quasiclassical approach
to the computation of single particle Green functions. 
In section \ref{sec:beyondq}, we briefly review
the diagrammatic and
statistical scattering approaches as the only methods so far developed
to compute mesoscopic fluctuations.  In the central sections
\ref{sec:field}-\ref{sec:fluct}, we introduce the aforementioned
field theoretical framework. In section \ref{sec:field},
we derive the effective action for a
diffusive SN-structure in the form of a supersymmetric nonlinear 
sigma-model. In section \ref{sec:saddle}, 
we obtain the saddle-point equations of the action and examine their
solution for some simple geometries.
As mentioned above, these saddle-point equations, obtained by a
stationary phase analysis of the effective action, are 
none other than the quasiclassical
equations of motion. In section \ref{sec:fluct}, we 
address the central issue of this paper, the behaviour of fluctuations
around the saddle-point solutions. The action displays a
spontaneous breaking of symmetry, whose massless, or Goldstone modes
are the diffusion modes of the system. The interaction of the diffusion
modes is incorporated naturally within this formalism, despite their
strong modification due to the proximity effect, and 
leads to mesoscopic fluctuations. We calculate in this section the 
renormalization of the spectrum of a quasi-1D SNS junction
due to such fluctuations. We also demonstrate
the spectral statistics of the SN
structure to be described at low energies by a modified version of a
universal Wigner-Dyson, or
random matrix theory. The field theoretic formalism will also allow
us to examine the onset at higher energies of non-universal
corrections which serve to destroy the correlations described by such a
universal model. In section \ref{sec:discuss}, we conclude with a
discussion.

%
%
%
%
\section{Andreev Reflection and the Proximity Effect}
\label{sec:andreev}
Consider a normal metal at mesoscopic length scales, that is, scales
much less than both $L_\varphi$ and $L_T$, where $L_\varphi$ is the
dephasing length due to electron-electron interactions and
$L_T=(T/D)^{1/2}$ sets the scale at
which the quantum mechanical coherence is cut off by thermal smearing
effects. The interest of such mesoscopic materials stems from the fact
that their physical behaviour is strongly influenced by effects of
large scale quantum interference.  Such effects manifest themselves in
both a variety of fluctuation phenomena and (non-stochastic) quantum
corrections to physical observables.

At the same time, the physics of bulk superconductors is also
determined by mechanisms of macroscopic quantum coherence.  For
example, the Cooper pairs forming a superconducting condensate
represent two-electron states whose phase coherence extends over a
(possibly macroscopic) scale set by the superconducting coherence
length.

\begin{figure}
\[\begin{array}{ccc}
\epsfig{file=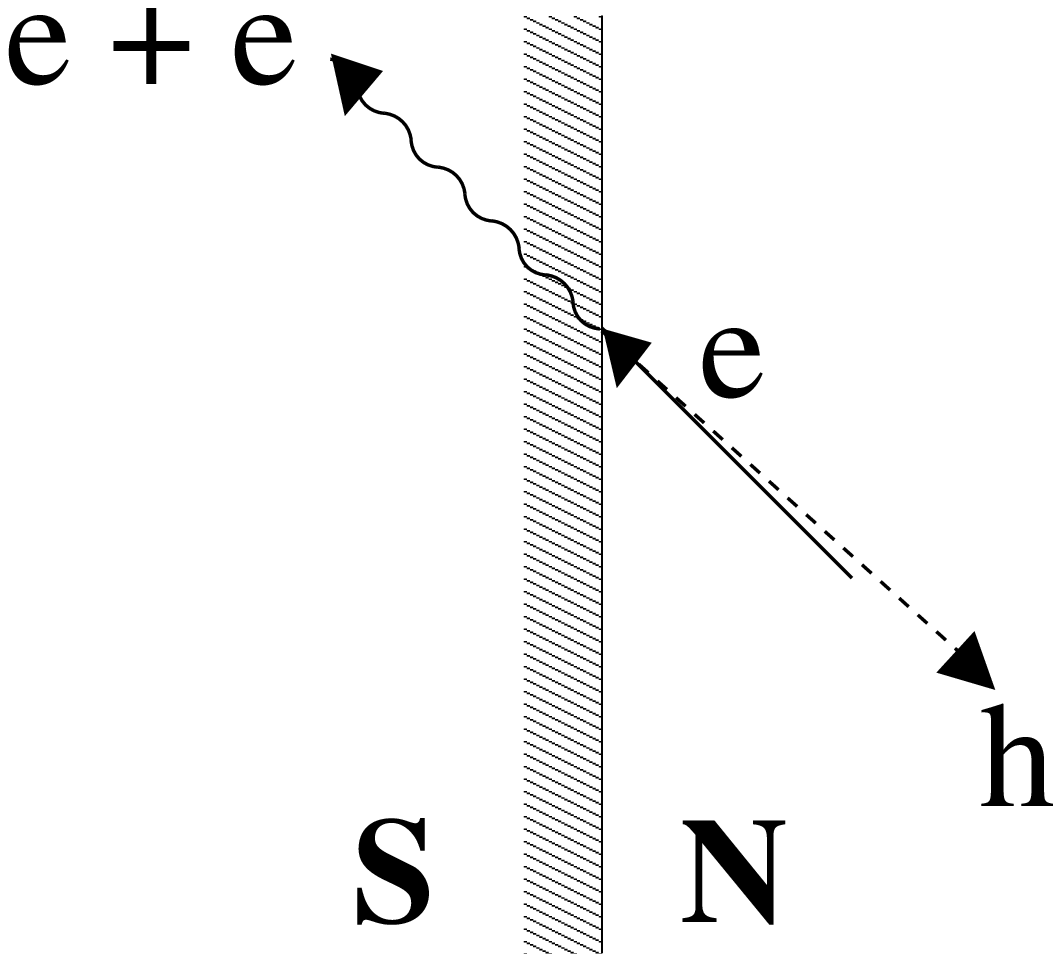,height=3.5cm} &\qquad \qquad&
\epsfig{file=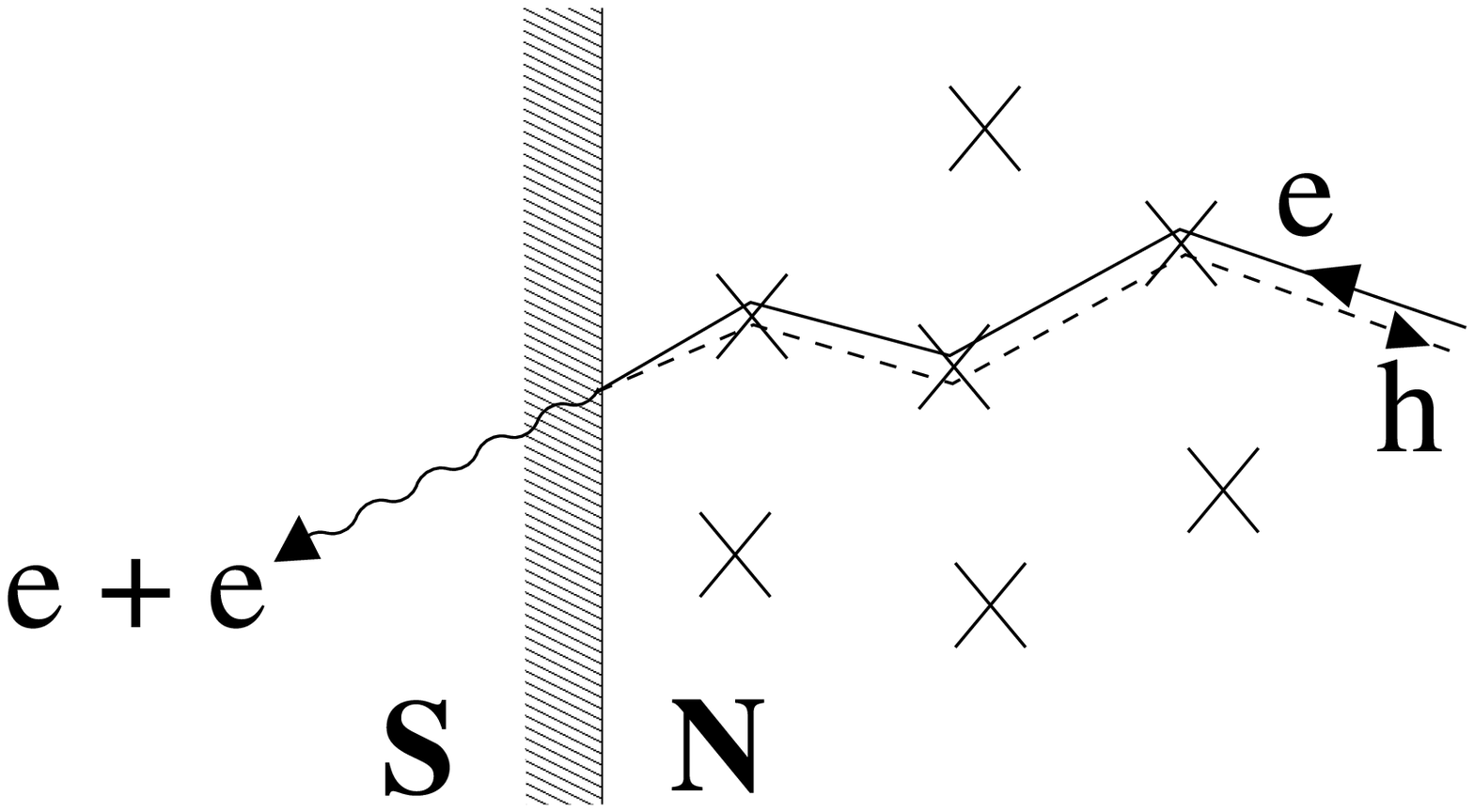,height=3.5cm} \\
{\em (a)} & &{\em (b)}
\end{array}\]
\caption{({\em a}) Andreev reflection of an electron at an SN
  interface,
and ({\em b}) a typical pair of Feynman paths that lead to a non-zero value of
  $\langle \psi^{\dagger}\psi^{\dagger}\rangle$. }
\label{fig:andreev}
\end{figure}

Given that the physics of both mesoscopic metals and bulk
superconductors is influenced by quantum coherence, it is appropriate
to expect that novel interference mechanisms arise when two systems of
this type are combined. This is indeed what happens and has led
to the continued interest in the physics of SN-hybrid systems.  A key
piece of information required for the understanding of large
scale manifestations of SN-quantum coherence is the manner in which normal
metals and superconductors exchange quantum phase information on a
{\it microscopic} level. The basic coupling mechanism between
a superconductor and normal metal is a form of interface scattering,
known as Andreev reflection~\cite{Andreev,Kulik}. In this process,
depicted in figure \ref{fig:andreev}({\em a}), an
electron at an energy below the superconducting gap, $\Delta$, strikes
the SN-interface. Due to its low excitation energy it represents a
forbidden quasi-particle state and is unable to enter the S-region.
Instead, however, it may be Andreev reflected off the boundary as a
hole. As a result two excess charges are left at the interface which
disappear into the superconducting condensate as a Cooper pair. 

The detailed physics of
Andreev scattering and its consequences for SN structures 
has been reviewed 
extensively in the literature
(see e.g. \cite{Beenakker97,Andreev,Tinkham,Imry,Likharev,Zaikin}).
Here we merely summarize some of its essential features that will be
of importance throughout:
\begin{itemize}
\item As opposed to ordinary specular reflection, Andreev-reflection
  represents a process of 'retro-reflection'. More precisely, 
 apart from a
  slight angular mismatch proportional to the excitation energy,
  $\epsilon$, of the electron above the Fermi energy, $\epsilon_F$,
the hole is reflected
  back along the trajectory of the incoming electron.
\item An electron with excitation energy $\epsilon$ 
  is scattered into a hole with energy $-\epsilon$.
\item The hole acquires a scattering phase $\pi/2-\varphi$, where
 $\varphi$ is the
  phase of the superconducting order parameter at the interface.
\end{itemize}
An important consequence of the existence of the Andreev scattering
mechanism is the formation of a Cooper-pair amplitude $\langle \langle
\psi^{\dagger} ({\bf x}) \psi^{\dagger} ({\bf x})\rangle\rangle$ in
the normal metal region. Here $\langle\langle \ldots\rangle \rangle$
not only represents the quantum mechanical expectation value but also
a {\it disorder average}. The creation of an average local pairing
amplitude can be heuristically understood from a simple semiclassical
consideration: Consider the creation of an electron somewhere at a point
${\bf x}$ inside a disordered metal adjacent to a superconductor (see
fig.~\ref{fig:andreev}({\em b})). Due to the presence of disorder, the electron
will propagate diffusively and may eventually strike the SN-interface
and be Andreev reflected. In general the newly created hole
may now diffuse along its own path. However, a particularly
interesting situation arises if the hole happens to propagate along
the path of the incoming electron back to the point of creation. As a
result we obtain a non-vanishing pairing field
amplitude $\langle \langle \psi^{\dagger} ({\bf x}) \psi^{\dagger} ({\bf
  x})\rangle\rangle$. The point is that during their propagation through
the disordered background both the incoming electron and the outgoing
hole accumulate a quantum mechanical scattering phase which
depends sensitively on microscopic details of the disorder. However,
owing to the fact that the two particles propagate along the same path
these phases cancel each other to a large extent. (For an excitation
energy $\epsilon=0$ the cancellation is, in fact, perfect.
For non-vanishing $\epsilon$ one obtains a phase mismatch
$\sim L_{\rm p}\epsilon /v_F \propto \epsilon L^2/ D$, where $L_{\rm
  p}$ is the length of the scattering path, $D$ the diffusion
constant, $v_F$ the Fermi velocity, $L$ the separation of ${\bf x}$
from the interface and we have used the fact that for diffusive
motion $L_{\rm p}/v_F \propto L^2/ D$.) Of course, while more generic path
pairs, where the electron and hole follow different paths, also contribute to
$\langle \psi^{\dagger} ({\bf x}) \psi^{\dagger} ({\bf x})\rangle$,
their contributions vanish upon disorder averaging due to their strong
phase dependence.

The non-vanishing of $\langle \langle \psi^{\dagger} ({\bf x})
\psi^{\dagger} ({\bf x})\rangle \rangle$ is the basic content of the
proximity effect. Besides its resilience against disorder, the pairing field
amplitude possesses a number of important features -- all of which are
related to the phase argument above -- that will be of importance for
all that follows:
\begin{itemize}
\item $\langle \langle \psi^{\dagger} ({\bf x}) \psi^{\dagger} ({\bf
    x})\rangle \rangle$ varies weakly as a function of ${\bf x}$. More
  precisely, it does not fluctuate on atomic scales but rather on
  scales set by $(D/\epsilon)^{1/2}$.
\item $\langle \langle \psi^{\dagger} ({\bf x}) \psi^{\dagger} ({\bf
    x})\rangle \rangle$ decays exponentially as a function of
  $\epsilon L^2/ D$. If either $T$ or the inverse dephasing time
  $\tau_\varphi^{-1}$ exceed $\epsilon$, the decay rate is set by
  these energy scales.
\item Quantitative expressions for the diffusive pairs of quantum
  paths entering the physics of the proximity effect are provided by
  so-called diffuson modes. Their meaning in the present context will
  become clear below.
\item The pairing field amplitude depends on the phases of the order
  parameters of the adjacent superconductors. If only a single
  superconducting terminal with constant phase, $\varphi$, is present, the phase
  dependence is simply $\sim \exp(i\varphi)$.
  In this case the phase is inessential
  and can be eliminated by means of a global gauge transformation.
  More interesting situations arise when more than one
  superconductor are present, in which case the phase sensitivity of
  the pairing amplitude provides the mechanism for the stationary
Josephson effect.
\end{itemize}

The non-vanishing of the pairing field amplitude heavily influences
the properties of the normal metal components of SN-systems.  Widely
known examples of proximity effect induced phenomena are the DC and AC
Josephson effect, which allow the possibility of supercurrent flows
through SNS-sandwiches. Another important phenomenon is the dependence
of the N-conductance on the phases of adjacent superconductors --
again triggered by the phase sensitivity of the proximity
amplitude~\cite{S+Nazarov,Petrashov}. However, as mentioned in the
introduction, the emphasis in this paper will be on a study of the
influence of the proximity effect on the single particle spectrum.

\subsection{Single-Particle Spectrum}

\begin{figure}
\begin{center} \epsfig{file=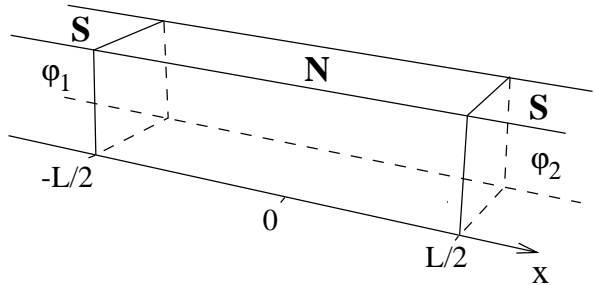,height=4cm}
\end{center}
\caption{The geometry of the SNS junction.}
\label{fig:sns}
\end{figure}

To understand the basic connection between the proximity effect and
the single particle spectrum, let us begin by considering the simple
geometry of an SNS-sandwich, shown in fig.~\ref{fig:sns}, 
where the N-layer is of width $L$,
of otherwise infinite extent and {\it clean}. This system was first
considered by Andreev~\cite{Andreev} who applied scattering theory to
the electron
wavefunction to show that the spectrum in the N region,
for trajectories at a fixed angle to the interface, is discrete
below the superconducting gap.  The 'Andreev levels' correspond to 
bound states with energies $\epsilon$ given by the quantization rule,
\begin{eqnarray}
\tan\left(\frac{\epsilon L}{|v_x|}
+{\rm sgn}(v_x)\frac{\Delta\varphi}{2}\right) 
= \frac{\sqrt{\Delta^2-\epsilon^2}}{\epsilon},
\label{qrule}
\end{eqnarray}
where 
$\Delta\varphi$ the phase difference across the junction and $v_x$
the component of the electron velocity normal to the interface. 
Eq.~(\ref{qrule}) then indicates an average Andreev level spacing of the
order of the inverse flight time, $|v_x|/L$, across the normal region.
A non-zero phase difference, $\Delta\varphi$, leads to
a shift in the levels so as to produce two separate
branches of the spectrum, for electrons and holes respectively.

%

To determine the observed spectrum, it is necessary to sum the pole
contributions to the DoS from bound states which arise from all
possible velocity directions, according to Eq.~(\ref{qrule}). Note
that non-zero contributions survive down to arbitrarily small energies
due to trajectories travelling close to parallel to the interface.
The energy dependence of the weights of these poles is so as to
produce a total DoS which is {\em linear} in energy $\epsilon$, at
energies $\epsilon \ll \Delta$.
 
The introduction of a finite concentration of impurities leads, upon
disorder averaging, to a smearing of the formerly sharp pole
structure. A less obvious outcome of this process is the
appearance of a sharp cut-off in the spectrum, the 'minigap', below
which the DoS vanishes entirely.  The minigap, $E_g$, is smaller than
the superconducting gap, $\Delta$, and depends on $\Delta\varphi$,
attaining its maximum for $\Delta\varphi=0$ and shrinking to zero as
$\Delta\varphi$ approaches $\pi$. In general $E_g$ depends on
$\Delta\varphi$ in a non-sinusoidal fashion \cite{Zhou}, although in
the limit of a short, diffusive junction, $L \ll \xi$, the dependence
becomes sinusoidal, $E_g = \Delta\cos(\Delta\varphi/2)$ \cite{KO}.

In the general case, the formation of a minigap in the metallic DoS
represents a highly non-trivial phenomenon. For example, it has been
shown \cite{Melsen} that the presence or absence of a gap depends on
the classical dynamical features of the metallic probe contacting the
superconductor: For samples with {\it integrable} classical dynamics
(such as the cubic system described above) there is no gap but rather
a DoS that vanishes linearly at the Fermi energy. The existence of
states all the way down to zero energy may be understood by means of
Bohr-Sommerfeld quantization arguments. By contrast, in systems with
{\it chaotic} classical dynamics, a gap opens whose magnitude may
depend on both the coupling strength to the superconductor and the
intrinsic classical transport time through the metal region.
 
\subsection{Failure of Semiclassics: Quantum Diffraction}
\label{sec:qdiff}
The prediction of a minigap provides a useful test for any analytic
approach to the physics of SN-systems: for example, an application of
standard approximation schemes to semiclassical formulae for the the
DoS fails to predict correctly a gap.

\begin{figure}
\[\begin{array}{ccc}
\epsfig{file=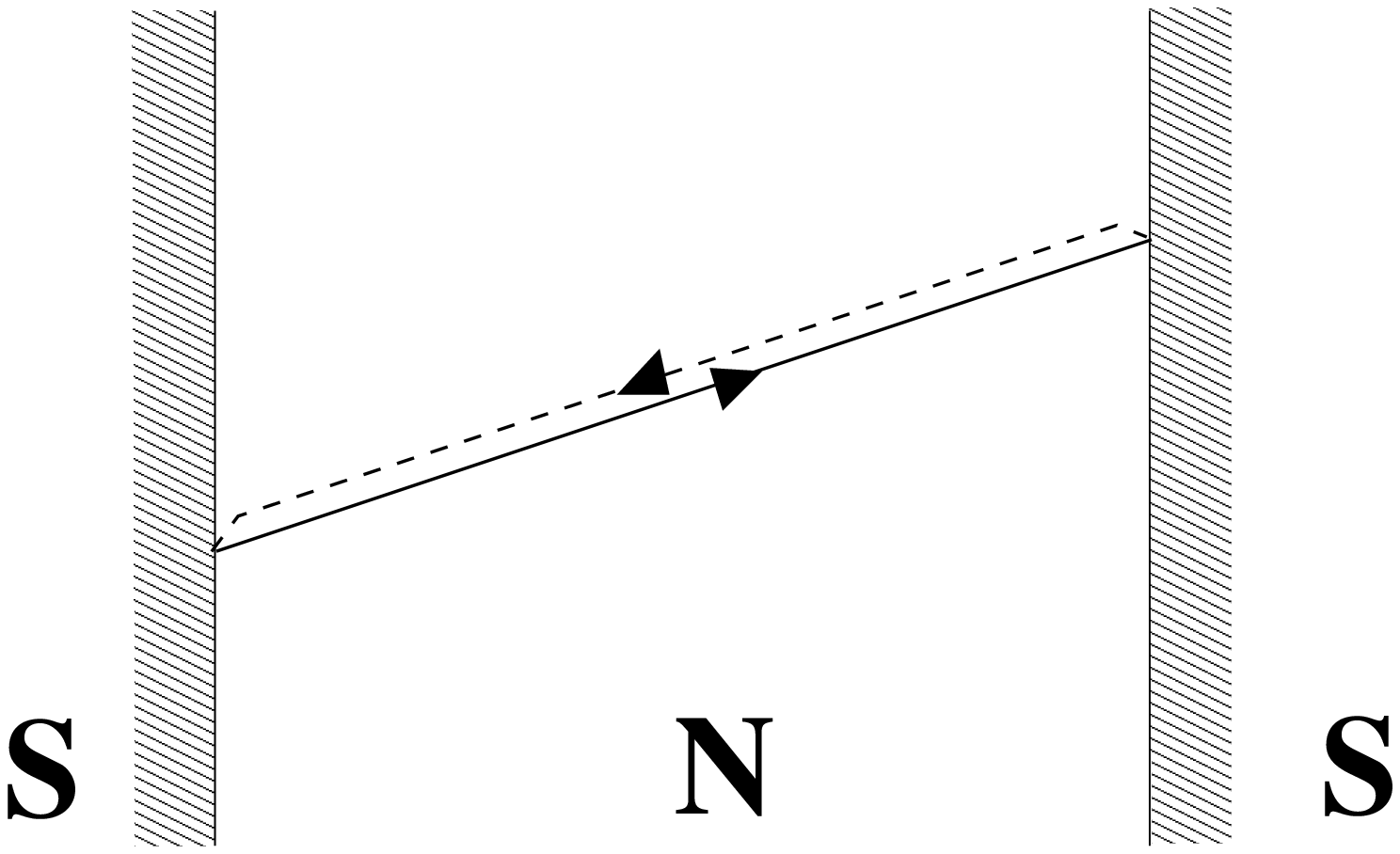,height=3.5cm} &\qquad &
\epsfig{file=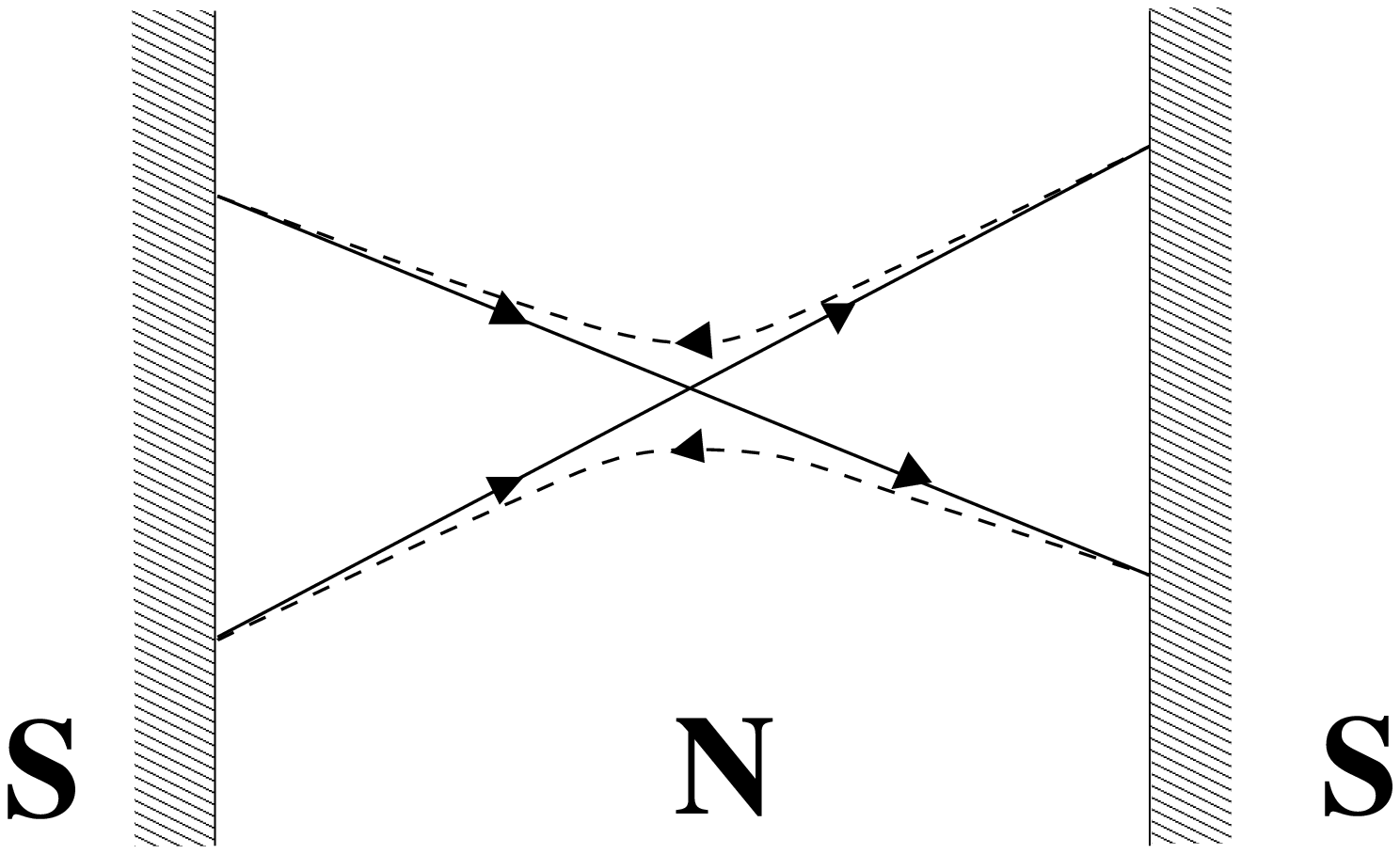,height=3.5cm} \\
{\em (a)} & &{\em (b)}
\end{array}\]
\caption{Trajectories that are ({\em a}) included and ({\em b}) not
  included within a semiclassical treatment.}
\label{fig:orbits}
\end{figure}

To
explain this point, let us consider the simple case of a diffusive
metallic cube of linear extension $L$ attached to a superconductor. In
this case the DoS gap is of width
$\simeq E_c$. In order to embed the previous heuristic path
arguments regarding the proximity effect into a quantitative
calculation of the proximity effect influenced DoS one might apply the
semiclassical Gutzwiller trace formula~\cite{Gutzwiller}, a powerful
computational tool often used in the analysis of the DoS and
correlations thereof. The Gutzwiller trace formula essentially states
that the averaged DoS can be obtained as a sum over all periodic
orbits with vanishing, or at least a disorder insensitive
action \cite{Argaman}. Small action contributions to trace
formulae are usually computed from the so-called diagonal
approximation. In the present context the diagonal approximation would
amount to counting Feynman paths such as the one depicted in
fig.~\ref{fig:orbits}({\em a}). (The smallness of the action of these
pairs of paths follows from the fact
that if the action of the electronic segment of the path (solid line)
has an energy dependent action $S(\epsilon)$, the action of the hole
segment (dashed line) will be $-S(-\epsilon)$. The two contributions
nearly cancel each other.) 

The problem with the standard diagonal approximation is that it fails
to predict a gap in the DoS, even if paths with multiple Andreev
scattering are taken into account.  In order to understand this
failure we first have to notice that small action path-pairs exist
which do not fall into the scope of the diagonal approximation (by
which we mean that they cannot be obtained as a superposition of two
identical segments, one electron- one hole-like).  A common feature of
these 'non-diagonal' path configurations is that they contain
'junction points' were the paths of electrons and holes split (cf.
fig.~\ref{fig:orbits}({\em b})). In order to understand the existence
of these splittings one has to keep in mind that the paths entering
the semiclassical picture do not correspond to rigorously defined
solutions of classical equations of motion but should rather be
thought of as objects that are smeared out (in configuration space)
over scales comparable with the Fermi wavelength~\cite{Aleiner}. As a
result two classically 'identical' paths may split and recombine at
some later stage, a process which is not accounted for by the diagonal
approximation. This splitting, as it is caused by the wave nature of
the electrons, is sometimes referred to as a quantum diffraction
phenomenon.

Note that the junctions appearing in fig.~\ref{fig:orbits}({\em b}) are
reminiscent of similar processes needed to generate weak localization
corrections to the conductance of normal metals~\cite{Aleiner}.
However, whereas weak localization corrections represent a correction
of ${\cal O}(g^{-1})$ to the classical conductance ($g\gg 1$ is the
dimensionless conductance), the diffraction corrections appearing in
the present context can by no means be regarded as small. In fact they
are as important as the leading order diagonal contributions which
implies that processes with up to an infinite number of 'junction points'
have to be taken into account. This fact not only explains the failure
of the diagonal approximation but also the difficulties encountered in
diagrammatic analyses of the proximity effect. The point is that each
of the 'legs' appearing in fig.~\ref{fig:orbits}{\em (b)} represents a
Cooperon. The perturbative summation of infinitely many Cooperons
represents a difficult problem, in particular in cases where the sytem
is truly extended in the sense that it cannot be treated within an
ergodic or zero mode approximation (for a perturbative analysis of the
zero-mode scenario, see Brouwer et al., \cite{Brouwer96b}).  
Fortunately there is an alternative
approach, the quasiclassical method reviewed below, which provides a
highly efficient tool for the effective summation of all interference
corrections contributing to the DoS and other physical observables.

%
%
%
%
\section{Quasiclassics}
\label{sec:quasi}

In order to prepare the discussion of quasiclassics we first need to
introduce its microscopic basis in general and the Gorkov Green
function in particular.

\subsection{The Gorkov Equations}
\label{sec:gorkov}
As long as interaction effects are neglected\footnote{In this paper the --
important -- r\^ole Coulomb interactions may play in SN-physics will
not be discussed.} 
the complete information on any SN-system is encoded
in its single particle Gorkov Green function, ${\cal
  G}^{\rm{r,a}}$. The latter is defined by the Gorkov
equations \cite{Gorkov}, whose matrix representation reads
\begin{eqnarray}
\label{gorkov}
&&\left(\matrix{\mu+\epsilon_\pm-
{1\over 2m}\left(\hat{\vc{p}}-
{e\over c}\vc{A}(\vc{r}_1)\right)^2-\hat V(\vc r_1)&
\Delta(\vc r_1)\cr -\Delta(\vc{r}_1)^*&\mu-\epsilon_\pm-{1\over
  2m}\left(\hat{\vc{p}}+{e\over c}\vc{A}(\vc{r}_1)\right)^2-V(\vc{r}_1) 
\cr}\right){\cal G}^{\rm r,a}(\vc{r}_1,\vc{r}_2)
\nonumber\\
&&\hspace{2.0cm}=\openone\delta^d(\vc{r}_1-\vc{r}_2),
\end{eqnarray}
where
\begin{eqnarray}
{\cal G}^{\rm r,a}=\left(\matrix{G^{\rm r,a}&F^{\rm r,a}
\cr {F^\dagger}^{\rm r,a}&
{G^\dagger}^{\rm r,a}}\right).
\end{eqnarray}
Here $G^{\rm r,a}$ and $F^{\rm r,a}$ represent the normal and
anomalous Green function, respectively, $\Delta(\vc{r})$
is the superconducting order parameter (which, in principle, has to be
determined self-consistently), $\epsilon_\pm=\epsilon\pm 
i0$, and $V$ is the impurity potential.

The equation above may be represented in a more convenient way by
introducing Pauli-matrices, $\sigma_i^{\rm ph}$, operating in the
two-component particle/hole space\footnote{
Recall the general definition of the Pauli matrices,
\begin{eqnarray*}
\sigma_1=\left(\matrix{0&1\cr 1&0\cr}\right),\quad 
\sigma_2=\left(\matrix{0&-i\cr i&0\cr}\right),\quad
\sigma_3=\left(\matrix{1&0\cr 0&-1\cr}\right).
\end{eqnarray*}}.  
Separating the order parameter into
its modulus, $|\Delta(\vc{r})|$, and phase, $\varphi(\vc{r})$,
Eq.~(\ref{gorkov}) takes the form
\begin{equation}
\left[\epsilon_F-{1\over 2m}\left(\hat{\vc{p}}-{e\over c}
\vc{A}(\vc r)\sigma_3^{\rm ph}\right)^2-
V(\vc{r})+\left(\hat{\Delta}(\vc{r}_1) +\epsilon_\pm\right)\sigma_3^{\rm ph}
\right]{\cal
  G}^{\rm r,a}(\vc{r}_1,\vc{r}_2)=\delta^d(\vc{r}_1-\vc{r}_2), 
\label{Gork2}
\end{equation}
where $\hat{\Delta} = \sigma_1^{\rm ph}|\Delta|
e^{-i\varphi(\vc{r}_1)\sigma_3^{\rm ph}}$.
The presence of an impurity potential, $V(\vc{r})$, makes
the solution of Eq.~(\ref{Gork2}) difficult.
However, as pointed out by Eilenberger \cite{Eilenberger} and Larkin
and Ovchinnikov \cite{Larkin}, a crucial simplification applies in the
case where the wavelength of the electrons, $\lambda_F$, is small
as compared to the characteristic scales over which the order parameter
$\Delta(\vc{r})$ and vector potential $\vc{A}(\vc{r})$ vary. Under
this condition one can resort to the 'quasiclassical approximation'.

\subsection{Quasiclassical Approximation}

The starting point of the quasiclassical approach is the observation
that the spatial structure of the Green function is comprised of
rapid oscillations, over a spatial scale of the Fermi wavelength,
modulated by a slowly fluctuating
background over longer scales. 
A quasiclassical analysis of the proximity effect involves
an averaging over the rapid variations of the Green function. 
At the same time, sufficient information is retained within the slower
modes of the Green function to provide a useful approximation to the
full consequences of the proximity effect. The advantage is a great
simplification of the corresponding kinetic equations.
As the derivation of the quasiclassical equations has been
reviewed extensively in the literature
(e.g. \cite{Rammer,Zaikin,Lambert98}), 
we restrict ourselves here
to a brief summary of the main results of the approach.

The quasiclassical, or Eilenberger, 
Green function, $g^{\rm r,a}(\vc{n},\vc{r})$, is
obtained from the Gorkov Green function by a) a Wigner transform,
b) an impurity average and c) an integral over the kinetic energy
variable. The precise definition reads
\begin{eqnarray}
g^{\rm r,a}(\vc{n},\vc{r}) = \frac{i}{\pi}\int d\xi\int d(\vc{r}_1-\vc{r}_2)
{\mathcal G}^{\rm r,a}(\vc{r}_1,\vc{r}_2) \exp(-i\vc{p}\cdot
(\vc{r}_1-\vc{r}_2)), 
\label{quasi}
\end{eqnarray}
where $\vc{r} =(\vc{r}_1+\vc{r}_2)/2$, $\xi_p = v_F(p-p_F)$,
$\vc n = \vc p/p$ and $p_F = m v_F$ is the Fermi momentum. The
application of this approximation to the Gorkov
equation, Eq.~(\ref{Gork2}) 
leads to the `Eilenberger equation' \cite{Eilenberger}:
\begin{eqnarray}
\vc{v}_F \cdot \vc{\nabla}_r g^{\rm r,a}(\vc n,\vc{r}) 
=i\left[\sigma_3^{\rm ph}(\epsilon_{\pm}+\hat{\Delta}(\vc{r}))
+\frac{i}{2\tau}\left\langle g^{\rm r,a}(\vc n',\vc r)
\right\rangle_{\vc n'},g^{\rm r,a}(\vc{n},\vc{r})\right],
\label{Eilen}
\end{eqnarray}
where $\tau$ is the elastic scattering time due to impurities. 
This equation essentially represents an expansion to
leading order in the ratio of $\lambda_F$ to
the scale of spatial variation of the slow modes of the Gorkov Green
function. 
It can be shown that the Eilenberger Green function obeys
the nonlinear normalization condition (see, for example, the discussions in
Refs.\cite{Zaitsev,Shelankov}) 
\begin{eqnarray}
g^{\rm r,a}(\vc n,\vc{r})^2=\openone.
\label{norm}
\end{eqnarray}
Eq.(\ref{Eilen}) represents an equation of Boltzmann-type which is
much simpler than the original Gorkov equation, but may still be difficult to
solve in general. However, significant further simplifications are
possible in the 'dirty limit'.
\subsection{Dirty Limit}
The 'dirty limit' is specified by the conditions $\ell=v_F \tau
\ll \xi$ (implying
that the dominant transport mechanism is diffusion) and
$\epsilon<\tau^{-1}$ (implying that 'time scales' $\epsilon^{-1}$ much
longer than the scattering time are explored). 
Under these
conditions, the dependence of the Green function on the angular
direction (represented by $\vc n$) is weak and one may 
expand in its first two spherical harmonics:
\begin{eqnarray}
g^{\rm r,a}(\vc{v}_F,\vc{r}) = g_0^{\rm r,a}(\vc{r})
+\vc n\cdot\vc{g}_1^{\rm r,a}(\vc{r})+\ldots,
\label{luder}
\end{eqnarray}
where 
$g_0^{\rm r,a}(\vc{r}) \gg \vc n \cdot \vc{g}_1^{\rm r,a}(\vc{r})$.
A systematic expansion of the Eilenberger equation in terms of $\vc
g_1$ then leads to a nonlinear and second-order equation for the
isotropic component,
\begin{equation}
D\vc{\nabla} (g_0^{\rm r,a}(\vc{r}) \vc{\nabla} g_0^{\rm r,a}(\vc{r})) +i
[\sigma_3^{\rm ph}(\epsilon_{\pm}+\hat{\Delta}(\vc{r})),
g_0^{\rm r,a}(\vc r)]=0,\hspace{0.5cm}g_0^{\rm r,a}(\vc r)^2=\openone,
\label{Usad3}
\end{equation}
known as the `Usadel equation' \cite{Usadel}.  
In order to specify a solution, one has to
supplement the equation with appropriate boundary conditions. The
analysis of the boundary behaviour of the equation becomes
somewhat technical. For this reason a more detailed discussion of the
boundary conditions
\begin{mathletters}
\begin{eqnarray}
\sigma(-) g_0 \partial_r g_0(-) = \sigma(+) g_0\partial_r g_0(+)&&,
\label{Kup1}\\
\sigma(\pm) g_0\partial_r g_0(\pm)=
\frac{G_T}{2}[g_0(+),g_0(-)],&&\qquad T\ll1,\label{Kup2}\\
 g_0(+)=g_0(-),&&\qquad T\simeq1,\label{Kup2s}
\end{eqnarray}
\end{mathletters}
has been made the subject of appendix \ref{sec:boundary}. Here $T\in
[0,1]$ is a measure for the transparency of the S/N interface,
$g_0(\pm)$ denotes the Green functions infinitesimally to the left
respectively right of the junction, and $G_T$ (cf. Eq.
(\ref{tunnel_cond})) is the tunnel conductance of the interface.

\subsection{Solution of the Usadel Equation}
Solutions of the Usadel equation with appropriate boundary conditions
have been derived for a vast number of geometries. At the same time
a systematic and
general solution scheme, based on an effective circuit theory, has been
constructed by Nazarov \cite{Nazarov}. Furthermore, a number of quasiclassical
predictions seem to be borne out well experimentally (see
e.g. \cite{Petrashov,Vegvar,Courtois,Charlat,Gueron,Kastalsky}.
In the field theoretic context
introduced below, the Usadel equation and its boundary condition will
reappear on the level of the mean field analysis in section
\ref{sec:saddle}. In addition explicit solutions for some simple geometries 
will be discussed in that section.

It is instructive to consider a diagrammatic
reinterpretation of the
Usadel solution in terms of a summation over real-space trajectories. 
Such a decomposition may be achieved by 
taking the solution of Eq. (\ref{Usad3}) in a normal region, with
the SN interface (at $x=0$, say) represented (for energies $\epsilon
\ll \Delta$) by a boundary, or source, term 
$\Delta \delta(x)$. An expansion of the solution in powers of $\Delta$
corresponds to a series of diffusive trajectories which include 
successive numbers of Andreev reflections at the interface.
Fig. \ref{fig:orbits} (b) provides an illustration of a trajectory
with four reflections, whilst 'starfish' trajectories with
arbitrary numbers of reflections are clearly possible. The reproduction of the
Usadel solution requires the summation of the full set of
trajectories, corresponding to diagrams to {\em all} orders. 
We see that the inherent difficulties of a diagrammatic treatment of 
the proximity effect, as discussed earlier, extend even to the
relatively simple task of reproducing quasiclassics.

%
%
%
%
\section{Beyond Quasiclassics}
\label{sec:beyondq}
The quasiclassical approach allows for the efficient
calculation of a wide spectrum of physical observables. In general, 
any observable that may be expressed in terms of a single
disorder averaged Green function is a candidate for quasiclassical
analysis. Note that, by extending the formalism so as to include
Keldysh-Green functions \cite{Keldysh}, 
observables that are commonly expressed in
terms of two-particle Green functions also become accessible. Important
examples are the conductance and other transport quantities. 
However, there are important classes of observables which do {\it not}
fall into the above category, thereby falling beyond the scope of
quasiclassics. The list of inaccessible quantities may be grouped roughly 
into four different categories:
\begin{itemize}
\item Physical observables which, by definition, are not expressible
  in terms of single particle Green functions. An example is given by
  the magnetic field dependence of the London penetration depth for
  a bulk superconductor, as studied by Larkin and Ovchinnikov
  \cite{Larkin}.  Its analysis requires the computation of the average
  of four momentum operators, $\langle p(0)p(t)p(0)p(t) \rangle$, a
  quantity that involves two- rather than one-particle Green
  functions. 
\item Higher order quantum interference corrections to single particle
  Green-functions. Very much like weak localization
  corrections of ${\cal O}(g^{-1})$ to the classical Drude conductance,
  the quasiclassical Green function represents the leading order term
  of a series expansion in powers of $g^{-1}$. The next to leading
  order contributions become important in cases where one is
  interested in quantum corrections of weak localization type or
  strong localization effects.
\item The quasiclassical approach (in its extension to include a
  Keldysh component) does not account for the corrections
  to two-particle Green functions due to the 
  interference of mutually time-reversed trajectories. An example
  property that is affected in this way is the conductivity, as we see
  below.
\item Most importantly, the quasiclassical approach does not allow
  for the study of mesoscopic fluctuation phenomena. The analysis of
  fluctuations requires the computation of disorder averages of two or
  more Green functions. Due to the impurity induced interference
  between different Green functions, quasiclassical techniques are
  inapplicable to these problems.
\end{itemize}
Given that these classes of problems 
cannot be addressed within
quasiclassics, it becomes necessary to seek some alternative approach.
Here, we briefly review two of perhaps the most
important theoretical tools currently established, namely, 
diagrammatics and the scattering matrix approach.

\subsection{Perturbative Diagrammatic Methods}
Microscopic diagrammatic methods have been applied to the study of
various SN-phenomena. The list of diagrammatic analyses includes 
computations of Josephson current fluctuations through SNS
junctions\cite{A+Spivak}, investigations of the phase sensitivity of
the N-conductance\cite{Spivak}, computations of universal
conductance fluctuations of SN-systems \cite{Takane} and more. 

\begin{figure}
\begin{eqnarray}
\begin{array}{ccc}
\epsfig{file=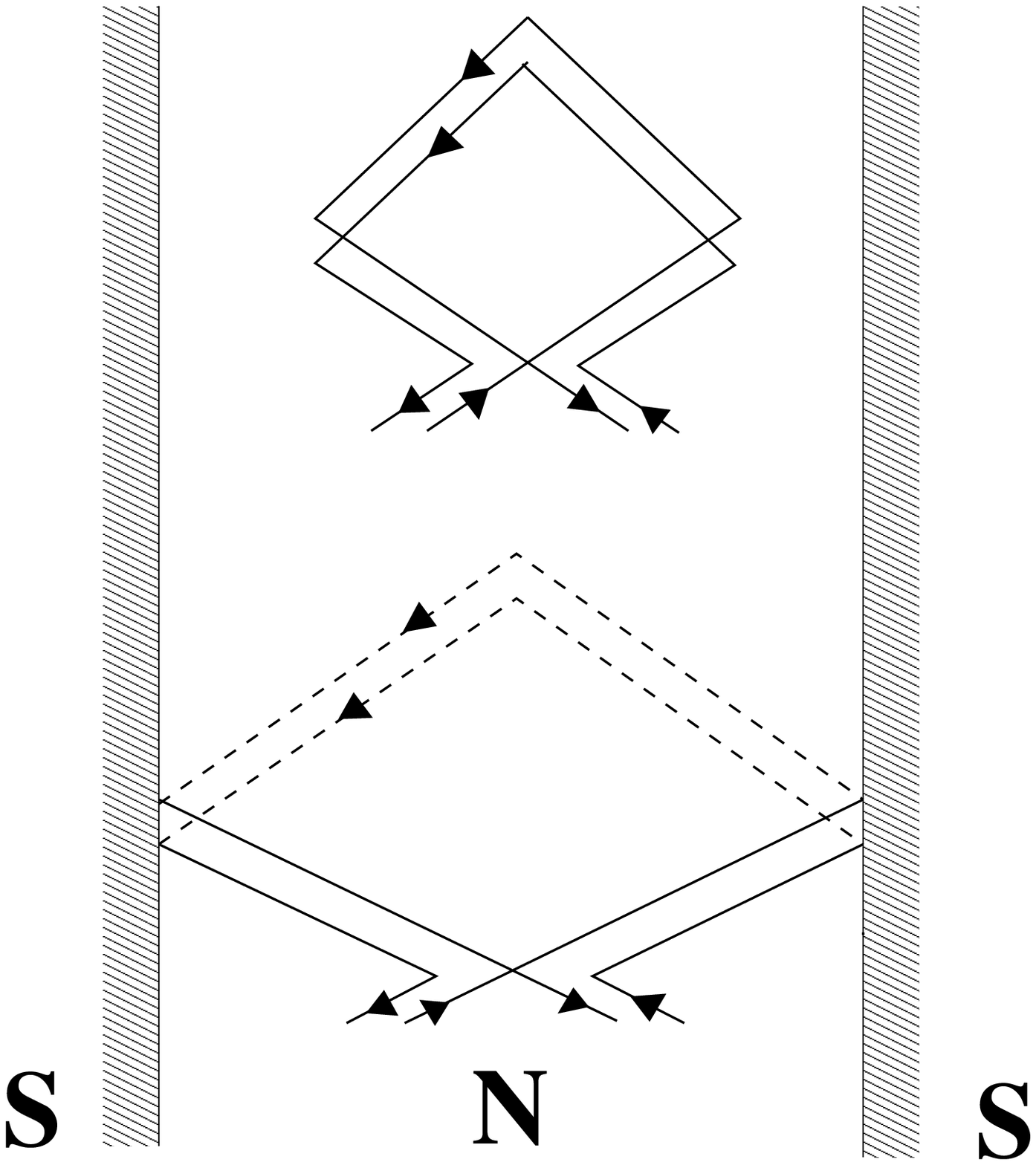,height=6cm}
&\qquad\qquad &
\epsfig{file=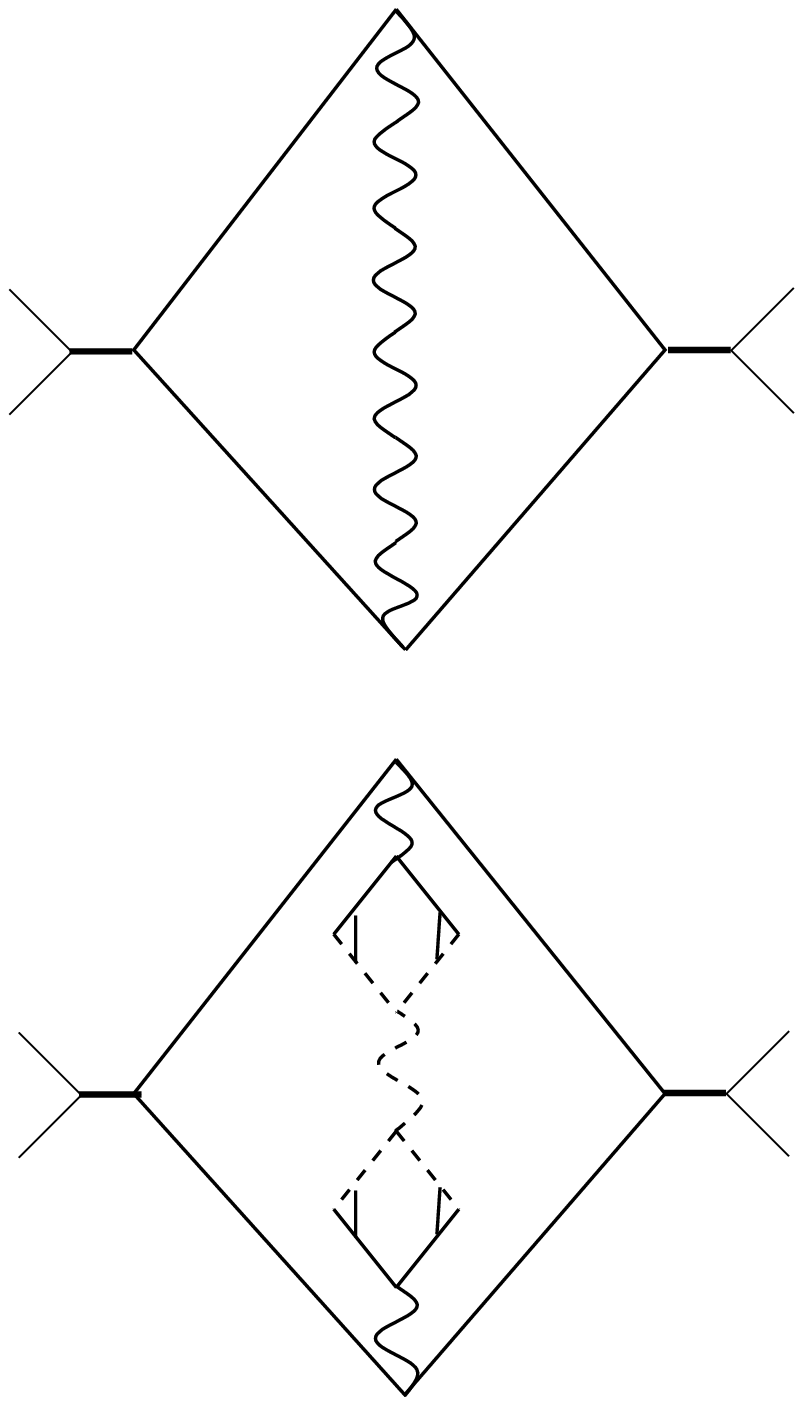,height=6cm} \\
({\em a}) & & ({\em b})
\end{array}
\nonumber\end{eqnarray}
\caption{({\em a}) shows the trajectories whose interference leads to the
first weak localization correction to the conductivity. ({\em b}) shows the 
corresponding diagrams, where the curvy line is the cooperon and the
triangles represent Andreev reflection processes.}
\label{fig:diff}
\end{figure}
\begin{figure}
\begin{center}
\epsfig{file=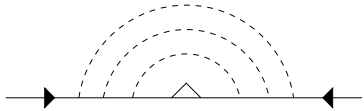,height=1.5cm}
\end{center}
\caption{First order renormalization of the Andreev scattering process
  necessary to include the proximity effect.}
\label{fig:renorm}
\end{figure}

The basic building block of diagrammatic analyses are the diffusion
modes both of diffuson and cooperon type. What makes these modes
different from their counterparts in pure N-systems is that they now
include Andreev scattering events, as represented in
fig.~\ref{fig:diff}. In the presence of the proximity effect, any of
the Andreev scattering vertices appearing in these modes is in turn to
be renormalized by further diffusion modes, as indicated in
fig.~\ref{fig:renorm}.  Note that these diagrams are the formal
representation of what in real space are the 'legs' of the
starfish-like structures appearing in fig.~\ref{fig:orbits}. The
problem with the diagrammatic approach is that in situations where the
proximity effect is {\it fully} established, the Andreev vertices
renormalize heavily, i.e. one has to sum self-consistently nested
series of the diagrams depicted in fig.~\ref{fig:renorm}. Another way
of putting this is to say that one has to perturbatively reconstruct
the solution of the Usadel equation, a difficult if not impossible
task.  In fact, an incomplete account of the proximity-induced
renormalization processes may lead to incorrect results: for example,
diagrammatic analyses of universal conductance fluctuations by Takane
et al. \cite{Takane} failed to reproduce correctly their surprising
insensitivity to external magnetic fields, as later demonstrated by
Brouwer et al. \cite{Beenakker97,Brouwer95}.  However, in cases where
the proximity effect is either suppressed or of secondary importance,
diagrammatic tools {\it can} be applied successfully to the study of
SN-systems.

In summary, it can be said that diagrammatics is applicable to the
perturbative analysis of SN-phenomena in cases with a weakly pronounced
proximity effect. As is usual with diagrammatic methods, non-perturbative
problems, such as localization, fine structure level statistics, and
so on, cannot be addressed.

\subsection{Multiple Scattering Formalism}
\label{sec:scatt}
Scattering theory provides a powerful theoretical tool for the
analysis of {\it quantum transport} through mesoscopic systems in
general, and SN-systems in particular. The reason for the efficiency
of the scattering theoretical formalism in the study of SN-systems is
not only its relative simplicity, but also the fact that the proximity
effect does not seem to lead to essential complications. Due to this
latter advantage, scattering theory has for a long time existed as 
the only tool for computing mesoscopic fluctuations under the
influence of the proximity effect. 

The starting points of the scattering approach are generalizations of the
standard multi-channel Landauer formulae for N-mesoscopic systems to
the SN-case. For example, in the case of a single N-sample attached
to a superconductor the conductance may be expressed as
\cite{BTK,Takane,Lambert91}
\begin{eqnarray}
  G_{NS} = \frac{2e^2}{h}{\rm tr}(1-S_{ee}^{\vphantom{\dagger}}
S_{ee}^{\dagger}+
  S_{he}^{\vphantom{\dagger}}S_{he}^{\dagger}) \label{scatt} 
\, =
  \frac{4e^2}{h}{\rm tr}S_{he}^{\vphantom{\dagger}}S_{he}^{\dagger},
\end{eqnarray}
where $S_{ee} (S_{eh})$ are the matrices describing the scattering
of electrons incoming from the normal metal to electrons
(holes). In a next step, the scattering matrices are expressed in
terms of a) the transmission matrices of the normal metal compound
(which are known in terms of their transmission eigenvalue
distribution functions \cite{Beenakker97,Beenakker93b}, 
and b) matrices describing the scattering off the superconductor.

The scattering theoretical approach is particularly powerful if the
observables of interest take the form of 'linear statistics', i.e.
quantities, $X$, that can be represented as
\[
X=\sum_n f(T_n),
\]
where $f$ is some function and $T_n$ is related to the $n$-th normal transmission
matrix eigenvalue. This is often but not always the case: for example,
in {\it time reversal invariant cases},
the conductance of the above
SN-system may be formulated as (see Beenakker, \cite{Beenakker92})
\begin{eqnarray}
G_{NS} = \frac{4e^2}{h}\sum_{n=1}^N \frac{T_n^2}{(2-T_n)^2}.
\end{eqnarray}
However, if time reversal invariance is broken, such a simple
representation is no longer possible and expressions involving not
only eigenvalues but also the diagonalizing matrices appear. It has
been shown by Brouwer et al. \cite{Brouwer96b} that, even under these
more complicated circumstances, scattering theory remains applicable.
What becomes necessary is to supplement the conventional transfer
matrix approach (by which we mean the derivation and solution of a
Fokker-Planck equation for the eigenvalues) by diagrammatic methods
accounting for the presence of the diagonalizing matrices.  Due to the
complex structure of the diagrammatic series appearing in the
SN-problem, only quantum dots (i.e. ergodic systems that can be
modeled in terms of a scattering matrix distributed through a single
Haar measure) rather than arrays thereof could be analyzed in this
way: we again encounter the notorious difficulties accompanying
perturbative approaches to SN-systems.

In passing we note that in some cases SN-quantum {\it dots}, chaotic
or disordered, can be modeled in terms of simple random matrix
theory. More precisely, random matrix techniques become
applicable if the proximity effect is suppressed. This happens if the
system is subject to a magnetic field (of the order of a few flux quanta
through the system), or if the phases of the adjacent
superconductors average to zero\footnote{The latter mechanism is
  rather subtle and it is not clear whether it can be realized
  in practice. The reason is that even minute phase fluctuations of
  ${\cal O}(g^{-1})$ invalidate the applicability of random matrix
  theory. However, systems with a phase-suppressed proximity effect are
  realized in {\it nature} as vortices in type II superconductors, see
  e.g Caroli et al. \cite{Caroli}. Here the vortex center has a 
  non-vanishing metallic DoS,
  an indication of proximity effect suppression.}. In spite of the
fact that the proximity effect is suppressed, the mechanism of Andreev
scattering remains active and manifests itself in the SN-random matrix
ensembles having symmetry properties that differ substantially from
the standard Wigner Dyson ensembles.

To summarize, statistical scattering theory represents a powerful tool
for the analysis of both mean and fluctuation characteristics of {\it
  global transport quantities}. Clearly, observables belonging to this
category are of outstanding importance from the experimental point of
view.  Nevertheless, problems remain for which one is interested in
observables that are local and/or microscopically defined. This
complementary class of quantities is inaccessible through phenomenological
scattering analyses. Thus at least one alternative theoretical tool
for the analysis of SN systems is called for.

%
%
%
%
\section{Field Theory for SN Systems}
\label{sec:field}

In the following central part of the paper we are going to introduce a
novel approach to the analysis of SN systems which is based on
field theoretical and, by construction, microscopic concepts. The
formalism will be applicable to observables that can be expressed in
terms of one or {\em products} of single particle
Green functions. If this criterion
it met, both mean values (including quantum corrections to
quasiclassical results) and mesoscopic fluctuations can be computed.
In a few exceptional cases, distribution functions can be obtained. It
will quickly turn out that the formalism is intimately related to each
of the approaches reviewed above: On the mean field level it
reproduces quasiclassics, perturbative fluctuations around the mean
field can be interpreted diagrammatically, and the connection to
scattering theory is established through general parallels between
$\sigma$-models and the transfer matrix approach \cite{B+Frahm}.

Prior to embarking on any kind of detailed discussion, let us briefly
outline the main conceptual steps of the construction of the field
theory:
\begin{enumerate}
\item Starting from the microscopic Gorkov Hamiltonian of an SN-system
  we will construct a generating functional for the disorder average
  of the product of a retarded and an advanced Green function.
  (Generalizations to products of more than two Green functions are
  straightforward.) The functional will be of a nonlinear $\sigma$-model
  type. Essentially it represents a supersymmetric extension of
  earlier (replicated) field theoretical approaches to bulk
  superconductors\cite{Oppermann}. In the present formulation, the
  order-parameter field will be imposed and not computed
  self-consistently (a common and mostly inessential simplification in
  the field of SN-systems\footnote{Self-consistent calculations of 
$\Delta(\vc{r})$
in mesoscopic SN-junctions, as achieved analytically by Zaikin \cite{Zaikin} in
  the clean case and numerically by several
  authors (e.g. Refs.~\cite{Kuprianov82,Bruder,Zhou98}) in the dirty case
  show that $\Delta$ is suppressed in the S
region near the interface.  It may also become non-vanishing in the N
region {\it if} an electron-electron interaction is included there. We
will neglect these effects as we do not anticipate that they would
have a significant impact on our results.}).
\item It will then turn out that the spatial inhomogeneity of the order
  parameter field poses a substantial problem: A straightforward
  perturbative evaluation of the field theory, by which we mean a
  perturbative expansion around any {\it spatially homogeneous}
  reference field configuration, is impossible. It goes without saying
  that this is nothing but the manifestation, in a field theoretical
  context, of the general perturbative
  difficulties characteristic to SN-systems.  
\item The way out will be to subject the field theory, prior to any
  perturbative manipulations, to a stationary phase analysis. Given
  what has been said under the previous item, it is clear that the saddle
  point configurations of the theory must be spatially inhomogeneous.
  More specifically it will turn out that the stationary phase
  equation of the theory is simply the Usadel equation. In other words,
  the quasiclassical approach to SN-systems turns out to be
  equivalent to the mean field level of the field theoretical formalism.
\item We then turn to the issue of fluctuations. Broadly speaking, two
  qualitatively different types of fluctuations will be encountered:
  Massive fluctuations around the mean field (giving rise to quantum
  corrections to quasiclassics) and a Goldstone mode. The latter will
  induce correlations between retarded and advanced Green functions
  and thereby mesoscopic fluctuations.
\end{enumerate}
To keep the discussion of the above hierarchy of construction steps
from being too abstract, 
the computation of correlations in the single particle
spectrum will serve as a concrete example of an application of the
theory.

Before turning to the actual construction of the field theory, it
should be noted that essential components of the machinery we are going to
discuss are not original, but have been introduced earlier: The general
supersymmetric field theoretical approach to N-mesoscopic systems has
been constructed by Efetov\cite{Efetov}. Oppermann\cite{Oppermann}, and
later Kravtsov and Oppermann\cite{Kravtsov}, introduced a
fermion-replicated $\sigma$-model description of bulk disordered
superconductors. As far as technical aspects are concerned, the
formalism we are deriving represents a supersymmetric extension of
Oppermann's model, tailored to the description of spatially
inhomogeneous structures. In its early stages, the construction of the
model follows a by now absolutely standard strategy. Essentially,
this requires an adaptation of Efetov's model to allow for the particular
structure of Gorkov Green functions. For this reason our presentation
of the early construction steps will be concise, but nevertheless self
contained.

\subsection{Field Integral and the Ensemble Average}

As is usual in the construction of field theories of mesoscopic systems,
the first construction step is to represent products of matrix
elements of Gorkov Green functions in terms of supersymmetric Gaussian
field integrals. To keep the discussion comparatively simple, 
we focus on the
case of two-point correlation functions $\langle {\cal
  G}^{\rm r}(\epsilon+\omega_+/2) {\cal G}^{\rm a}(\epsilon - \omega_+/2)
\rangle$ and choose, as a specific example, the quantity 
\begin{equation}
\label{correlation_function}
\left\langle {\rm tr}_{{\rm ph},\vc r}\left({\cal
   G}^{\rm r}(\epsilon+\omega_+/2)\right) {\rm tr}_{{\rm ph},\vc r}
  \left( {\cal G}^{\rm a}(\epsilon - \omega_+/2)\right)\right\rangle,
 \end{equation}
 which appears in the computation of the fluctuations of the DoS,
 \footnote{Note that the correlation function
   $R(\epsilon,\omega_+)$ differs in two respects from the analogous
   quantity in $N$-mesoscopic systems: (i) The mean DoS,
   $\langle \nu(\epsilon) \rangle$, will not, in general, be constant.
   Hence, $R(\epsilon,\omega_+)$ may explicitly depend on the center
   coordinate $\epsilon$, a fact that can be remedied by an unfolding
   procedure (see e.g. \cite{Bohigas}). (ii) as opposed
   to $N$-systems, correlation functions such as $\langle {\cal G}^{\rm r}
   {\cal G}^{\rm r} \rangle$ are non-trivial in the sense that they do
   not equal the product of averages. Both aspects (i) and (ii) will be
   commented on later in more detail.}
 \begin{equation}
 \label{R}
   R(\epsilon,\omega_+)=\frac{1}{\langle \nu(\epsilon) \rangle^2}
 \langle\delta \nu(\epsilon+\omega_+/2)\delta
 \nu(\epsilon-\omega_+/2)\rangle.
 \end{equation}
 Here $\delta \nu = \nu - \langle \nu \rangle$,
 $\nu(\epsilon) = -\frac{1}{2 \pi} {\rm Im\,} {\rm tr}_{{\rm ph},\vc
   r}\left({\cal G}^{\rm r}(\epsilon)\right)$ and ${\rm tr}_{{\rm
     ph},\vc r}$ denotes a trace with respect to 
 both position and particle hole index.

 In order to represent objects of this kind in terms of Gaussian field
 integrals, we first introduce a 16-component vector field, $\psi=
 \{\psi_{\lambda, s, \alpha, t} (\vc r) \},\; \lambda, s, \alpha, t
 =1,2$, with complex commuting (anticommuting) components $\alpha=1\,
 (\alpha=2)$. The significance of the two-valued indices $\lambda, s,
 \alpha, t$ is summarized in the following table:
 \begin{center}
   \begin{tabular}{{|c||c|c|}}\hline
 index & significance & abbreviation\\ \hline
 $\lambda$ & advanced/retarded & ar \\
 $s$       & particle/hole     & ph \\
 $\alpha$  & boson/fermion     & bf \\
 $t$       & time reversal     & tr \\ \hline
   \end{tabular}\\
 \bigskip
 \end{center}
 Apart from the index $s$ accounting for the $2 \times 2$ matrix
 structure of the Gorkov Green function, all other indices are standard
 in supersymmetric approaches to disordered systems. For a discussion
 of their significance we refer to Efetov's book\cite{Efetov}. We next
 introduce the action
 \begin{eqnarray}
 \label{psi_action}
 S = -i\int \bar{\psi}\left[\mu
 -\frac{1}{2m}\left(\hat{\vc{p}} - \frac{e}{c} \vc{A} \sigma_3^{\rm ph} \otimes
   \sigma_3^{\rm tr} \right)^2-V
 +\sigma_3^{\rm ph}\otimes
 \left(\tilde{\Delta}+\epsilon+
 {\omega_+\over 2}\sigma_3^{\rm ar}\right)
 \right]\psi ,
 \end{eqnarray}
 where 
 \begin{eqnarray}
 \label{delta_def}
 \tilde{\Delta}(\vc{r}) = \Delta \sigma_1^{\rm ph}
 \otimes\sigma_3^{\rm tr} 
 \exp(-i\varphi(\vc{r}) \sigma_3^{\rm ph} \otimes
 \sigma_3^{\rm tr}),
 \end{eqnarray}
 and the Pauli matrices $\sigma_i^{\rm ar}, \sigma_i^{\rm ph},
 \sigma_i^{\rm bf}$, and $\sigma_i^{\rm bf}, \; i=1,2,3$ operate in the two
 dimensional spaces of $\lambda,\alpha,s,t$-indices respectively. The
 fields $\psi$ and $\bar \psi$ are related to one another via 
 \[
 \bar \psi = \left(\sigma_1^{\rm tr} \otimes E_{11}^{\rm bf} +  i\sigma_2^{\rm
   tr} \otimes E_{22}^{\rm bf}\right)\psi,
 \]
 where the matrices $E_{ij}^{\rm x}, \; {\rm x} = {\rm ar, ph, bf, tr}$,
 are defined as 
 $\left(E_{ij}^{\rm x}\right)_{i'j'} = \delta_{ii'} \delta_{jj'}$, and
 the indices $i',j'$ refer to the space ${\rm x}$. 

 As in analogous theories of $N$-systems, the action, $S$, can be
 employed to represent correlation functions in terms of Gaussian field
 integrals. Specifically, the expression  
 (\ref{correlation_function}) takes the form
 \begin{eqnarray}
 \label{cf_psi}
 && \left\langle {\rm tr}_{{\rm ph},\vc r}\left({\cal G}^{\rm
       r}(\epsilon+\omega_+/2)\right) {\rm tr}_{{\rm ph},\vc r}\left(
     {\cal G}^{\rm a}(\epsilon -
     \omega_+/2)\right)\right\rangle=\nonumber\\ 
 &&\hspace{1.0cm}=
\frac{1}{16}\int {\cal D}(\psi,\bar \psi)e^{-L[\psi,\bar\psi]} \int
\bar\psi_{1} \sigma_3^{\rm bf}\otimes \sigma_3^{\rm ph} \psi_{1}
\;\int \bar\psi_{2} \sigma_3^{\rm bf}\otimes \sigma_3^{\rm ph}
\psi_{2},
\end{eqnarray}
where the indices on the $\psi$-fields refer to the ar-space, and all
other indices are summed over.

After this preparation -- which essentially has comprised of an extension of
existing supersymmetric framework to account for the additional
ph-structure -- we may
proceed in strict analogy to standard procedures:
\begin{itemize}
\item Firstly, averaging over Gaussian-distributed disorder, with the 
correlation function
  \[
  \langle V(\vc r) V(\vc r') \rangle = \frac{1}{2 \pi \nu_n \tau}
  \delta (\vc r - \vc r'),
  \]
  where $\nu_n$ denotes the DoS of a bulk N system, generates the
  quartic contribution to the action,
  \[
  S_{\rm int} = \frac{1}{4\pi\nu_n\tau}\int (\bar{\psi}\psi)^2.
  \]
\item Next, $S_{\rm int}$ is decoupled by introducing a $16\times16$
  Hubbard-Stratonovich matrix field,
  \[
  \exp(-S_{\rm int})=\int {\cal D}Q\;\exp\left[-\frac{1}{2\tau}\int
      \left(\bar{\psi}Q\psi-\frac{\pi\nu_n}{4}{\rm
          str}Q^2\right)\right],
  \]
  where 'str' denotes the supersymmetric extension of a matrix 
trace\footnote{We use the convention that ${\rm str}A = {\rm tr}A_{\rm bb}-
{\rm tr}A_{\rm ff}$.}.
\item In a third step we integrate over the $\psi$-fields to arrive at
  the $Q$-represented action,
  \begin{eqnarray}
    S[Q] = \frac{1}{2}{\rm str}_{r}\ln{\mathcal G}^{-1}
      -\frac{\pi \nu_n}{8\tau} \int  {\, \rm str}Q^2,
    \label{free2}
  \end{eqnarray}
  where
  \begin{eqnarray}
    {\mathcal G}^{-1} = \mu -\frac{1}{2m}\left(\hat{\vc{p}} -
      \frac{e}{c} \vc{A}
      \sigma_3^{\rm ph} \otimes \sigma_3^{\rm tr} \right)^2
    +\sigma_3^{\rm ph}\otimes \left(\tilde{\Delta}(\vc{r})+\epsilon+
      {\omega_+\over 2}\sigma_3^{\rm ar}\right)+\frac{i}{2\tau}Q
  \end{eqnarray}
\end{itemize}
and ${\rm str}_{r}$ denotes a trace extending over both internal and
spatial degrees of freedom.  The next step in the standard
construction of the $\sigma$-model is to subject the action $S[Q]$ to
a saddle point analysis. As we will see shortly, the presence of a
superconductor with spatially inhomogeneous order parameter will
necessitate substantial modifications to the standard mean field
scenario. In order to gain some insight into the structure of the mean
field equations, the next section will be devoted to the study of the
comparatively simple case of a bulk disordered superconductor.
However, prior to specializing the discussion, let us make some
general remarks as to the structure of the stationary phase equations.

Varying the action (\ref{free2}) with respect to $Q$ generates the stationary
phase equation,
\begin{eqnarray}
{\bar Q}(\vc{r})={i\over \pi\nu_n}{\cal G}(\vc{r},\vc{r}).
\label{sadd1}
\end{eqnarray}
Before embarking on the explicit computation of solutions to
Eq. (\ref{sadd1}) -- which represents a
16-dimensional matrix equation -- it is convenient
to elucidate further its structure. In
fact, a striking simplification arises from the fact that the
$Q$-independent part of the kernel ${\cal G}^{-1}$ is diagonal in all
indices save the ph-indices. 

In the {\it standard case}, that is, no superconductivity and hence no
ph-indices, the diagonality is complete and one may start out
from an ansatz for $\bar Q$ which is fully diagonal. Exploiting
the fact that the energy-difference between the Green functions, $\omega_+$, is
typically small in comparison with all other energy scales of the
system, one might be tempted to argue that ${\cal G}^{-1}$ is not
only diagonal but even approximately proportional to the unit matrix
in the internal indices. As a consequence one might assume that $\bar
Q$ is proportional to the unit matrix as well. This, however, is
wrong. The infinitesimal imaginary increment contained in $\omega_+$
gives rise to a phenomenon of spontaneous symmetry breaking in the
sense that the saddle point solution in the retarded sector differs
from the one in the advanced sector (see e.g. \cite{Wegner,Efetov}). More
precisely, the saddle point solution of the field theory for
$N$-systems reads\footnote{Note that 
$\sigma_3^{\rm ar}$ is commonly denoted by $\Lambda$ in the literature.} 
\[
\bar Q_{\rm N}=\sigma_3^{\rm ar}.
\]

What kind of modifications arise in the presence of
superconductivity? Firstly, the kernel ${\cal G}^{-1}$ is no longer
fully diagonal. It contains a non-trivial matrix structure in
ph-space\footnote{One might argue that the off-diagonality may be
  removed via a global unitary transformation, at least in the case of
  a spatially homogeneous order parameter. This, however, would
  contravene the conditions enforced by analyticity on
  the structure of the imaginary increments contained in the action,
  which, as we saw in the N-case, play a crucial role in determining
  the structure of the solution.}. Thus, the simplest ansatz for a
saddle point solution is diagonal in all indices save the ph-indices.
Secondly, we may expect that, as in the N-case, the structure of the
solution depends on the infinitesimal increments added to $\omega_+$.
An inspection of Eq. (\ref{psi_action}) tells us that the role played by
the matrix $\sigma_3^{\rm ar}$ in the standard case will now be taken
over by $\sigma_3^{\rm ph} \otimes \sigma_3^{\rm ar}$. Finally, the
solution in the N-case, $\bar Q_{\rm N}=\sigma_3^{\rm ar}$, was fully
universal in the sense that it did not depend on any energy
scale. Since the 'perturbations' arising in the action due to the
presence of the superconductor -- most notably the order parameter --
are weak in comparison with the dominant energy scale, $\mu$, it is
sensible to assume that the {\it eigenvalues} of the saddle point
solution will still be $\pm 1$.

Starting from the comparatively simple example of a bulk
superconductor\cite{Oppermann} we will next confirm these suppositions
by explicit calculation.

\subsection{Example: A Bulk Superconductor}
\label{sec:bulk}
In this section we assume the order parameter to be spatially
constant. The resulting saddle point equation has previously been
discussed in\cite{Oppermann}. Specifically, we assume $\tilde \Delta
(\vc r) \equiv -|\Delta| \sigma_2^{\rm ph}$, corresponding to a constant gauge
$\varphi=\pi/2$, and assume that the vector potential vanishes, 
$\vc{A}=0$.  Homogeneity of the gap function implies homogeneity
of the saddle-point solution ${\bar Q}(\vc{r})=\bar{Q}$. We next
introduce the ansatz
\[
\bar Q = \vc q \cdot \vc{\sigma}^{\rm ph},
\]
where $\vc{\sigma}^{\rm ph} = (\sigma^{\rm ph}_1,\sigma^{\rm ph}_2,
\sigma^{\rm ph}_3)^T$ is a vector of
Pauli matrices operating in ph-space. 
The components of the vector $\vc q$ are diagonal matrices
which are trivial in all but the ar-space,
\begin{eqnarray}
\vc q =\left(\begin{array}{cc}
\vc q_+&\\
&\vc q_-
    \end{array}\right)  \begin{array}{c}
{\rm r}\\
{\rm a}
  \end{array}.
\label{arstruct}
\end{eqnarray}
and normalized to unity, $\vc q_\pm^T \cdot \vc q_\pm =
\openone$.  To proceed it is convenient to adopt an elegant
parametrisation for the Green function suggested by
Eilenberger\cite{Eilenberger}. First note that in momentum
representation ${\cal G}^{-1}$ can be written as
\[
{\cal G}^{-1}(\vc p)= -\zeta(p) + i \vc w \cdot \vc{\sigma}^{\rm ph},
\] 
where 
\begin{eqnarray*}
\zeta(p)&=&\frac{p^2}{2m}-\mu,\\
\vc{w}&=&{\vc{q}\over 2\tau}-i\vc{r},\qquad \vc{r}=
(\epsilon+i0\sigma_3^{\rm ar})\hat{e}_3+i\Delta\hat{e}_1,
\end{eqnarray*}
and $\omega_+$ has been set to zero. Using the fact that $(\vc
w \cdot \vc{\sigma}^{\rm ph})( \vc w \cdot \vc{\sigma}^{\rm ph}) = 
\vc w \cdot \vc w \equiv |\vc
w|^2$, it is a straightforward matter to show that
\begin{eqnarray}
\bar{\cal G}(\vc{p})={1\over 2}\sum_{s=\pm 1}{\openone^{\rm
    ph}+s\hat{\vc{w}}\cdot
\vc{\sigma}^{\rm ph}\over -\zeta(p)+is|\vc{w}|},
\label{gbar}
\end{eqnarray}
where $\hat{\vc{w}}=\vc{w}/|\vc{w}|$.  Performing the trace over
momenta, and making use of the relations
\begin{equation}
\label{sum_vs_integral}
\sum_{\vc p} f\left(\frac{p^2}{2m}-\mu\right) \simeq {\rm Vol}\cdot\nu_n \int d\zeta
f(\zeta),
\end{equation}
where 'Vol' is the system volume, as well as
\begin{eqnarray}
\int d\zeta \sum_{s=\pm 1}{s^n\over -\zeta+is|\vc{w}|}=\cases{0&$n=0$,\cr
-2\pi i + {\cal O}({\rm max \,}(\Delta,\epsilon)/\mu) &$n=1$,\cr}
\label{grga}
\end{eqnarray}
the saddle point equation (\ref{sadd1}) takes the simple
form 
\begin{eqnarray*}
&&  {i\over \pi\nu_n}{\cal G}(\vc{r},\vc{r}) = {i\over \pi\nu_n {\,\rm
  Vol}}\sum_{\vc p} {\cal G}(\vc p) = \hat\vc w \cdot \vc \sigma
\\
&&\qquad \Leftrightarrow \vc{q}=\hat{\vc{w}}.
\end{eqnarray*}
which is solved by 
\begin{mathletters}
\begin{eqnarray}
\vc{q} &=& {i\vc{r} \over \sqrt{R}}, \\
R &=& |\Delta|^2-\epsilon^2. 
\label{qsc}
\end{eqnarray}
\end{mathletters}
To achieve the correct analyticity of $\vc{q}$, we take the branch
cut in the square root in $R$ to be along the negative real axis.
Introducing the parameterization
\begin{eqnarray}
\vc{q}=\left(\sin\theta_s,0,\cos\theta_s\right),
\label{param}
\end{eqnarray}
we find 
\begin{eqnarray}
\cos\theta_s&=&{|\epsilon|\over |R|^{1/2}}\cases{\sigma_3^{\rm ar}, &
$|\epsilon|> |\Delta|$,\cr 
-i {\rm sgn}(\epsilon), &$|\epsilon|<|\Delta|$,\cr}\nonumber \\
\sin\theta_s&=&{i|\Delta|\over |R|^{1/2}}\cases{\sigma_3^{\rm ar}, 
&$|\epsilon|>
|\Delta|$,\cr 
-i {\rm sgn}(\epsilon), &$|\epsilon|<|\Delta|$.\cr} 
\label{bulk}
\end{eqnarray}
In the particular limit of zero order parameter, the solution
collapses to a ph-diagonal one, $\vc{q}=\hat{e}_3\sigma_3^{\rm
  ar}\leadsto \bar Q = \sigma_3^{\rm ph} \otimes \sigma_3^{\rm ar}$. This
result is consistent with the conventional saddle-point equation of
the normal conductor (cf. the remarks made towards the end of the
preceding section): For vanishing order parameter, the particle/hole
extension simply generates two copies of the normal Hamiltonian.
The effect of a non-vanishing value of $\Delta$ is to induce a
rotation of the saddle-point in the ph sector. In the
extreme limit of $\epsilon\rightarrow0$ with a finite gap, the
saddle point rotates as
far as $\vc{q} = {\rm sgn}(\epsilon)\hat{e}_1$, or $\theta_s ={\rm
  sgn}(\epsilon)\pi/2$.

The local DoS can be computed from $\vc q$
as\footnote{To derive Eq.~(\ref{DoSexp}) one starts out from the functional
  representation of the local DoS, 
\[
{\rm tr}_{\rm ph}\langle {\cal G}^{\rm r}(\epsilon;\vc r,\vc r)\rangle=
\frac{1}{4}\int {\cal D}(\psi,\bar \psi) e^{-L[\psi,\bar\psi]}
  \bar\psi_{1}(\vc r) \sigma_3^{\rm bf}\otimes \sigma_3^{\rm ph} 
\psi_{1}(\vc r).
\]
After the Hubbard-Stratonovich transformation the preexponential terms
take the form of a functional expectation value $\sim \langle {\rm str\;
  }(Q \sigma_3^{\rm bf}\otimes \sigma_3^{\rm ph} \otimes E_{11}^{\rm
  ar})\rangle_Q$ which, upon evaluation in the saddle point approximation
leads to Eq.~(\ref{DoSexp}).}
\begin{eqnarray}
 \langle \nu(\vc{r}) \rangle = -\frac{1}{2\pi} {\rm Im \,}
  {\rm tr}_{\rm ph} \langle{\cal G}^{\rm r}(\epsilon;\vc r,\vc r)\rangle
 = \frac{\nu_n}{8}{\;\rm Re\;}\langle {\rm str\;
  }(Q \sigma_3^{\rm bf}\otimes \sigma_3^{\rm ph} \otimes E_{11}^{\rm
  ar})\rangle_Q\stackrel{Q\rightarrow \bar Q}{\longrightarrow} \nu_n
{\rm Re}[q_+(\vc{r})]_3.
\label{DoSexp}
\end{eqnarray} 
Here $\langle \dots \rangle_Q$ denotes the functional expectation
value of the field $Q$ and the final expression is obtained by
evaluating the functional integral at its saddle point value.
For the bulk case, this gives a superconducting DoS, $\nu_s$, of 
$\nu_s = \nu_n {\rm Re}\cos\theta_s$, which leads by Eq. (\ref{bulk}) to the
familiar BCS form, 
\[
\nu_s = \Theta(|\epsilon|-|\Delta|)
\frac{|\epsilon|}{(|\Delta|^2 -
\epsilon^2)^{1/2}}.
\]
Note that below the gap, 
$|\epsilon|<|\Delta|$, the loss of antisymmetric ar structure
of $\vc{q}$ in Eq. (\ref{bulk}) is reflected in a zero DoS.

We may next ask whether the saddle point solution, $\vc{q}$, is
unique. Anticipating that all saddle point configurations must share
the eigen{\it value}-structure of $\bar Q$, a general ansatz probing
the existence of alternative solutions reads
\begin{equation}
\bar Q \rightarrow T \bar Q T^{-1},
\label{alternative}
\end{equation}
where $T$ is some rotation matrix. In order to understand the
structure of the resulting saddle point {\it manifold}, it is
essential to appreciate that there are three parametrically different
energy scales in the problem.
\begin{itemize}
\item The asymptotically largest scale in the problem is the chemical
  potential. The existence of the large parameter $\mu/E$, where $E$
  may be any other scale involved, stabilizes the eigen{\it value}
  structure of the matrix $Q$. (This follows ultimately from the
  structure of the pole integral (\ref{grga}) - see also the
  corresponding discussion in \cite{Eilenberger}.) 
  Thus, as long as we
  are not interested in corrections of ${\cal O}(E/\mu)$,
  configurations of the type of Eq. (\ref{alternative}) exhaust the field
  integration domain of interest.
\item The next largest scales are $\Delta$ and/or $\epsilon$. Amongst
  the configurations parameterized by Eq. (\ref{alternative}), there are
  some that are massive in these parameters and some that are not.
\item The smallest scale is $\omega_+$. Its  physical significance is
  that of an inverse of the time scales at which we are probing
  correlations. With regard to correlation functions in $\omega_+$,
  field  fluctuations that are massive in the intermediate parameters
  $\epsilon$ and $\Delta$ are clearly of little if any relevance.
\end{itemize}
After these preparatory remarks, it should be clear that we will be
concerned mainly with fluctuation matrices $T$ that still lead to
solutions of the saddle point equation up to corrections $\propto
\omega_+$. An inspection of the action (\ref{free2}) tells us that
such $T$ have to fulfill the condition $[T,\vc \sigma^{\rm ph}]=0$. On
the other hand the matrices $T$ must not commute with $\bar Q$, as
otherwise they would be ineffective. Taking these two facts together,
we see that the most relevant fluctuations, $T$, around $\bar Q$, are
those that generate rotations in ar-space: For $\omega_+=0$ any
configurations fulfilling the above conditions again represent
solutions of the saddle point equation.  In other words, the $T$'s
generating these configurations are Goldstone modes.

Before extending the discussion to spatially inhomogeneous problems
and the impact of the existence of Goldstone modes, let us comment on
a mathematical aspect of the construction of the theory. We have seen
that, for $\Delta \not= 0$, the saddle point solution $\bar Q$
differed substantially from the standard saddle point $\sigma_3^{\rm
  ph} \otimes \sigma_3^{\rm ar}$ of the bulk metallic phase.  This
raises the question of whether the superconducting saddle point,
$\vc{q}\cdot\vc{\sigma}^{\rm ph}$, and $\sigma_3^{\rm ph} \otimes
\sigma_3^{\rm ar}$ are both contained in the field manifold of the
nonlinear $\sigma$-model. Clearly this question will be of concern as
soon as we deal with SN-{\it hybrid}-systems, and, in fact, the answer
is negative. However, it turns out that the problem can be surmounted
by analytic continuation of the parameter space spanning the $Q$-field
manifold. Since the discussion of the manipulations needed to access
both saddle points is inevitably somewhat technical, it has been
deferred to an appendix and may be skipped by readers who are not
interested in details of the formalism.

Our final objective will be to describe SN-systems rather than bulk
superconductors. What makes the analysis of SN-systems technically
more involved is that the action is manifestly inhomogeneous: In our
comparatively coarse modelling, where the order parameter is
externally imposed, the N-component of an SN-system will be
characterized by a sudden vanishing of the order parameter. Within a
more accurate description, based on a self-consistently determined
order parameter field, the situation would be even more complicated.
As in the preceding section, the SN-action
may also be subjected to a mean field approach. However, due to the imposed
inhomogeneities in the order parameter, the stationary phase
configurations will in general no longer be spatially uniform. At
first sight it may not be obvious how solutions to the spatially
inhomogeneous stationary phase problem may be found. The correct
strategy for this problem is again prescribed by the
existence of the threefold hierarchy of energy scales, discussed above.
Before going into details, let us give a brief account of the forthcoming
construction steps:

\begin{enumerate}
\item We will first employ the general ansatz 
   \begin{equation} 
\label{Q_general}
Q(\vc r) = T(\vc r) \sigma_3^{\rm ph} \otimes \sigma_3^{\rm ar} T^{-1} (\vc
r)
   \end{equation}
   akin to the one used in the bulk case. Eq. (\ref{Q_general})
   implies that the $Q$-matrices have an eigenvalue structure set by
   the matrix $\sigma_3^{\rm ph} \otimes \sigma_3^{\rm ar}$ thereby
   automatically solving the saddle point problem with regard to the
   highest energy scale $\mu$ (cf. the corresponding remarks made
   above).
\item In a second step we substitute the above ansatz into
  (\ref{free2}) and derive a 'medium-energy' effective action that
  contains no energies higher than $\epsilon$ and/or $\Delta$. Thirdly
  we will perform a second stationary phase analysis thereby
  determining those field configurations (\ref{Q_general})
  that extremise the medium energy action. 
\item By accounting for fluctuations around these configurations, we will
  finally be able to explore the low energy physics on scales $\omega_+$.
\end{enumerate}
Beginning with the derivation of the 'medium energy action', we
now formulate this program in more detail. 

\subsection{Gradient expansion and 'medium energy action'}

In constructing the effective medium energy action, it is again
crucial to exploit the existence of a scale separation in energy.
Anticipating that the relevant field configurations $T(\vc r)$
fluctuate weakly as a function of $\vc r$, we first borrow a
parameterization of the kinetic energy operator that has previously been
used in constructing the quasiclassical equations of
superconductivity\cite{Eilenberger}
\begin{equation}
  \label{kinetic_operator}
\frac{1}{2m}\left(\hat{\vc{p}} - \frac{e}{c} \vc{A}
      \sigma_3^{\rm ph} \otimes \sigma_3^{\rm tr} \right)^2 \simeq 
\frac{1}{2m}p^2+\frac{1}{m}\vc{p}\cdot \hat{\vc{q}},  
\end{equation}
where 
\[
\hat{\vc{q}} = -i\vc{\partial} - \frac{e}{c} \vc{A}
      \sigma_3^{\rm ph} \otimes \sigma_3^{\rm tr}.
\]
The idea behind Eq.~(\ref{kinetic_operator}) is that the slowly
fluctuating entities in the action, most notably $Q$, effectively
do not vary on scales of the Fermi wavelength. Thus, it makes
sense to decompose the momentum operator into two parts, $\hat{\vc{p}} =
\vc{p} + \hat{\vc{q}}$, where the 'fast' component, $\vc p$, has eigenvalues of
order of the Fermi momentum, $p_F$, and so can be treated as a c-number with
regard to slowly varying structures. The 'slow' component, $\hat{\vc{q}}$,
accounts for both slow spatial variations and the magnetic field. For
a more substantial discussion of (\ref{kinetic_operator}) we refer to
the original literature\cite{Eilenberger}.

We next substitute (\ref{Q_general}) and (\ref{kinetic_operator}) into
the action (\ref{free2}) to obtain 
\[
S[Q]=\frac{1}{2}{\rm str}_{r,p} {\rm ln\;}
\left[\underbrace{\mu-\frac{1}{2m}p^2
    +\frac{i}{2\tau}\sigma_3^{\rm ph}\otimes \sigma_3^{\rm
      ar}}_{\displaystyle (G_0)^{-1}} +
  V_1 + V_2 + T^{-1}[V_1 + V_2,T]\right],
\]
where
\[
V_1 = -\frac{1}{m}\vc{p} \cdot \vc{q},\;\;\;
V_2=\sigma_3^{\rm ph}\otimes \left(\tilde{\Delta}(\vc{r})+\epsilon+
      {\omega_+\over 2}\sigma_3^{\rm ar}\right)
\]
and ${\rm str}_{r,p}$ denotes a trace of internal indices, the 'fast'
$p$'s and the spatial coordinate. We next expand to lowest
non-vanishing order in the 'slow' operators $V_i$. As will become
clear from the structure of the resulting series, the small parameters
of the expansion are $(l/L)^2$, for $V_1$, and $\epsilon\tau, \Delta\tau$
for $V_2$.  Here $L$ denotes the typical scale at which the matrices $T$
fluctuate. To lowest order we obtain
\[
S[Q]\rightarrow \frac{1}{2}{\rm str}_{r,p}\left[
G_0 T^{-1}[V_2,T]\right]-
\frac{1}{4}{\rm str}_{r,p}\left[G_0(V_1+T^{-1}[V_1,T])\right]^2.
\]
Note that there is no contribution at first order in $V_1$. The reason
is that $V_1$ is linear in the vectorial fast momentum $\vc p$,
whilst $G_0$ is even in $\vc p$. Thus, the trace over fast momenta
annihilates this contribution.

To prepare the tracing out of the fast momenta, we next formulate some
useful identities describing the behaviour of the 'fast' Green
function $G_0$. The following relations can be proved
straightforwardly by explicitly performing the momentum integrations
(cf. Eq.~(\ref{sum_vs_integral})) and using some Pauli-matrix
algebra. 
\begin{itemize}
\item $G_0$ in its momentum representation may be written as
(cf. Ref.~\cite{Eilenberger} and Eq.(\ref{gbar}))
\[
G_{0}(p)=\frac{1}{2}\sum_{s=\pm 1}\frac{1+ s\sigma_3^{\rm ph}\otimes
  \sigma_3^{\rm ar}}
{-\zeta(p)
  + s \frac{i}{2\tau}}.
\]
\item The momentum trace over a single Green function becomes
\[
\sum_p G_0(p) = {\rm const.\;}\cdot \openone-i\pi\nu_n ({\rm Vol})
\sigma_3^{\rm ph}\otimes \sigma_3^{\rm ar}.
\]
\item Further, if operators $\hat{\vc{A}}$ and $\hat{\vc{B}}$ vary slowly in
 space, then
\[
\sum_p{\rm str\;}\left[G_0(p)\, \vc{p}\cdot\hat{\vc{A}}\, G_0(p)\,
  \vc{p}\cdot\hat{\vc{B}} \right] = 
\frac{m^2}{4}({\rm Vol}) 2\pi  D\nu_n \sum_s
{\rm str\;}\left[(1+s \sigma_3^{\rm ph})\hat{\vc{A}} \cdot (1-s
  \sigma_3^{\rm ph}) \hat{\vc{B}}
    \right].
\]
\end{itemize}
An application of these identities to the effective action above leads
to
\[
S \rightarrow S_1+S_2,
\]
where 
\[
S_1=-i \frac{\pi \nu_n}{2} \int \; {\rm str\;}\left[Q \sigma_3^{\rm
    ph}\otimes \left(\tilde{\Delta}(\vc{r})+\epsilon+
      {\omega_+\over 2}\sigma_3^{\rm ar}\right) \right],
\]
and, setting $\hat{\cal O}= \hat{\vc{q}} + T^{-1}
[\hat{\vc{q}}, T] = T^{-1} \hat{\vc{q}} T$, 
\[
S_2=-\frac{1}{4} \pi D \nu_n \int  \; {\rm
  str\;}\left[(1+\sigma_3^{\rm ph}
  \otimes\sigma_3^{\rm ar} ) \hat{\cal O}(1-\sigma_3^{\rm ph}
  \otimes\sigma_3^{\rm ar} ) \hat{\cal O}\right].
\] 
By using the identity
\begin{eqnarray*}
&&{\rm str\;}\left[(1+\sigma_3^{\rm ph} \otimes \sigma_3^{\rm ar}) \hat{\cal
    O}(1-\sigma_3^{\rm ph} \otimes \sigma_3^{\rm ar}) \hat{\cal
    O}\right] =
-\frac{1}{2} {\rm str\;}\left[[\hat q,Q]^2\right],\\
\end{eqnarray*}
we obtain
\[
S_2=-\frac{\pi D \nu_n}{8}  \int  \; {\rm
  str\;}\left[\tilde \partial Q\tilde \partial Q \right],
\]
where 
\begin{equation}
  \label{covariant}
  \tilde \vc{\partial}=\vc{\partial} - i\frac{e}{c} \vc{A}
      [\sigma_3^{\rm ph} \otimes \sigma_3^{\rm tr},\;]
\end{equation}
represents the covariant derivative. Putting everything together we obtain our
final result for the 'medium energy action'
\begin{eqnarray}
S[Q]=-{\pi\nu_n\over 8}\int {\rm str}\left[D(\widetilde{\partial}Q)^2
+4iQ\sigma_3^{\rm ph}\otimes\left(
\hat{\Delta}+
\epsilon+{\omega_+\over 2}\sigma_3^{\rm ar}\right)\right].
\label{ea}
\end{eqnarray}
Correlation functions are now obtained by substituting the action
(\ref{ea}) into a functional integral over all $Q$-fields subject to
the constraint $Q^2=\openone$:
\begin{eqnarray}
\int {\mathcal D}Q (\cdots) e^{-S[Q]},
\end{eqnarray}
where ${\mathcal D}Q$ denotes the invariant measure on the manifold
of matrices $Q^2=\openone$. For instance, the correlation function
(\ref{correlation_function}) takes the form (cf. Eq.~(\ref{cf_psi}))
\begin{eqnarray}
\label{cf_Q}
&&  \left\langle {\rm tr}_{{\rm ph},\vc r}\left({\cal G}^{\rm
        r}(\epsilon+\omega_+/2)\right) {\rm tr}_{{\rm ph},\vc r}\left(
      {\cal G}^{\rm a}(\epsilon -
        \omega_+/2)\right)\right\rangle=\nonumber\\
&&\hspace{1.0cm}= -\left(\frac{\pi \nu_n}{4}\right)^2 \int
  {\cal D}Q \int {\rm str\,}\left(Q E_{11}^{\rm ar}\sigma_3^{\rm
    bf}\otimes \sigma_3^{\rm ph}\right) \int {\rm str\,}\left(Q E_{22}^{\rm
        ar}\sigma_3^{\rm 
    bf}\otimes \sigma_3^{\rm ph}\right) \;e^{-S[Q]}.
\end{eqnarray}
In the limit $\Delta \rightarrow 0$, the functional integral
represents a superposition of two independent copies of normal metal
$\sigma$-models, one corresponding to the particle, the other to the
hole sector. Due to the decoupling of these two components, the
ph-structure becomes meaningless. For $\Delta \not=0$, the situation
is more interesting. Given the spatially inhomogeneous structure of
the action, an exact computation of correlation functions -- in the
sense of a complete integration over the nonlinear field manifold --
is in general not feasible. Under these circumstances, the first and
seemingly straightforward approach one might try is a perturbative
one. Yet, as usual with perturbative approaches in SN-physics,
straightforward perturbation theory does not work here.

To understand the origin of the difficulties let us introduce the
parameterization 
\begin{equation}
\label{exponential}
Q=e^W (\sigma_3^{\rm ph} \otimes \sigma_3^{\rm ar}) e^{-W},
\end{equation}
where $[W,\sigma_3^{\rm ph} \otimes \sigma_3^{\rm ar}]^+=0$.  The
parametrisation (\ref{exponential}) is frequently used in perturbative
analyses of the $\sigma$-model. In {\it standard} (N) applications of the
$\sigma$-model, its substitution into the action leads to a series
\begin{equation}
\label{W_action}
S[Q] \rightarrow S[W] \equiv S^{(2)}[W] +  S^{(4)}[W] + S^{(6)}[W] +
\dots
\end{equation}
where $S^{(2n)}[W]$ denotes the contribution of $2n$-th order in $W$.
The functional can then be evaluated by expanding perturbatively
around the second order contribution $\exp(-S^{(2)}[W])$ and applying
Wick's theorem. The resulting Taylor series converges rapidly due to
the fact that contributions $S^{(2n)}[W]$ are multiplied by large
coupling constants $g_n\gg1$ (all of which are parametrically of the
same order.) 

In the case $\Delta \not=0$ the situation is more complicated. The
point is that contributions $S^{(m)}[W]$, $m$ being {\it odd}, arise
from the perturbative expansion of the vertex $ \sim{\rm
  str}(Q\sigma_3^{\rm ph} \hat{\Delta})$. In particular, a non-vanishing
contribution of {\em first} order in $W$ emerges. The presence of this term
invalidates perturbation theory. (This can be seen formally by means
of a simple power counting argument: In the expansion of
$\exp(-S^{(1)}[W])$, each $W$ is multiplied by a large coupling
constant $g_1$. On the other hand, the Wick contraction of {\it two}
$W$'s, gives a factor $g_2^{-1}$. Thus, the perturbative series
expansion of $\exp(-S^{(1)}[W])$ diverges in the parameter
$g_1^2/g_2\gg 1$.).

To get some idea of how these problems can be overcome, it is helpful
to understand the physical origin of the divergences arising in
perturbation theory. To this end let us consider the disorder average
of the ph-block of the Gorkov Green function, ${\cal
  G}^{12}(\epsilon)$, as a simple example of a quantity that strongly
couples to the divergence of the $W$-perturbation series. When
expressed in terms of the functional integral, the average $\langle
{\cal G}^{12}(\epsilon) \rangle$ takes the form 
\[
\langle {\cal G}^{r,12}(\epsilon;\vc{r},\vc{r})\rangle 
\sim \left \langle {\rm
    str}\left(Q(\vc{r}) E_{11}^{\rm ar}\otimes E_{12}^{\rm ph}\otimes
    \sigma_3^{\rm bf}\right)\right\rangle_Q \sim \langle {\rm
    str}(W(\vc{r}) X)\rangle_W
\] 
where $\langle \dots \rangle_{Q/W}$ stands for functional integration
in the $Q$- respectively $W$-representation of the theory and $X$ is the
fixed matrix, $X=E_{11}^{\rm ar}\otimes E_{12}^{\rm ph}\otimes
    \sigma_3^{\rm bf}$.

Suppose now, we intended to evaluate this functional expectation value
perturbatively. To lowest order in $W$ we would obtain

\[
\langle {\cal G}^{r,12}(\epsilon;\vc{r},\vc{r}) \rangle 
\sim \left \langle {\rm str\,}(W(\vc r)
X){\,\rm str\,}\int d\vc{r}' (W(\vc{r}') \hat \Delta(\vc{r}'))
\right \rangle_0 \sim
\int d\vc{r}' K(\vc{r},\vc{r}')\Delta(\vc{r}'), 
\] 
where $\langle \dots \rangle_0$ stands for functional integration
weighted by the quadratic action $S^{(2)}[W]\sim \int {\rm str\,
  }(WK^{-1} W)$. The kernel, $K$, governing $S^{(2)}[W]$ is the
familiar diffusion pole $K^{-1} \sim D\partial^2 + i\epsilon$. Thus we
see that the first correction to the 'anomalous' Green function ${\cal
  G}^{12}(\epsilon)$ is proportional to the order parameter and --
owing to the spatially long ranged behaviour of the diffusion pole --
stretches far into the normal metal. Moreover, since the
characteristic energy scale of the diffusion pole is ${\rm
  max}(\epsilon, E_c)$, we see that the correction is of ${\cal
  O}(\Delta/{\rm max}(\epsilon, E_c))$, which, for sufficiently strong
order parameter/coupling between N and S, exceeds unity. Remembering
that the characteristic scale of the dimensionless quasiclassical
Green-function {\it is} unity we have to conclude that the
perturbation series resulting from a naive $W$-expansion of the
functional integral does not converge. To understand both the reason
for this failure and the particular form of the first order
correction, it is instructive to compare with the type of divergencies
that appear within diagrammatic perturbation theory.
In diagrammatic analyses, the correction to 
first order in $\Delta$ to the
anomalous Green function is indeed given by a single diffusion mode.
The real-space representation of this term has already appeared in
fig.~\ref{fig:andreev}({\em b}), while the corresponding diagrammatic
representation has also appeared as fig.~\ref{fig:renorm}.
Noting that this correction is only the first contribution to
what becomes upon summation a full representation of the 
proximity effect,
the origin of the problem becomes clear: By perturbatively
expanding around $\sigma_3^{\rm ph} \otimes \sigma_3^{\rm ar}$ we have
chosen the metallic limit of the Gorkov Green function, ${\cal
  G}^{a,r} = \pm i \sigma_3^{\rm ph}$ as a reference point. The superconductor,
however, drives the adjacent normal metal region to a state that is
far from conventionally metallic. In order to force a description of
the system in terms of a 
perturbation theory around the metallic limit, we have to
pay the price of an infinite perturbation series. Even worse, due to
the effective spatial inhomogeneity of each perturbative contribution,
arising from the space dependence of the diffuson, summation of the 
series becomes impossible.
To summarize, the
considerations above tell us that perturbative approaches based on
spatially constant reference configurations are doomed to fail and
that the origin of the problem is the spatially inhomogeneous
manifestation of the proximity effect.

%
%
%
%
\section{Stationary Phase Analysis}
\label{sec:saddle}

Given what has been said at the end of the previous section, the correct
strategy for overcoming the problems arising in perturbation theory 
becomes apparent: Prior to any perturbative attempts,
it is preferable to seek a solution to the 
stationary phase equation $\delta S[\bar
Q]/\delta \bar Q=0$. Due to the spatial inhomogeneity of the problem,
no uniform solutions $\bar Q = {\rm const.}$ will be found. Once a
solution $\bar Q$ has been obtained, both perturbative and
non-perturbative evaluation schemes may be safely {\it superimposed}.
The reason is that, by construction, no linear terms appear when the
action is expanded around $\bar Q$. 

We find it convenient to formulate the stationary phase analysis in a
gauge where the phase dependence of the order parameters at the
S/N-boundaries has been eliminated, at the expense of introducing a
vector potential in the bulk N-region. To be specific, we perform the
gauge transformation
\begin{eqnarray}
Q(\vc{r})\to \exp\left[{i\over 2}\left(-{\pi\over 2}+\varphi(\vc{r})
\right)\sigma_3^{\rm ph}\otimes\sigma_3^{\rm tr}\right] Q(\vc{r}) 
\exp\left[-{i\over 2}\left(-{\pi\over 2}
+\varphi(\vc{r})\right)\sigma_3^{\rm ph}\otimes
\sigma_3^{\rm tr}\right],
\end{eqnarray}
where, within the superconducting region, $\varphi(r)$ is equal to the phases
of the order parameter, as in Eq.~(\ref{delta_def}), and in the normal
region can be chosen arbitrarily. 
Inserting the gauge transformed field into the action we obtain
\begin{eqnarray}
S[Q]=-{\pi\nu_n\over 8}\int {\rm str}\left[D(\widetilde{\partial}Q)^2
+4iQ \vc{Y}\cdot \vc{\sigma}^{\rm ph}\right],
\end{eqnarray}
where
\begin{eqnarray}
\vc{Y} = i\Delta(\vc{r}) \hat{e}_1+\left(\epsilon
+\omega_+\sigma_3^{\rm ar}/2\right)\hat{e}_3,
\end{eqnarray}
and the vector potential entering the covariant derivative has
been transformed by
\begin{eqnarray}
\frac{e}{c} \vc{A}&\to& \frac{e}{c} \vc{A}-{1\over 2}\partial\varphi
\label{superfl}
\end{eqnarray}
Note that the right-hand side 
of Eq.~(\ref{superfl}) may be interpreted as $-2m$ times the 
superfluid velocity. 
To find the stationary phase equation, we introduce a small variation 
\[
Q\rightarrow e^{\delta W} Q e^{-\delta W}\simeq Q + [\delta W,Q]
\]
into the action and demand vanishing of the contribution at first
order in $\delta W$. A straightforward calculation then yields the
equation
\begin{eqnarray}
D\widetilde{\partial}_i(\bar{Q}\widetilde{\partial}_i \bar{Q})-
i\left[\bar{Q},\vc{Y}\cdot\vc{\sigma}^{\rm ph}\right] = 0.
\label{Usad}
\end{eqnarray}
The first step to analysing the general set of solutions of this equation
is to specify that the solution is as simple as possible,
i.e. as diagonal as possible. Noting that the equation is diagonal in
ar-,tr- and bf-space (in bf it is even trivial), we see that, as in
the case of a bulk superconductor, a sufficiently general ansatz reads
\begin{eqnarray}
  \label{usadel_ansatz}
&&\hspace{0.5cm} \vc q_\pm = 
\vc q_\pm^{\,1} E_{11}^{\rm tr} + \vc q_\pm^{\,2} E_{22}^{\rm tr},
\hspace{0.5cm} \vc q \cdot \vc 
  q = \openone_{\rm bf,ar,tr},
\end{eqnarray}
where $\vc q_\pm$ refers to the retarded/advanced blocks defined in 
Eq.~(\ref{arstruct})
and $\vc q_\pm^{\,1,2}$
are vectors of complex numbers (i.e. structureless in bf-space). The
restriction of the saddle point equation to the blocks $q_\pm\equiv
\vc q_\pm\cdot \vc\sigma^{\rm ph}$ now
reads
\begin{eqnarray}
D\widetilde{\partial}_i(q_\pm \widetilde{\partial}_i  q_\pm)-
i\left[ q_\pm,\vc{Y}_\pm \cdot\vc{\sigma}^{\rm ph}\right] = 0,
\label{Usad2}
\end{eqnarray}
where 
\begin{eqnarray*}
\vc{Y}_\pm = i\Delta(\vc{r}) \hat{e}_1+\left(\epsilon
\pm\omega_+/2\right)\hat{e}_3.
\end{eqnarray*}
Comparing (\ref{Usad2}) with (\ref{Usad3}) and identifying $ q_\pm$
with $g_0^\pm$, we identify the stationary
phase equation of the nonlinear $\sigma$-model as the Usadel equation.
One consequence is that we are immediately able to write down the
boundary conditions at a (perfectly transmitting) 
SN interface, by direct analogy with the
Kuprianov and Lukichev relations, eqns.~(\ref{Kup1}) and (\ref{Kup2}):
\begin{mathletters}
\begin{eqnarray}
\sigma q_\pm\partial_\perp q_\pm\Big|_{x+} &=& \sigma q_\pm\partial_\perp q_\pm
\Big|_{x-}, \\
\label{Kupbc}
q_\pm\Big|_{x+} &=& q_\pm \Big|_{x-},
\end{eqnarray}
\end{mathletters}
the former of which implies current conservation at the interface.
Here
$\partial_\perp$ is the normal derivative across the (planar) boundary, and
$x\pm$ denotes a space point infinitesimally to the left/right of
a boundary point $x$. Note that the normal state conductivities,
$\sigma$, in the left and right region may differ. 

In passing we note that, although we have stated above and will use
further the relations for a perfectly transmitting interface, in
general we {\em need not} keep to such a restriction within this
formalism. For instance, we could have equally well employed the
following conditions in the limit of {\em small} transparency, again
by analogy with eqns.~(\ref{Kup1}) and (\ref{Kup2}):
\begin{mathletters}
\begin{eqnarray}
\sigma q_\pm\partial_\perp q_\pm\Big|_{x+} &=& \sigma q_\pm\partial_\perp q_\pm
\Big|_{x-}, \\ 
&=& \frac{G_T}{2} [q_\pm \Big|_{x+},q_\pm \Big|_{x-}],
\end{eqnarray}
\end{mathletters}
where $G_T$ is the tunnel conductance of the junction, as given by
Eq.~(\ref{tunnel_cond}). By modelling the tunnel barriers
microscopically (as was done, e.g. in Ref.\cite{IWZ}) these boundary
conditions can in fact be re{\it derived} within the $\sigma$-model formalism.

The coincidence of the stationary phase equation of the $\sigma$-model
with the Usadel
equation, which, as mentioned in the introduction, 
was first observed by Muzykantskii and
Khmelnitskii \cite{Muz2} in a different context, 
has fundamental consequences for all that follows:
\begin{itemize}
\item Already on the level of the saddle point equation, the
  $\sigma$-model contains all the information that is otherwise
  obtained quasiclassically. In particular, solutions of the equation
  can in most cases of interest be {\it imported} from the extensive
  literature on Usadel equations for SN-systems.
\item The facts that a) the solutions of the Usadel equation for the
  retarded and the advanced Green function are different and b) the
  $\omega_+=0$ action is isotropic in ar-space, imply that we encounter a
  situation of spontaneous symmetry breaking: The mean field does not
  share the symmetries of the action and a Goldstone mode, operating
  in ar-space, will appear.
\end{itemize}
These two observations suffice to dictate the further strategy: One
first has to solve or import a solution of Eq.~(\ref{Usad}). Then the
solution $\bar Q = {\rm diag \,}(q_+,q_-)$ is substituted back into
the action and fluctuations around the block diagonal solution are
introduced via, $\bar Q \rightarrow T \bar Q T^{-1}$. In analysing
fluctuations, the main emphasis will be on exploring the r\^ole of the
Goldstone mode. However, before we proceed to the actual formulation of
this program, it is worthwhile to stay for a moment at the mean field
level and to acquire some familiarity with the Usadel equation and
the structure of its solution.

\begin{figure}
\begin{center} \epsfig{file=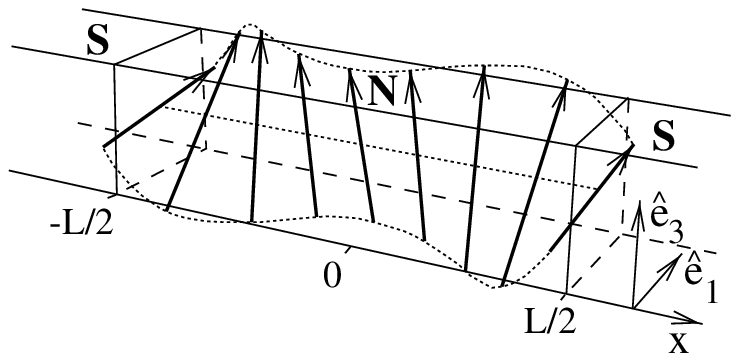,width=10cm}
\end{center}
\caption{
  Schematic plot of the retarded component (real part) of the  
  saddle-point solution, $\vc{q}_+$, for an SNS-junction with
  coincident phases of the order parameters.}
\label{fig:twist}
\end{figure}

We first note that the different 'sectors',  $\vc q_\pm^{\,1,2}$, of
the solution vector are not independent but rather connected to each
other via symmetry relations.
\begin{enumerate}
\item The general relation (cf. Eq. (\ref{gorkov}))
\begin{eqnarray}
{\mathcal G}^{\rm A}(\vc{r},\vc{r}') = \sigma_3^{\rm ph}
({\mathcal G}^{\rm R}(\vc{r}',\vc{r}))^{\dagger}\sigma_3^{\rm ph}
\end{eqnarray}
implies
\begin{eqnarray}
\vc{q}_- = {\rm diag}(1,1,-1)(\vc{q}_+)^*.
\label{arsymm}
\end{eqnarray}
\item Taking the transpose of the Usadel equation in the tr-sector 1,
  we obtain
\[\vc q^{\,2} = {\rm diag\,}(1,-1,1) \vc q^{\, 1}.
\]
\end{enumerate}
As for the spatial behaviour of the solution, some remarks may be made
in general. Deep in a superconducting region, the large $\sigma_2^{\rm
  ph}$-component of $\vc
Y$ enforces an approximate equality of $\vc q \simeq \hat e_1$.
Conversely, deep in a normal metal, $\vc q$ will be aligned with
$\hat e_3$. The Usadel equation describes a smooth interpolation
between these two limits, where the gradient term inhibits strong
spatial fluctuations (cf. fig.~\ref{fig:twist}).

In order to say more about the spatial structure of the solution to
the Usadel equation, we have to restrict the discussion to specific
examples. Here we will consider two simple prototype systems,
representative of the wide classes of systems with a) quasi infinite,
and b) compact normal metal region. Since we consider a quasi-1D
geometry in each case, we denote by $x$ the position variable perpendicular
to the interface.

\begin{figure}
\begin{center} \epsfig{file=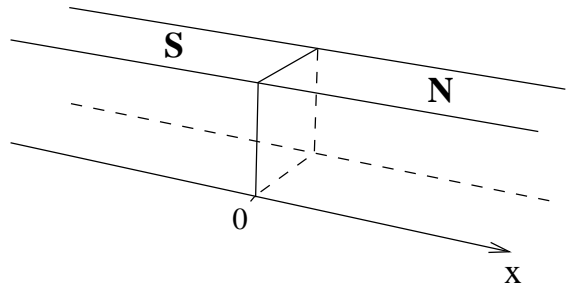,height=4cm}
\end{center}
\caption{The geometry of the infinite, quasi-1D SN junction.}
\label{fig:sn}
\end{figure}

\begin{figure}
\begin{center} \epsfig{file=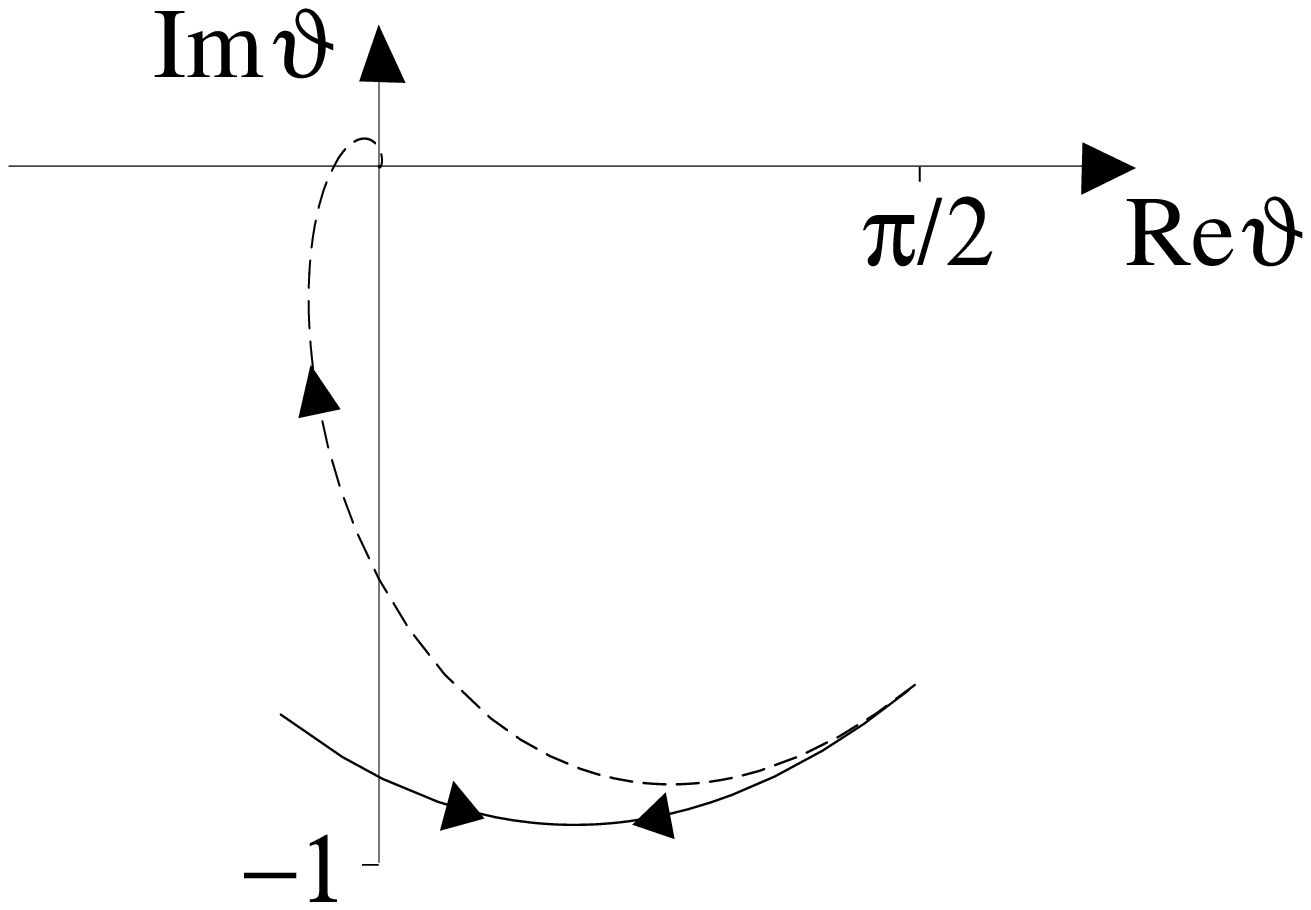,height=4.5cm}
\end{center}
\caption{Dotted line: the trajectory of the retarded component of 
$\theta$ as a function of 
  position for an infinite SN junction with coincident phases and
  $\epsilon/\Delta=0.6$ and $\gamma=0.1$. The trajectory starts at
  $x=0$ and approaches
  the origin as $x\to\infty$.
Solid line: The
  same for an SNS system, with $L/\xi = 5$, starting at $x=-L/2$ and
  ending at $x=L/2$. The trajectory reverses
  direction at the centre of the N region, $x=0$.}
\label{fig:spiral}
\end{figure}

{\it Infinite SN-junction:}
Consider the  model system shown in 
fig.~\ref{fig:sn}. The normal metal and superconductor
regions occupy $x>0$ and $x<0$ respectively, so that the gap function is 
modeled by $\Delta(x) = |\Delta| \Theta(-x)$. The system is quasi
one-dimensional in the sense that its constant width is comparable
with the elastic mean free path (i.e. there are many conducting
channels but no diffusive motion in the transverse direction.) We assume
that no external magnetic field is present. Furthermore, since there is
only one superconducting terminal, an elimination of the  phase of the
order parameter does not induce a gauge potential and we can globally
set $\vc A=0$.
The analytic solution of the
corresponding Usadel equation is reviewed in Appendix~\ref{app:sn}.
Due to the global absence of a vector potential, the spatial rotation of the
vector $\vc q$ takes place in the 1-3-plane only. Hence, it
can be parameterized as (cf. the analogous form for a bulk
superconductor, Eq.(\ref{param}))
\[
\vc{q}(x) = (\sin\theta(x),0,\cos\theta(x)).
\]
In fig.~\ref{fig:spiral} we have plotted the curve in the complex
plane that is traced out by $\theta(x)$ upon variation of $x$.

\begin{figure}
\begin{center} \epsfig{file=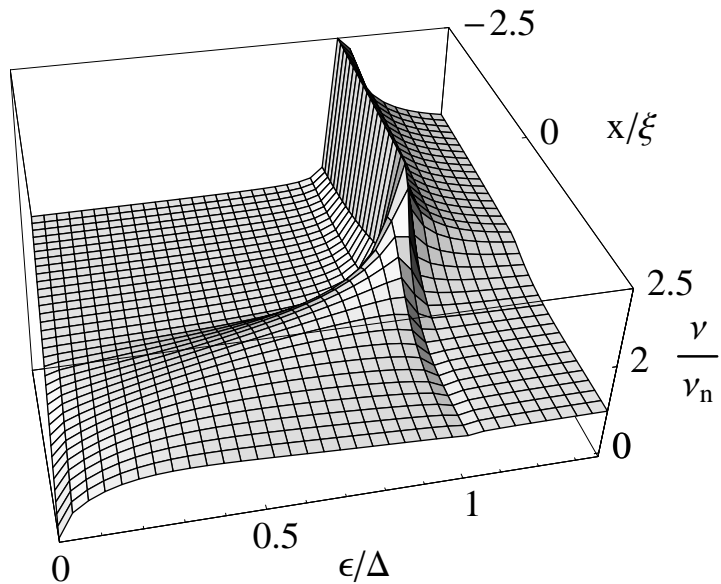,height=8cm}
\end{center}
\caption{Local DoS of the infinite SN junction
as a function of both energy and position for $\gamma=0.1$.}
\label{fig:3Dsn}
\end{figure}

Fig.~\ref{fig:3Dsn} shows the local DoS $\nu(x) = \nu_n {\rm
  Re}\cos\theta(x)$ (cf. Eq.~(\ref{DoSexp})) obtained for a particular
value of the material parameter $\gamma=\nu_n
\sqrt{D_n}/(\nu_s \sqrt{D_s})$, where $D_{n,s}$ are the diffusion constants
in the N,S-region. Note that the deeper one proceeds
into the S-region, the more the DoS approaches the characteristic
BCS-form. Due to the proximity effect, the structure of the subgap
($\epsilon<|\Delta|$) DoS in the N-region remains non-trivial. Only in
the asymptotic limit $x\rightarrow -\infty$, the region of suppressed
DoS shrinks to zero and 'normal' behaviour is restored. More
  precisely, substantial alteration of the DoS induced by the
  proximity effect is restricted spatially to a region of several diffusion
  lengths from the interface into the normal metal, or several
  (dirty) superconducting coherence lengths into the superconductor. 

\begin{figure}
\begin{center} \epsfig{file=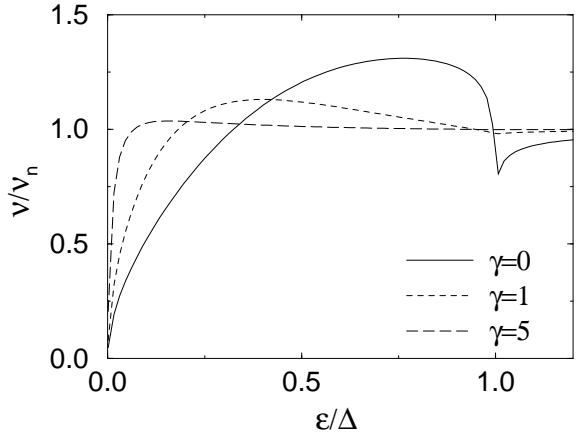,height=6cm}
\end{center}
\caption{The effect of $\gamma$ on the energy dependence of the
  DoS, at $x=1.5\xi$.}
\label{fig:gammasn}
\end{figure}

Fig.~\ref{fig:gammasn} shows how variation in $\gamma$ affects the
DoS. In particular, we may take the `rigid' limit $\gamma\rightarrow
0$, for which the bulk superconducting value of the Usadel angle is
imposed at the interface,
and retain a non-trivial
structure in the spectrum.

For a more comprehensive discussion of the local DoS of the 
system above we refer the reader to e.g. Ref.\cite{Bruder}.

{\it SNS-junction:} As an example representative for the class of
SN-systems with compact metallic region we next discuss the quasi
one-dimensional SNS-junction displayed in fig.~\ref{fig:sns}.
As opposed to the SN-system, the physics of the SNS-system {\it
  does} depend on the phases of the order parameters in the
superconducting terminals. For this reason SNS-systems with an in-built
possibility to tune the phases of the superconductors are sometimes
referred to as Andreev interferometers. In Appendix \ref{app:sn} the
quantitative solution of the Usadel equation is discussed explicitly
for the comparatively simple case $\Delta\varphi\equiv \varphi_1 -\varphi_2
=0$. 

An analytic solution to the general case, $\Delta\varphi\neq0$, is
also possible, although cumbersome. In the limits of a very short ($L
\ll L_{\epsilon}$) and very long ($L \gg L_{\epsilon}$) junction,
separate approximation schemes to the general solution may be
employed: the short junction case has been treated by Kulik and
Omelyanchuk \cite{KO} (for $L \ll \xi, L_{\epsilon}$) and Likharev
\cite{Likharev,Likharev76} (for $\xi \ll L \ll L_{\epsilon}$), while
the long junction case ($L \gg \xi,L_{\epsilon}$) has been treated by
Zaikin and Zarkov \cite{Zaikin,Z+Zarkov}.  At this point we restrict
ourselves to a discussion of a few qualitative characteristics of the
solutions obtained in various regimes.

The most conspicuous feature of the system is the appearance of a
minigap, $E_g$. As mentioned in section \ref{sec:andreev}, the precise
form of the gap depends on the phase difference $\Delta \varphi$. Its
maximum width is taken for $\delta \varphi=0$, while this width tends
to $\Delta$ and $E_c$ for short and long (as defined above) junctions,
respectively.  With growing phase difference, $E_g$ shrinks until, at
$\Delta \varphi=\pi$, it closes altogether\cite{Zhou} (apart from a
'microgap' of the width of the single particle level
spacing\cite{A+Zirn} -- the latter is induced by the general
phenomenon of level repulsion in disordered metals).

\begin{figure}
\begin{center} \epsfig{file=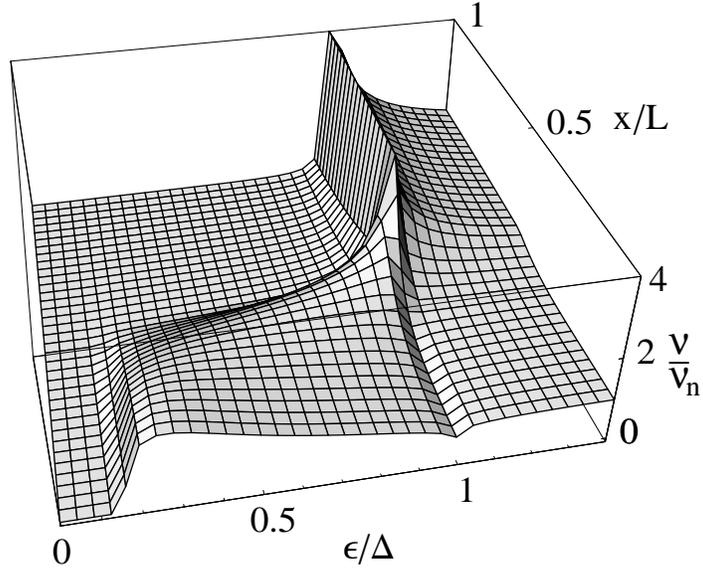,height=8cm}
\end{center}
\caption{The local DoS for the SNS-junction
as a function of both energy and position, for 
$L/\xi=5$, $\gamma=0.1$ and $\varphi=0$.}
\label{fig:3Dsns}
\end{figure}

We take as illustrative the case of an SNS-junction with coincident
phases, and of arbitrary width.  In this case, the solution for
$\vc{q}$ again lies in the 1-3-plane and is
parametrized, as before, by the angle $\cos\theta=\vc{q}\cdot
\hat{e}_3$.  Fig.~\ref{fig:spiral} shows the trajectory of the
retarded $\theta$ in the complex plane as a function of
position, and fig.~\ref{fig:3Dsns} 
shows the local DoS as a function
of both energy and position. Note, as compared to the infinite SN
case (fig.~\ref{fig:3Dsn}), the appearance of the minigap, $E_g$,
which is below the superconducting gap and independent of position.

\begin{figure}
\begin{center} \epsfig{file=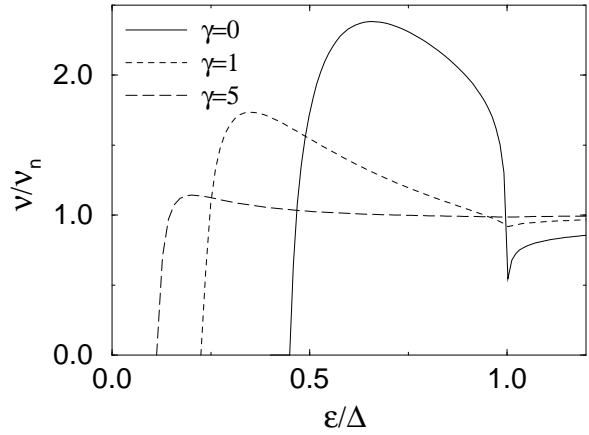,height=6cm}
\end{center}
\caption{The effect of $\gamma$ on the energy dependence of the
local DoS at the centre of the junction. Here $L/\xi=3$.}
\label{fig:gammasns}
\end{figure}

\begin{figure}
\begin{center} \epsfig{file=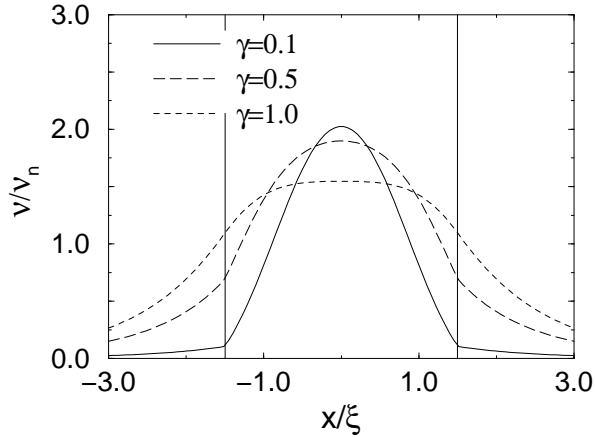,height=6cm}
\end{center}
\caption{The effect of $\gamma$ on the
position dependence of the DoS for the SNS-junction, with $\epsilon/\Delta =
  0.5$ and $L/\xi=3$.}
\label{fig:well}
\end{figure}

Figs.~\ref{fig:gammasns} and \ref{fig:well} show how variation of
the material parameter $\gamma$ affects the local DoS. Further, 
fig.~\ref{fig:well} shows how the local DoS at the SN-interface decreases
as $\gamma$ is reduced (e.g. as 
the ratio of the disorder concentrations in the S- and the
N-region is lowered.)

%
%
%

\section{Fluctuations} 
\label{sec:fluct}
Having discussed the mean field content of the theory, we now turn to
the central issue of this paper, that of mesoscopic fluctuations. The
generality of the field theoretic
machinery we have been discussing is such that it may be
 employed to analyse the majority of
fluctuating observables in SN-systems. 
Our strategy of extending the conventional $\sigma$-model formalism
allows us to take full advantage of the versatility of an approach that has
been greatly developed in the study 
of fluctuation phenomena in purely N-mesoscopic systems.
Yet in order not to diversify too much, we focus here on the
discussion of the specific example of spectral fluctuations. In
fact, the status of spectral fluctuations is slightly higher than that
of an ordinary example, since a vast number of
fluctuation phenomena are directly or indirectly related to
fluctuations in the single particle spectrum (see Ref.
\cite{Efetov,Weidenmueller}  for 
review).

A specific issue is the nature of fluctuations
in the local DoS around the mean values (displayed in figs.~\ref{fig:3Dsn} 
and \ref{fig:3Dsns}). To characterize these
fluctuations quantitatively, we employ the correlation function
(\ref{R}). Representing the correlation function in terms of Green
functions (cf. Eq.~(\ref{correlation_function})), we see that we need
to compute functional expectation values of the type specified in 
Eq.~(\ref{cf_Q}). An
evaluation of Eq.~(\ref{cf_Q}) on the mean field
level discussed previously leads to a vanishing fluctuation component,
$\langle {\cal G}^{\rm r} {\cal G}^{\rm a} \rangle-
\langle {\cal G}^{\rm r}\rangle \langle{\cal G}^{\rm a} \rangle$,
due to the fact that  
the mean field configuration, $\bar Q$,
is diagonal in the ar-indices. In other words,
there is no connection between advanced and retarded components and
the functional evaluation of the product of Green functions equals the
product of the averages. Consequently the physics
of mesoscopic fluctuations is contained entirely in
fluctuations around the block
diagonal mean field configuration: $\bar Q \to T \bar Q T^{-1}$.  
At first sight the analysis of the
fluctuation degrees of freedom, $T$ -- after all a 16-dimensional
matrix -- seems to be a complicated task. Fortunately the totality
of fluctuations may be organized into three separate types, each
with its own physical significance. Such a  
classification scheme leads to a substantial simplification of the
analysis. To be specific, we distinguish between fluctuation
matrices, $T$, that
\begin{itemize}
\item[a)] are {\it diagonal in the space of advanced and retarded
  components}. These types of fluctuations still do not give rise to
  correlations between retarded and advanced Green functions.
  Nonetheless they are of physical significance: Quantum corrections
  to the Usadel solution are described by fluctuations of this type.
  We elaborate on these effects in section \ref{sec:afluct}.
\item[b)] are {\it non-diagonal in ar-space but are proportional to
    unity in ph-space, $[T,\vc \sigma^{\rm ph}]=0$}. Fluctuations of this type
    {\it do} induce correlations between different Green functions and
    thereby mesoscopic fluctuations. They will be discussed in detail
    in section \ref{sec:bfluct}.
  \item[c)] {\it fulfill neither of the conditions a) and b)}. Whereas the
    physical significance of these fluctuations is less clearly
    identifiable than the one of the a- or b-type fluctuations, they
    are nevertheless of importance. The reason is that the c-type
    fluctuations tend to destroy at sufficiently high energies 
    the correlations that derive from mesoscopic b-type fluctuations.
\end{itemize}
The three fluctuation types are summarized in the following table:
\begin{center}
  \begin{tabular}{{|c||c|c|c|}}\hline
type & structure in ar space & ph space & lead to\\ \hline
a & diagonal & non-diagonal &  corrections
      to Usadel Green function
\\
b & non-diagonal  & $\propto \openone^{\rm ph}$ & mesoscopic fluctuations\\
c & non-diagonal & non-diagonal & destruction of b-type fluctuations \\ \hline
  \end{tabular}\\
\bigskip
\end{center}
Besides the criteria a)--c), further restrictions to be imposed on the
fluctuation matrices follow from two fundamental symmetries of the
model:
Firstly, general convergence criteria\cite{Efetov} enforce
the condition
\begin{equation}
\label{Usym}
  T^\dagger = \eta T^{-1} \eta^{-1},
\end{equation}
where 
\[
\eta = E_{11}^{\rm bf} \otimes \sigma_3^{\rm ph}
\otimes \sigma_3^{\rm ar} + E_{22}^{\rm bf}.
\]
Secondly, the tr-space structure of the model implies\cite{Efetov}
\begin{equation}
  \label{Tsym}
  T^T=\tau T^{-1} \tau^{-1},
\end{equation}
where 
\[
\tau=E_{22}^{\rm bf} \otimes i\sigma_2^{\rm tr} +
  E_{11}^{\rm bf} \otimes \sigma_1^{\rm tr}.
\]
For future reference we note that it is often convenient to represent
both the matrices $T$ and the above symmetries in terms of the {\it
  generators} of the fluctuations:
\begin{equation}
  \label{exp_para}
  T = \exp(W),
\end{equation}
where the generators $W$ are subject to the constraints
\begin{equation}
\label{Usym_W}
  W^\dagger = - \eta W \eta^{-1}
\end{equation}
and
\begin{equation}
  \label{Tsym_W}
  W^T=-\tau W \tau^{-1}.
\end{equation}
The above scheme generally classifies the various types of fluctuation
corrections to the quasiclassical picture of dirty superconductivity.
Beginning with the a-modes, we now turn to a more comprehensive
discussion of these fluctuations.  Although we do not provide by any
means a comprehensive survey of the full diversity of effects, we will find 
even in simple geometries of SNS-structures
a significant range of fluctuation types and associated phenomena. In
order to prevent the subdivision of these phenomena into excessively many
classes, we assume throughout that the spatial extent of the N
region is sufficiently large that $E_c \ll \Delta$. In this case, the
minigap of the SNS-junction with zero phase difference,
$\Delta\varphi=0$,
lies at energy $E_c$. The significance
of this restriction will be discussed further below,
in section \ref{sec:discuss}.


\subsection{a-Type Fluctuations: Quantum Corrections to the
  Quasiclassical Theory}
\label{sec:afluct}
In this section, fluctuations of type a) around the Usadel saddle
point will be considered. After specifying the general structure of
these fluctuations, we will exemplify their effect by discussing the
quantum corrections to the quasiclassical picture of single particle
properties of SNS-structures. Other types of SN-structures can
straightforwardly be subjected to an analogous analysis.

Fluctuations of type a) are diagonal in ar-space. Since the saddle
point $\bar Q$ is also ar-diagonal, the full effect of the a-type
fluctuations may be studied by considering just one of the ar-sectors
of the model action (\ref{ea}). For example, we may consider the retarded
sector describing the behaviour of a single, disorder-averaged retarded
Green function.  The restricted action is given by
\begin{eqnarray}
S_{\rm ret}[Q^{11}]=-{\pi\nu_n\over 8}\int {\rm str}\left[D(\widetilde{\partial}Q^{11})^2
+4i Q^{11}\left(
i|\Delta| \sigma_1^{\rm ph}+
\epsilon \sigma_3^{\rm ph}\right)\right],
\label{ea_ret}
\end{eqnarray}
where the eight dimensional matrix field $Q^{11}$ denotes the rr-block
of $Q$ and the parameter $\omega$ (which is meaningless for a single
Green function) has been dropped. To keep
the notation simple, we will henceforth (until the end of the section)
denote $Q^{11}$ again by $Q$.

Following the general philosophy of our classification, we organize
the field $Q$ into a saddle point contribution $\bar Q$ (which
is given by the solution of the retarded Usadel equation) and a-type
fluctuations around it:
\begin{eqnarray}
Q =R Q_fR^{-1},\qquad Q_f=T \sigma_3^{\rm ph} T^{-1},
\label{parQ}
\end{eqnarray}
where $R$ represents the inhomogeneous rotation
parametrising the saddle-point,
\begin{eqnarray}
\bar Q=\vc{q}\cdot\vc{\sigma}^{\rm ph} = R\sigma_3^{\rm ph} R^{-1},
\end{eqnarray}
and the rotation matrices $T$ generate the a-fluctuations. More
precisely, $T\in {\bf G}/{\bf H}$, where ${\bf G}$ is the group of
eight dimensional matrices subject to the constraints (see
Eq.~(\ref{Usym})),
\begin{eqnarray}
\label{Usym_a}
  T^\dagger = \eta_a T^{-1} \eta_a^{-1},\qquad \eta_a = E_{11}^{\rm bf}
  \otimes \sigma_3^{\rm ph} + E_{22}^{\rm bf}, 
\end{eqnarray}
and Eq.~(\ref{Tsym}).  The subgroup ${\bf H}\subset {\bf G}$ is
defined through ${\bf H} = \{ h\in {\bf G}|[h,\sigma_3^{\rm
  ph}]=0\}$\footnote{Note that there is some freedom in parameterizing
  fluctuating field configurations. For example, as an alternative to
  Eq.~(\ref{parQ}), one might let the fluctuation matrices $T$ act
  from the 'outside',
  \[
  Q = T \bar Q T^{-1}.
  \]
  As we will see shortly, there are situations where this
  parameterization is advantageous. In general, however, it creates
  unwanted problems. To see this, interpret the $T$'s as 'rotations'
  acting on the unit vector $\vc q$ appearing in $\bar Q = \vc q \cdot
  \vc\sigma^{\rm ph}$. Clearly, there are rotations that are
  ineffective (namely those that rotate $\vc q$ around itself) and
  should be excluded from the configuration space of the $T$'s. In
  practice, however, it is difficult to disentangle these rotations
  from the relevant ones. For example, parameterizing the matrices $T$
  in terms of some kind of spatially fixed coordinate systems,
  $T=T(\theta_1,\dots )$, where $\theta_i$ are rotation angles around
  certain fixed axes, one finds that the effective action
  $S[\theta_1,\dots]$ contains unphysical divergencies. These are due
  to the fact that some 'directions' in the parameter space spanned by
  the $\theta$'s correspond to ineffective rotations, thereby being
  energetically costless. The way to remove these spurious degrees of
  freedom is to introduce a 'moving' coordinate frame which, by
  construction, only parameterizes rotations around axes perpendicular
  to $\vc q$. This is exactly what the representation (\ref{parQ})
  achieves. Due to $T\in {\bf G}/{\bf H}$, the $T$'s contain only
  degrees of freedom that {\it effectively} modify the matrix
  $\sigma_3^{\rm ph}$.
}.  Note that these symmetry relations imply that the
matrices $T$ span Efetov's eight dimensional coset space of
orthogonal symmetry. In other words, save for the presence of the order
parameter (and the different coupling of magnetic fields), the
action (\ref{ea_ret}) is identical to that of a conventional
$\sigma$-model for an advanced and a retarded normal metal Green
function.

In order to understand more fully the effect of the fluctuation
matrices $T$, we first have to analyze their commutation behaviour
with the different contributions to the action (\ref{ea_ret}), i.e.
the magnetic field, the energy term and the order parameter.
Surprisingly, it turns out that there is a subset of $T$'s which not
only commute through the order parameter (no matter what its phase)
but also are {\em insensitive} to magnetic fields. For reasons that
will become clear below, we will call these matrices the {\it
  C-modes}. In the limit $\epsilon\rightarrow 0$, the C-modes become
completely massless. This implies that for low energies these modes
need a special, or {\em non-perturbative}, treatment.

\subsubsection{The C-Modes: Non-Perturbative Corrections to
  Quasiclassical Green Functions}

Amongst the set of a-type matrices, $T$, consider the subset $T_C\equiv
\exp(W_C)$, subject to
the additional constraints, 
\begin{enumerate}
\item $[\sigma_1^{\rm ph},W_C]=0$ (order parameter $1$-component  commutes
  through), and 
\item $[\sigma_3^{\rm ph} \otimes \sigma_3^{\rm tr},W_C]=0$ (no
  coupling to magnetic field). 
\end{enumerate}
In combination with eqns.~(\ref{Usym_W}) and (\ref{Tsym_W}), this gives
altogether four constraints and the non-trivial statement is that a
set of generators $W_C$ obeying all of them actually exists. These are the
C-modes. Before turning to the actual construction of these modes, let us
qualitatively discuss some of their general properties.

First note that the conditions 1. and 2. above imply that the C-modes
further commute with the 2-component of the order parameter. This 
follows from the observation that
the phase twist needed to interpolate between the 1- and the
2-component of the order parameter is equivalent to the appearance of
a magnetic gauge field which (see 2.) is invisible to the $W_C$'s.
Being insensitive to both magnetic fields and order parameters with
arbitrary phase positioning, the C-modes merely couple to the
gradients and the energy term in the action (\ref{ea_ret}).  In the limit of
small $\epsilon$, they become completely massless. More precisely, for a
{\it spatially constant} $T_C^0\in {\bf G}_C$, where ${\bf G}_C$ is the
subgroup of ${\bf G}$ fulfilling the extra constraints 1. and
2., $S[T_C^0\bar Q \left(T_C^0\right)^{-1}] 
\stackrel{\epsilon \rightarrow 0}{=}S[\bar Q]$.

\begin{figure}
\begin{center} \epsfig{file=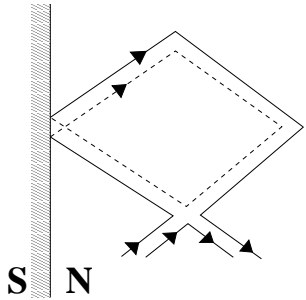,height=4cm}
\end{center}
\caption{Semiclassical illustration of an interfering path
  configuration contributing to the C-mode corrections to the
  DoS. Compare the relative orientation of the arrows with those
  appearing in figure \ref{fig:andreev} (b).}
\label{fig:classC}
\end{figure}

Physically, the C-modes represent modes of quantum interference in
SN-systems which survive magnetic fields. Within a different
formalism, these modes have for the first time been noticed
in Ref.~\cite{Brouwer95a}. Subsequently various physical phenomena caused
by their presence have been discussed in the literature:
\begin{itemize}
\item In the initial paper, Ref.~\cite{Brouwer95a} mentioned above, it was 
  observed that in SN-systems weak localization corrections to the
  conductance may survive magnetic fields. The quantum interference
  process responsible for that effect is associated with the C-modes.
\item In Ref.~\cite{A+Zirn}, an SN-system 
 subject to a magnetic field, but not exhibiting a
  minigap (due to suppression of the proximity effect by the magnetic
 field), was considered. In this case the single-particle DoS
  vanishes at the chemical potential on a scale set by the mean level
  spacing. The existence of this 'micro-gap' is also an effect 
caused by the C-modes.
\item Feigl'man and Skvortsov\cite{Feiglman1} discuss the effect of
  C-modes on the transport behaviour of vortices in moderately clean
  type II-superconductors.
\item The above phenomena relate to mean single particle properties.
  For the sake of completeness, we here mention some manifestations of
  C-mode fluctuations in two-particle properties: The level statistics
  of SN-quantum dots in a magnetic field falls into a symmetry class
  that is different from any of the standard Wigner-Dyson classes.
  Referring to a classification scheme due to Cartan, the SN/magnetic
  field symmetry class has been termed 'class C'\cite{A+Zirn}. As in standard
  mesosocopic systems, these level fluctuations can also be
  associated with channels of microscopic quantum interference. Whereas
  Wigner-Dyson fluctuations in diffusive N-systems are caused by 
  'diffuson' and 'Cooperon' modes, the class C fluctuations are
  connected to the modes specified above, and hence the name 'C-modes'. A
  $\sigma$-model formulation of the C-mode spectral statistics of
  random matrix ensembles was presented in Ref.~\cite{Frahm}.

  C-type level statistics in vortices has recently been
  microscopically derived 
  by Skvortsov et al. \cite{Feiglman2}. Thermal
  transport carried by C-modes through the core 
of superconductor vortices is considered
  in a recent paper by Bundschuh et al. \cite{Zirnbauer98}.
\end{itemize}
Being effective already on the level of single particle properties,
the C-modes must originate from interference processes between
particles and holes. However, they cannot be identical with the modes
displayed in figure \ref{fig:andreev} (b), since the latter are field
sensitive. A typical type-C path configuration is displayed in figure
\ref{fig:classC}. In the analysis below we will
derive quantitative expressions for processes of
this type.

After these general remarks, we next turn to the analysis of the
C-mode contribution within the $\sigma$-model formalism. Specifically,
we will discuss the effect of these modes on the mean density of
states of the SNS-geometry displayed in figure \ref{fig:sns}.

Before turning to the core of the discussion, let us make a technical
remark which will have some impact on the organization of the remainder
of the section: The invariance of the action under spatially constant
C-transformations , $S[T_C^0\bar Q \left(T_C^0\right)^{-1}]
\stackrel{\epsilon \rightarrow 0}{=}S[\bar Q]$ suggests an interpretation
of these modes as a global {\it symmetry} of the action. In particular,
it is more natural to let them act on $\bar Q$ from the outside:
$Q = T_C\bar Q T_C^{-1}$ (compare with the {\it inside} representation of
general a-fluctuations, Eq. (\ref{parQ}), and the footnote on 
p.\pageref{Usym_a}). Of course it is possible to
forcefully contrive an inside parameterization
for the C-fluctuations,
via $T_C\bar Q T_C^{-1} = R \tilde T_C \sigma_3 \tilde T_C^{-1} R^{-1}$, where the
unitarily transformed $\tilde T=R^{-1} T_C R$. In practice, however, this
representation is inconvenient and, more seriously, makes it difficult
to separate the C-mode fluctuations from general fluctuations around
the saddle point. These considerations imply
that, in general, it is difficult to treat the C-modes and the rest of
the a-type fluctuations simultaneously. {\it Physically}, however,
these problems are of little significance: Below it will be shown
that the minimum price in energy associated with a non-C-fluctuation is
of ${\cal O}(E_c)$. This implies that two regimes with qualitatively
different fluctuation behaviour exist:
 \begin{itemize}
 \item Low energies, $\epsilon \ll E_c$, where the C-modes are
   relevant, whereas the other a-fluctuations can safely be ignored, and
 \item High energies $\epsilon > E_c$, where {\it all} fluctuations
   have a comparable action of ${\cal O}(E_c)$ and it is pointless to
   carefully distinguish between the different types (C or other).
 \end{itemize}
Below we will discuss these cases separately. Although our analysis
is not applicable to the crossover regime of intermediate energies, we
do not expect qualitatively remarkable phenomena to arise there.

{\it Low energies, $\epsilon \ll E_c$:} To quantitatively analyse the
fluctuation physics in this regime, we need first to derive
an effective action for the C-modes. Fluctuations other than C are ignored.

To this end, we use that $[T_C,\sigma_1^{\rm ph}]=[T_C,\sigma_2^{\rm
  ph}]=0$ and represent the $Q$-field as
\begin{equation}
\label{QC}
Q=q_3 Q_C,\qquad Q_C = T^{\vphantom {-1}}_C \sigma_3^{\rm ph} T_C^{-1}.
\end{equation}
Substituting this parameterization into Eq.~(\ref{ea_ret}), we
obtain the desired action
\begin{eqnarray}
S_C[Q_{C}]=-{\pi\nu_n\over 8}\int {\;\rm str\;}\left[Dq_3^2 (\partial Q_C)^2
+4iq_3\epsilon Q_C \sigma_3^{\rm ph}\right].
\label{ea_C}
\end{eqnarray}
Among the general set of C-type fields $Q_C$ , there is a
particular mode $Q_C^0=T^{0\vphantom {-1}}_C \sigma_3^{\rm ph}
(T^0_C)^{-1}$ which not only has C-symmetries but also is spatially
constant. Substituting this 'zero-mode' into (\ref{ea_C}), we obtain
the  action 
\begin{eqnarray}
S_C^0[Q_C^0]=-{i\tilde{s}\over 2} {\;\rm str\;}\left[ Q_C^0 \sigma_3^{\rm ph}\right],
\label{ea_C0}
\end{eqnarray}
where 
\begin{equation}
  \label{s_def}
  \tilde s = \pi\epsilon \nu_n \int q_3.
\end{equation}
Note that deep in N, where $q_3 = 1$, $\tilde s$ = $\pi\epsilon/
\bar{d}$, where $\bar{d}$ is the mean level spacing, 
coincides with the standard parameter
$s$ \cite{Efetov} commonly employed in the literature on spectral
correlations in metals.

In order to give the above zero mode action some physical
significance, it has to be shown that it is energetically gapped
against the action of the higher (spatially fluctuating) field
configurations of C-symmetry. The spectrum of fluctuating C-modes can
be determined at various levels of accuracy. For our purposes, it
suffices a) to demonstrate that a ground state gap exists and b) to
coarsely estimate its magnitude. To do so, we first note that the mode spectrum
is essentially determined by the gradient operator appearing in
Eq.~(\ref{ea_C}). Integrating by parts, the latter can be rewritten as
\begin{equation}
\label{diff_C}
\sim \int q_3^2 {\; \rm str\;}\left[Q  D q_3^{-2}\partial(q_3^2
  \partial) Q \right].
\end{equation}
The rationale behind this reformulation is that $D
q_3^{-2}\partial(q_3^2 \partial)$ may be regarded as a differential
operator which is Hermitian with respect to the scalar product
$\langle f,g\rangle \equiv \int q_3^2 f g$. Being Hermitian with
compact support (taking $q_3 \rightarrow 0$ in S), the spectrum of the
differential operator is discrete. To estimate the spacing $\Delta E$
between the zero eigenvalue of the spatially constant eigenmode and
the first excited eigenvalue, we use the fact that the range of
support of the operator is set by $L$, the extension of N. Standard
reasoning for the eigenvalue structure of one-dimensional Hermitian
differential operators with compact support then leads to the estimate
$\Delta E \sim D/L^2 = E_c$. Notice that the actual spatial structure
of the excited eigenfunctions may be complicated. For example, unlike
with standard applications of $\sigma$-models to N-systems, typical
eigenmodes of the action obey neither Neumann nor Dirichlet boundary
conditions, but rather exhibit more complicated edge behaviour which,
in principle, may be derived from Eq.~(\ref{diff_C}) once $q_3$ is
known. For our purposes, entering this discussion will turn out to be
unnecessary.

The considerations above show that for energies $\epsilon \ll E_c$,
the action is governed by the spatially constant C-mode\footnote{A
  closer analysis, similar to that presented in section
  \ref{sec:cfluct} with regard to the destruction of correlations by
  c-type fluctuations, shows that the range of stability of the C-zero
  mode is in fact limited by $\epsilon \sim \bar{d}\sqrt{g} \ll
  E_c$.}.
It is reasonable to ask for which physical applications such low
energies may be expected to play a r\^ole.  For zero phase difference
between the superconductor terminals, the minigap is of ${\cal
  O}(E_c)$ and, with regard to the DoS, the C-mode fluctuations are
expected to be of little importance. To actually make visible the
impact of these fluctuations on the DoS, we concentrate here on the
case of a junction close to $\pi$-phase difference, i.e. a junction
where the gap is nearly but not completely closed: $E_g \ll E_c$. Note
that previous studies of the C-type fluctuations having concentrated
on the limiting case where the proximity effect is {\em totally}
suppressed ($E_g \to 0$).  It is worth remarking that, while we have
limited here our comments to issues surrounding the DoS, {\it subgap}
properties of SNS-junctions, such as the Josephson coupling, may well
be affected by the C-fluctuations, even under the broader conditions
of a fully-established proximity effect. However, the analysis of
these phenomena lies beyond the scope of this paper.

Focusing on the range of applicability of the action (\ref{ea_C0}),
we notice that for {\it very} small energy parameters, so that $\tilde
s = {\cal O}(1)$, the zero mode must be treated in a non-perturbative
manner.  Very much as in the study of N $\sigma$-models at low energy
scales, the fluctuations become unbounded as $\tilde s \rightarrow 0$.
Rather than perturbatively expanding in terms of the generators $W_C$,
it then becomes necessary to integrate over the entire manifold of
matrices $T_C$.  Non-perturbative analyses of this type have
previously been applied to the study of a random matrix
ensemble\cite{Frahm} of non-proximity effect SN-structures and of
normal core excitations of vortices in superconductors
\cite{Feiglman2}.  Here we discuss how the C-fluctuations affect the
low energy behaviour of the Gorkov Green function in SNS-junctions.

To this end, we first need to specify a global parameterization of the
matrices $T_C$. It is a straightforward (if lengthy) matter to show
that a general $8\times8$-matrix $T_C = \exp (W_C)$, subject to both
the general constraints (\ref{Usym_W}) and (\ref{Tsym_W}) and 1. and 2.,
can be parameterized as
\begin{eqnarray}
  \label{TC}
  &&T_C = v u a,\\
&&\qquad a = \exp\left(\frac{i\theta}{2} E_{22}^{\rm bf} \otimes \sigma_1^{\rm
    ph} \otimes \sigma_1^{\rm tr}\right),\nonumber\\
&&\qquad u=\exp \left(iy E_{22}^{\rm bf}\otimes\sigma_3^{\rm tr}
\right)\otimes \openone_{\rm
  ph},\nonumber\\
&&\qquad v=\exp \left(
  \begin{array}{cc}
&\lambda - \mu \sigma_3^{\rm tr}\\ 
\mu + \lambda \sigma_3^{\rm tr}&
  \end{array}\right)_{\rm bf}\otimes \openone_{\rm ph},\nonumber
\end{eqnarray}
Where $\lambda$ and $\mu$ are Grassmann variables. 
For the (invariant) measure associated to the integration over the
matrix $Q_C$ we obtain (cf. the analogous but more difficult
calculations of integration measures in \cite{Efetov})
\[
\int dQ_C (\dots) = \int_0^{2\pi} \frac{dy}{2\pi}\int_0^\pi \frac{d\theta
\sin\theta}{2\sin^2(\theta/2)}\int d\lambda d\mu (\dots).
\]
Substituting the parametrisation (\ref{TC}) into the action
(\ref{ea_C0}) we obtain
\begin{equation}
  \label{ea_C0np}
  S_C^0[Q_C^0] = -2i \tilde s (\cos\theta-1).
\end{equation}
As an example, we apply this action to a calculation of the C-mode
corrections to the local DoS. Substituting the zero mode integration
over $Q_C$ for the functional expectation value in Eq.~(\ref{DoSexp}) and
performing the (trivial) integrations over Grassmann variables and $y$,
we obtain
\begin{eqnarray*}
&&\nu = \nu_n {\, \rm Re\,} \left\{q_3 \left (1 -\frac{1}{2}\int_{0}^{\pi}
  d\theta \sin\theta
e^{2i \tilde s (\cos\theta-1)}\right)\right\}\\
&&\hspace{0.5cm}=\nu_n {\, \rm Re\,} \left\{q_3 \left (1
  -\frac{1-e^{-4i\tilde s}}{4i\tilde s} \right)\right\}\\
&&\hspace{0.5cm}=\nu_n {\, \rm Re\,} \left\{q_3\left (1-\frac{\sin(4\tilde
    s)}{4\tilde s} + \frac{1-\cos(4\tilde s)}{4i\tilde s} \right)\right\}.
\end{eqnarray*}
The last line tells us that for small energies, $\tilde s
\rightarrow 0$, the DoS {\it always} (i.e. including  the case of a
$\pi$-junction) vanishes on a scale set by the
mean level spacing. This is the
DoS 'micro-gap' that has been discussed previously in
Refs.\cite{A+Zirn,Frahm,Feiglman1,Feiglman2}. 
Moreover, for general $\tilde s$, possessing real and
imaginary components, the DoS is not only determined by
${\rm Re\; }q_3$ but also by the imaginary component, ${\rm Im\;
  }q_3$, of the Usadel solution. 
Finally, $\tilde s$ contains the Usadel solution in an
integrated form, that is, the C mode introduces some non-local influence
on the local DoS by the Usadel solution at different points of the
system. The corrections vanish algebraically as
$\tilde{s}^{-1}$. For ${\rm Im\;} \tilde s>1$ the (oscillatory) factors
containing the exponentiated parameter $\tilde s$ can be neglected and
we obtain the simplified result
\begin{eqnarray}
\nu \simeq \nu_n \left( {\rm Re \;} q_3 + \frac{1}{4}{\; \rm
    Im\;}\frac{q_3}{\tilde s}\right).
\label{nonosc}
\end{eqnarray}
Note that the definition (\ref{s_def}) implies that for the {\it global}
DoS ($\sim \int \nu$) the algebraic corrections vanish. 
For a diagrammatic interpretation of this correction to the Usadel
DoS, cf. Ref.~\cite{A+Zirn}.

\begin{figure}
\[\begin{array}{ccc}
\epsfig{file=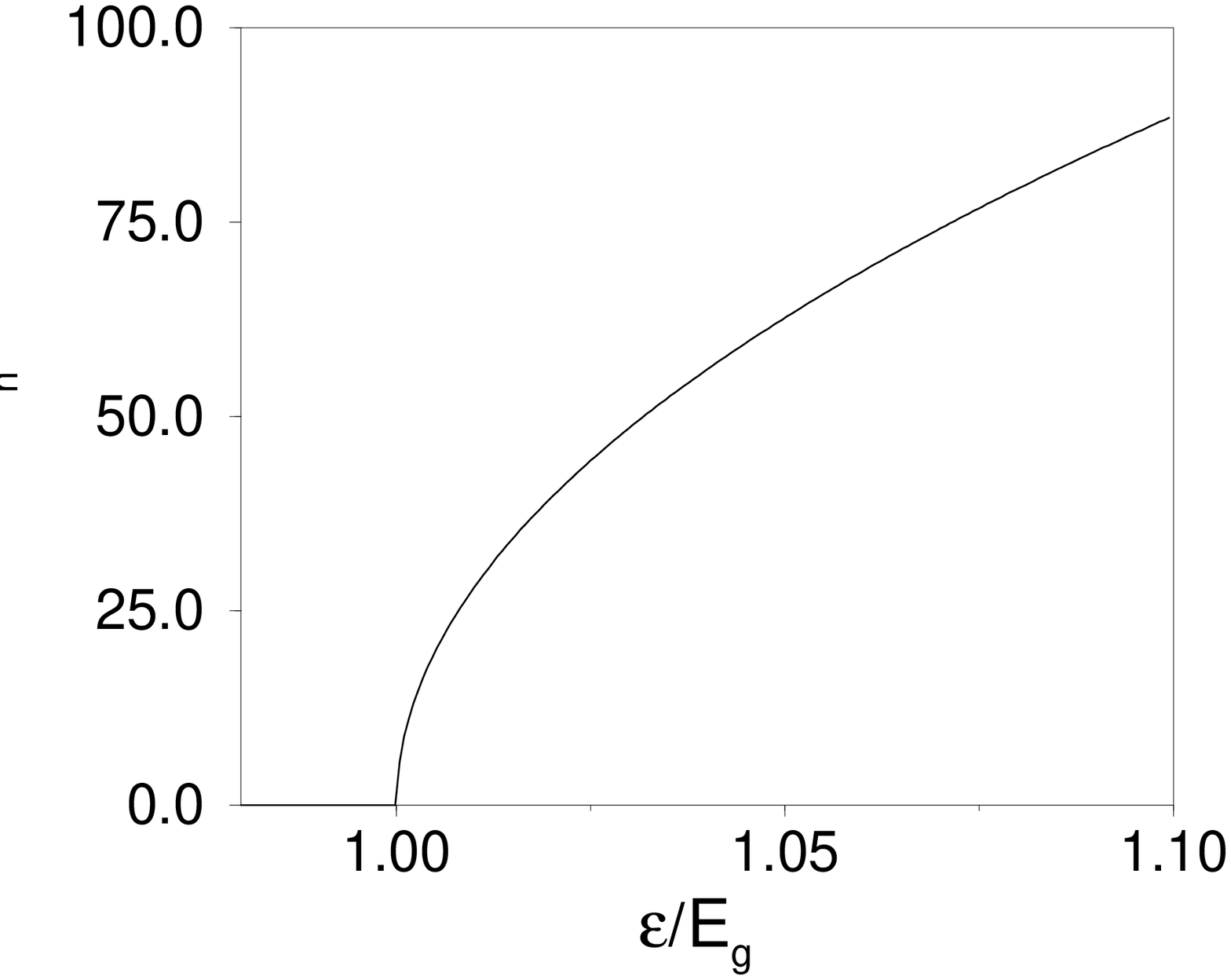,height=5cm} &\qquad \qquad&
\epsfig{file=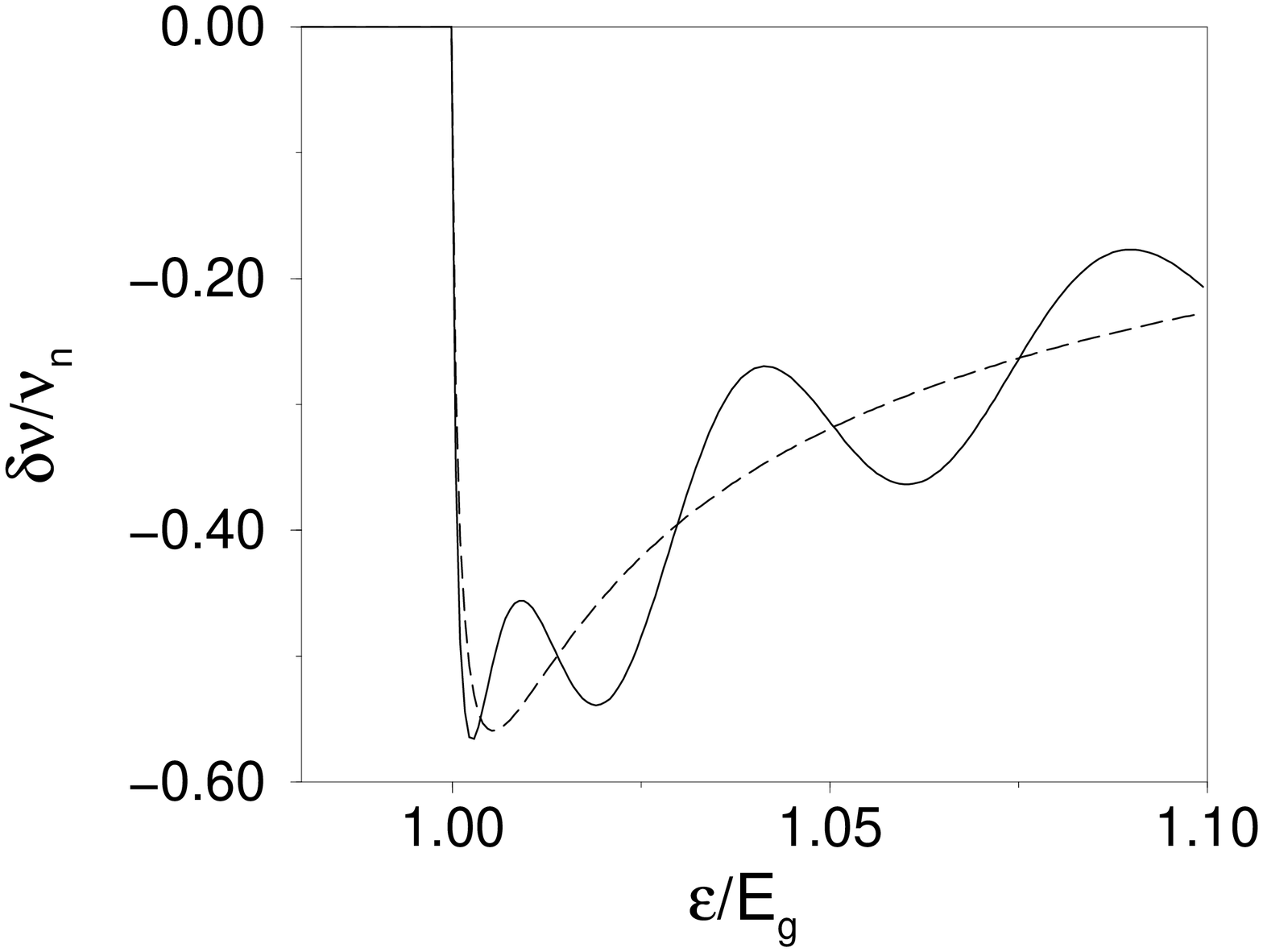,height=5cm} \\
{\em (a)} & &{\em (b)}
\end{array}\]
\caption{({\em a}) Quasiclassical DoS and ({\em b}) the C-mode
  correction 
  for $\Delta\varphi = \pi-0.0025$ and $g=5$, at the centre of the
  junction and in the vicinity of the
  minigap. The dotted line
  represents the non-oscillatory part of the correction, as given by 
Eq.~(\ref{nonosc}).}
\label{fig:DoScorr}
\end{figure}

For $\Delta \varphi = \pi-0.0025$ and $g=5$, 
the quasiclassical DoS and the corrections to it
are displayed in figure \ref{fig:DoScorr}. We display here the DoS in
the vicinity of the minigap, at an energy regime similar to that considered by
Zhou et al. \cite{Zhou98}. For this set of
parameters, the quasiclassical DoS displays a very strong peak, which
dies down for $(\epsilon-E_g)/E_g = {\cal O}(1)$. 

{\it Large energies, $\epsilon > E_c$:} For large energies, the
isolated 0-mode action is no longer of significance. The energy of
{\it all} modes, C- or other, is larger than $E_c$. In particular,
spatially fluctuating configurations (with a 'kinetic' energy cost of
${\cal O}(D/L^2 = E_c)$) need to be taken into account, too.  Since
the energy associated with all these fluctuations is parametrically of the
same order, separating the C-modes from the rest becomes pointless. In
the next section we discuss the corrections to quasiclassics at
energies larger or comparable to $E_c$ arising from a perturbative
treatment of all a-type fluctuations.

\subsubsection{Perturbative Corrections to the Quasiclassical Green Function}
\label{sec:a_sns_pert}
In the following we consider the impact of a-type fluctuations on the
'high' energy ($\epsilon \sim E_c$) behaviour of the average Gorkov Green
function. To keep our discussion simple, we limit consideration in
this section to a SNS geometry where time-reversal symmetry is
maintained (i.e. where the phase difference between the two
superconductors, $\Delta\varphi$, is zero.)  The violation of
time-reversal symmetry through the variation of the phase of the order
parameter across the junction will not change our discussion
qualitatively.

Specifically, the questions we are going to address are
\begin{itemize}

\item Do quantum corrections lead to the suppression of the minigap in the 
normal region?

\item If so, does the minigap edge remain sharp, or are states introduced 
at all energy scales below the gap?

\end{itemize}
As in the previous sections, the theory developed in this section 
may also be straightforwardly generalized to other types of
geometries and observables.

For large energies $\epsilon \sim E_c$, it is convenient to parameterize the totality of a-type
fluctuations as in Eq.~(\ref{parQ}). Due to the comparatively large energy
cost associated with fluctuations around the Usadel saddle point, it is
sufficient to expand the action to low orders in terms of the
generators $W$ of the fluctuation matrices $T$. Substituting the
parameterization (\ref{parQ}) into the action (\ref{ea_ret}), we
obtain 
\begin{eqnarray}
S[Q]=-{\pi\nu_n\over 8}\int {\rm str}\left[D(\widetilde{\partial}Q_f)^2
+4i\epsilon R^{-1}\sigma_3^{\rm ph}R Q_f\right],
\label{ea_sn}
\end{eqnarray}
where $\widetilde{\partial}=\partial+[R^{-1}\partial R,\;]$. Note that
the above action does not contain the superconducting order parameter.
This is accomplished by demanding that the non-C-type fluctuations
obey Dirichlet boundary conditions at the NS-interface. As a
consequence these modes are spatially varying with a minimum
fluctuation energy of ${\cal O}(E_c)$, which justifies their
perturbative treatment. As for the C-modes, these do not couple to
the order-parameter anyway. In principle, their treatment is difficult
because, as mentioned above, they a) fulfill mixed boundary conditions
different from Dirichlet or Neumann and b) are difficult to separate
from the complementary set of a-type fluctuations.  However, for large
energies we believe these complications to be physically irrelevant:
The spectrum of {\it all} fluctuating modes is discrete with a typical
spacing of ${\cal O}(E_c)$. For energies $\epsilon$ comparable with
$E_c$, all modes need to be summed over.  Under these
conditions, the detailed structure of boundary conditions and/or
eigenvalues of individual modes becomes largely inessential; What
matter are the global features of the energy spectrum associated with
the fluctuations, most importantly, the typical mean energy spacing.
For this reason, we feel justified in ignoring the different boundary
behaviour of the C-modes and to globally impose Dirichlet boundary
conditions (thereby correctly modelling the typical spacing between
consecutive eigenmodes).  We believe that this simplification does not
lead to qualitative errors.

To obtain the perturbative expansion of the action, we employ the
exponential parameterization $T=\exp(W)$ and expand the generators $W$
in terms of  ph-Pauli matrices, $W=w_1
\sigma_1^{\rm ph}+w_2\sigma_2^{\rm ph}$ (so that $[W,\sigma_3^{\rm
  ph}]_+=0$). For zero phase difference, the Usadel solution encoded in the rotation
matrices $R$ can be parameterized in terms of a single angle
$\theta$ (cf. Eq.~(\ref{angparam})). The rotation matrices $R$
mediating between $\sigma_3^{\rm ph}$ and the Usadel saddle point 
then take the simple form 
\begin{eqnarray}
R=\exp\left[-i\frac{\theta}{2} \sigma_2^{\rm ph}\right],\qquad 
(\partial R)R^{-1}=-{i\over 2}\sigma_2^{\rm ph}\partial\theta.
\end{eqnarray}
Substituting these expressions into the action and expanding up to
second order in $w_i$, it is a straightforward matter to show that
\begin{eqnarray}
S[W]=\pi\nu_n \int {\rm str}_0 \left[D\left((\partial w_1)^2+(\partial
w_2)^2 -(\partial\theta)^2 w_1^2\right) -2i\epsilon\cos\theta\left(w_1^2+w_2^2
\right)\right]+O(W^3).
\label{actionw}
\end{eqnarray}
Here and in the following, the notation '${\rm str}_0$' 
represents a supertrace over all degrees of freedom except for the
ph-components, which have been traced over.
The absence of terms at first order in $W$ is assured by the expansion around
the saddle-point configuration of $Q$. 

To eliminate the term in $\sim (\partial \theta)^2$, we make use of the
fact that the Usadel equation (\ref{diff1}) possesses the first integral,
\begin{eqnarray}
D(\partial\theta)^2-4i\epsilon(\cos\theta-\cos\theta(0))=0,
\end{eqnarray}
where we have used the fact that in the middle of the junction, $\partial
\theta =0$. Substituting this result into Eq.~(\ref{actionw}), we obtain 
\begin{eqnarray}
S[W]=\pi\nu_n \int {\rm str}_0\left[D\left((\partial w_1)^2+(\partial w_2)^2
\right)+2i\epsilon
w_1^2(2\cos\theta(0)-3\cos\theta)-2i\epsilon w_2^2\cos\theta\right].
\label{actw2}
\end{eqnarray}
To compute corrections to the DoS, we substitute the exponential
parameterization into the functional representation (\ref{DoSexp}) to
find
\begin{eqnarray}
\label{DoSq}
\nu(x)&=&{\nu_n\over 8} {\rm Re}\left\langle {\rm str}\left(R
\sigma_3^{\rm bf}\otimes \sigma_3^{\rm ph} R^{-1} Q\right) 
\right\rangle_Q\nonumber \\
&=&\nu_n {\rm Re}\cos\theta\left[1+\frac{1}{2}\left\langle{\rm str}_0\left(
\sigma_3^{\rm bf}(w_1^2+w_2^2)\right)\right\rangle_W\right]\nonumber\\
&=&\nu_n {\rm Re}\cos\theta\left[1+\frac{1}{\pi\nu_n L_\perp^{d-1}}
\left(\Pi_1(x,x) - \Pi_2(x,x)\right)\right],
\end{eqnarray}
where $x$ is the coordinate along the junction and $L_{\perp}$ is
spatial extent of the N region in all other directions.  The last line
in Eq.~(\ref{DoSq}) is obtained by an application of Wick's
theorem\cite{Efetov} to the Gaussian expectation values $\langle
\sigma_3^{\rm bf}w_i^2 \rangle_W$. The 'propagators' $\Pi_i$ play the
r\^ole of generalized diffusion poles. They are defined through
\begin{eqnarray}
  \label{propagator}
  \left[-D\partial_x^2 + 2i\epsilon(2\cos \theta(0) - 3\cos
  \theta(x))\right]\Pi_1(x,y)&=&\delta(x-y),\nonumber\\ 
  \left[-D\partial_x^2 - 2i\epsilon\cos
  \theta(x)\right]\Pi_2(x,y)&=&\delta(x-y).
\end{eqnarray}
Without going into details we remark that the relative minus sign
between $\Pi_1$ and $\Pi_2$ in (\ref{DoSq}) derives from the
different symmetries of the matrices $w_1$ and $w_2$ under matrix
transposition (cf. Eq. (\ref{Tsym_W})).

We will not proceed any further analytically. In order to
quantitatively evaluate the $\Pi$-dependent corrections to the DoS,
one would have to compute the generalized diffusion poles
(\ref{propagator}). Due to the presence of the spatially varying terms
$\sim \cos\theta$, a general solution of the differential equations is
difficult\footnote{In fact, relatively standard techniques
  \cite{Morse} may be employed for the solution of Eq.
  (\ref{propagator}). This follows from the fact that, upon
  substitution of the Usadel solution Eq. (\ref{SNSsol}), these
  equations are classifed as 'Lam\'{e}' equations \cite{Magnus}. In
  comparatively simple situations, such as asymptotically large
  energies $\epsilon \gg E_c$, infinite SN- rather than finite
  SN-systems, and so on, analytical solutions are available.  However,
  in order not to diversify the discussion unnecessarily we do not
  elaborate on these cases.}. Nonetheless, quite a few characteristic
properties of the DoS corrections can be deduced from (\ref{DoSq})
simply by inspection:

For {\it asymptotically large energies $\epsilon \gg E_c$}, one
expects no influence of the superconductor on the normal
metal. Indeed, in that limit, $\cos
\theta \to 1$ implying that a) the Usadel DoS becomes metallic and b)
$\Pi_1 - \Pi_2 \to 0$, i.e. no quantum corrections to the DoS.

For {\it intermediate energies} just above the minigap edge $E_g$, $\cos
\theta$ varies smoothly as a function of position. In this regime
Eq. (\ref{DoSq}) gives corrections of ${\cal O}(g^{-1})$ to the DoS whose
quantitative evaluation is difficult.

Finally, let us consider {\it subgap energies}, $\epsilon < E_g$
(remaining of course outside the regime $\epsilon \ll E_c$).
Here, according to the quasiclassical
analysis, the DoS vanishes, implying that $\cos\theta$ is purely
imaginary and the effective action~(\ref{actw2}) 
purely real.  As a consequence, the propagators $\Pi_i$ are real, too,
and the DoS, as computed according to (\ref{DoSq}), {\em vanishes}
identically below the quasiclassical minigap edge. In other words,
the perturbative inclusion of first order quantum corrections does not give
rise to the appearance of states below the quasiclassical edge. The
vanishing of ${\rm Re\,}(\cos \theta)$ actually suffices to demonstrate
that the robustness of the gap pertains to {\em all} orders of perturbation
theory.

\begin{figure}
\[\begin{array}{ccc}
\epsfig{file=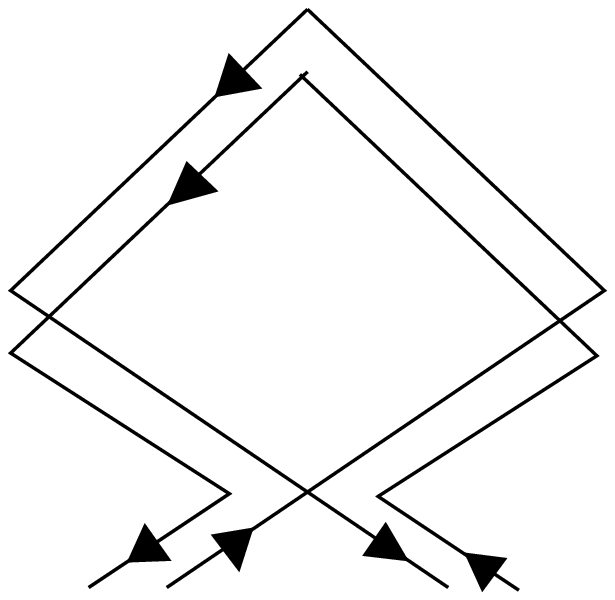,height=5cm} &\qquad \qquad&
\epsfig{file=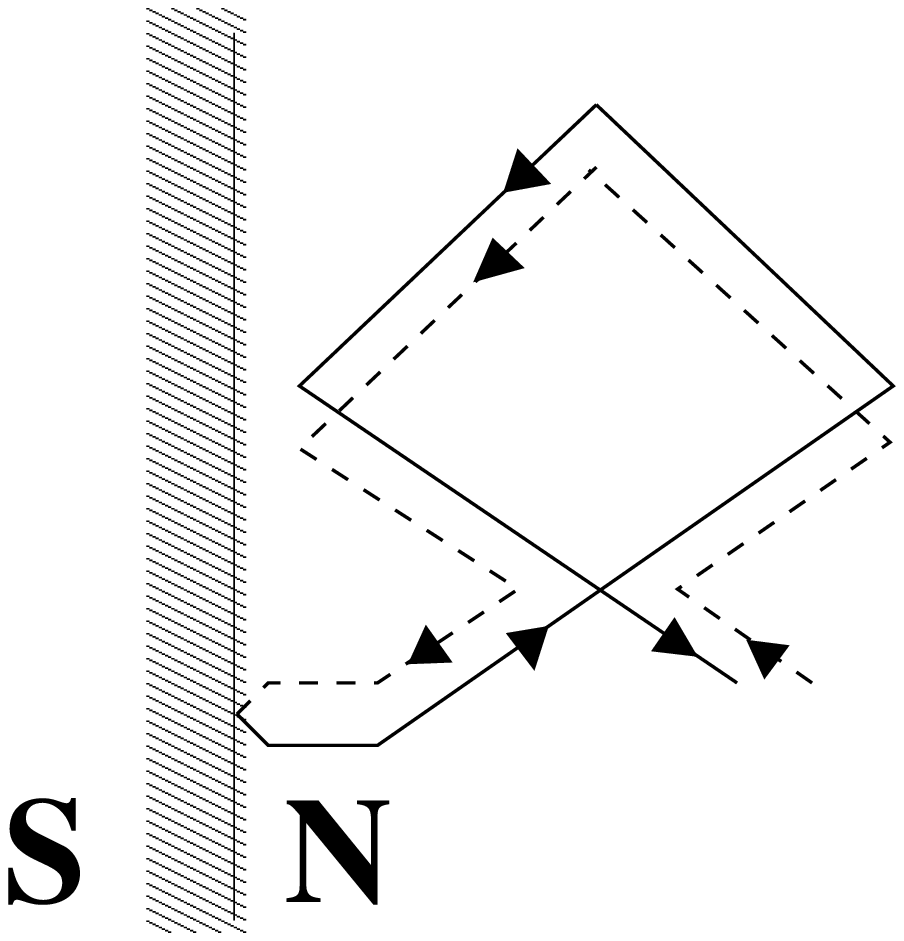,height=5cm} \\
{\em (a)} & &{\em (b)}
\end{array}\]
\caption{Renormalisation of the diffusion constant, $D$, by
  interference of ({\em a})
  trajectories with their time-reversed counterparts, and ({\em b})
  particles and holes.}
\label{fig:Dren}
\end{figure}

This conclusion presents something of a puzzle: Taking into account
quantum corrections, the above result indicates that the minigap
remains fixed at energy $E_g$. The latter is determined by the {\it
  bare} value of the diffusion constant, $D=v_Fl/d$. However, the
intuition afforded by the one-parameter scaling theory of
localization~\cite{Abrahams} suggests that observables such as the
tunneling DoS should depend only on the value of $D$ renormalized by
weak localization corrections. In bulk normal metallic samples, weak
localization corrections (to two-particle properties) stem from
mechanisms of quantum interference between trajectories connected
through a time reversal operation.  (see fig.~\ref{fig:Dren}({\em
  a})). In the present case, weak localization effects can arise due
to the interference of particles and holes (see
fig.~\ref{fig:Dren}({\em b})). Since holes bear similarity with time
reversed particles, there is no conceptual difference to the above
N-interference mechanism, and one expects a standard renormalization
of the diffusion constant (albeit already on the level of one-particle
properties). Yet, according to the analysis above, the minigap edge, a
function of the unrenormalized $D$, is robust against perturbative
quantum corrections in the particle/hole channel.

In fact, the absence of weak localization corrections to the minigap
{\em edge} signals the failure of the perturbation theory.  To
properly identify quantum weak localization corrections to the
diffusion constant, and therefore the minigap edge, it is necessary to
renormalize the saddle-point equation itself. This situation parallels
that encountered in the study of the renormalization of the gap in a
dirty bulk superconductor where quantum corrections (in the Cooper
channel) lead to a renormalization of the gap
equation (see e.g. Ref.~\cite{Smith}). In the present case, weak
localization corrections to the minigap edge are obtained within a
renormalization group procedure. Since, operationally, this procedure
is somewhat technical, its description has been made the
subject of appendix~\ref{app:renorm}.

The renormalization group procedure described in
appendix~\ref{app:renorm} may be employed safely down to energy scales
in excess of $E_c$. 
However, at energy scales in the vicinity of the
minigap, the Cooperon propagator depends sensitively on the geometry
and it becomes necessary to include the additional flow in $R$ (the
matrix rotating to the RG-affected saddle point of the theory),
coupled to that in $D$, as the cutoff is lowered towards
$E_g$. Although the 
manner in which such renormalisation processes are included
self-consistently lies beyond the scope of this paper, the outcome
of the RG procedure can be summarized as follows:
Treating quantum interference correction within a RG-scheme leads to a
{\it shift} of the minigap edge. The overall structure of the gap
edge (e.g.the non-analytic behaviour of the DoS in the vicinity of the
DoS) is maintained. In particular, no states are found below the
(renormalized) gap edge.     

These findings leave us with the question whether indeed, no states
exist below the (renormalized) gap edge or whether the
computation simply has not been accurate enough. Although a
quantitative analysis of this question is beyond the scope of this
paper, we believe that the second option is the correct one: To find
states below the minigap, one must account for contributions to the
action which cannot be accessed by a perturbative shift of the
inhomogeneous saddle-point. Contributions of this kind have been
identified in bulk normal conductors as soliton-like configurations of
the $Q$-matrix fields, and have been associated with a rare class of
states which are described as ``anomalously'' or ``nearly'' localized
within the metallic phase~\cite{Muz,Falko,Mirlin}. Poorly contacted to
the superconductor, these states are able to exist at energies below
the minigap and generate contributions to the average DoS
exponentially small in $g$. Although we see the SN system as a useful
and challenging arena in which to investigate the localization
properties of such rare states, their consideration lies beyond the
scope of this paper.

\begin{figure}
\begin{center} \epsfig{file=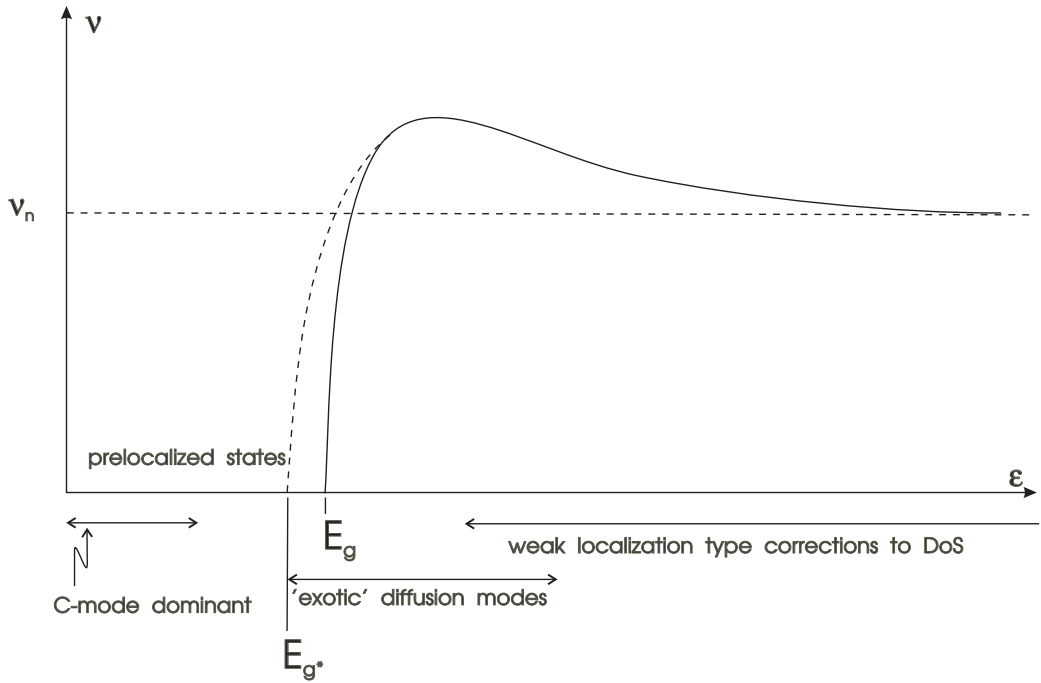,height=9cm}
\end{center}
\caption{Schematic indication of the r\^ole of the various DoS
  corrections. $E_{g^\ast}$ denotes the renormalized gap edge.}
\label{fig:a_type}
\end{figure}

Before leaving this section, let us in summary list the -- admittedly
diverse -- set of a-type fluctuation mechanisms renormalizing the
single-particle properties of mesoscopic SN-structures (see figure
\ref{fig:a_type}):
\begin{itemize}
\item For energies $\epsilon\gg E_c$, the N-component behaviour is largely
  metallic. However, the presence of the superconductor is exerted in
  terms of massive quantum corrections to the DoS and other single
  particle properties. The larger $\epsilon$ is, the smaller are the
  corrections.
\item Energies just above the minigap edge $E_g$ are the most
  difficult to analyse. Quantum corrections to quasiclassics are
  carried by diffusion type modes which -- due to the pronounced
  energetic and spatial inhomogeneity of the DoS -- are difficult to
  treat analytically. By perturbatively including such corrections, one
  obtains corrections to the DoS above the gap. Both the position
  of the gap and the vanishing of the subgap density of states remain
  unchanged.
\item By embedding the diffusive modes into an iteration of
  RG-analyses and solutions of renormalized Usadel-type mean field
  equations, one arrives at a shifted minigap edge. The
  non-analyticity of the gap is maintained -- that is, there are no smooth
  DoS 'tails' leaking downwards out of the sharp edge.
\item Presumably, 'nearly localized' subgap states can be found with a
  probability that is exponentially small in the metallic conductance $g$.
\item Eventually, for energies $\epsilon \ll E_c$, the fluctuation physics is
  governed by the C-mode whose impact on various physical
  observables (for {\em non}-proximity effect SN-structures) has already
  been discussed in the literature.
\end{itemize}
It is important to question whether the above corrections can be made
experimentally visible. As far as the DoS is concerned, the answer
must be a conservative one: the chances are that it will be impossible
to separate the high energy $1/g$-corrections from the Usadel
background.  Furthermore, for good metals ($g\gg 1$), finding nearly
localized subgap states will also be difficult, since, as shown in
Refs.~\cite{Muz,Falko,Mirlin}, disorder configurations leading to
nearly localized states are exponentially rare.  Thus, as far as the
mean DoS is concerned, the above fluctuation contributions will
probably be hard to detect. However, the primary purpose of this
section has been to demonstrate that a variety of interference
mechanisms adding to the standard quasiclassical picture exist {\it in
  principle}. If and to what extent these fluctuations give rise to
observable changes in single particle properties other than the DoS
(e.g. the Josephson coupling characteristics) represents a subject of
future research.

We now leave the issue of the renormalization of single particle
properties and turn to the discussion of correlations between more
than one Green function, as described by the b- and c-type
fluctuations. 

\subsection{b-Type Fluctuations: The Goldstone Mode}
\label{sec:bfluct}
In this section we discuss the class of fluctuations around the Usadel
saddle point which above has been denoted by 'type b)'. Unlike the
a-fluctuations, fluctuations of type b) induce {\it correlations}
between different Green functions.  What makes the b-fluctuations
particularly important is their Goldstone mode character: In the limit
of vanishing energy {\it difference} between the considered Green
functions, these modes become truly massless, a signature for the
presence of pronounced mesoscopic fluctuations.

Consider the action~(\ref{ea}) in the simple case $\omega_+=\vc A=0$.
Obviously, any transformation $\bar Q \rightarrow T_0 \bar Q T_0^{-1}$
leaves the action invariant provided that $[T,\sigma_i^{\rm ph}]=0$.
Among the group of matrices ${\bf G_0} \equiv \{T=T_0\otimes
\openone^{\rm ph}\}$, there is a subgroup ${\bf H_0} \subset {\bf
  G_0}$, $[{\bf H_0},\sigma_3^{\rm ar}]=0$ which not only leaves the
action invariant, but also the saddle point solution $\bar Q$ itself.
As a consequence, fluctuation matrices contained in the subgroup ${\bf
  H_0}$ are completely ineffective and do not couple to the theory.
However, the elements $T_0$ of the  coset space ${\bf G_0}/{\bf
  H_0}$ {\it do} generate non-trivial transformations of the diagonal
saddle point. Moreover, in the limit $T_0={\rm const.}$, these
transformations do not alter the action -- they are Goldstone modes.

Being Goldstone modes, the effective action of the $T_0$'s can only
contain gradient terms and mass terms induced by sources of symmetry
breaking, such as finite $\omega_+$ and $\vc A$. The
actual structure of the action depends crucially on its behaviour
under time reversal. For the sake of simplicity, we focus here on the
two pure symmetry cases:
\begin{itemize}
\item[i)] {\it Orthogonal symmetry}: The action is time reversal
  invariant, $\vc A=0$.
\item[ii)] {\it Unitary symmetry}: Time reversal invariance is broken,
  $|\vc A|L\Phi_0^{-1}\gg g^{-1/2}$.
\end{itemize}
Here, $\Phi_0$ is the flux quantum. Note that $\vc A$ denotes the
vector potential {\it with} account for the phase difference between
the superconducting terminals, so that phase differences $\Delta
\varphi \Phi_0^{-1} \gg g^{-1/2}$ suffice to drive the system into the
unitary symmetry class. The reason that $g^{-1/2}$ appears as a
measure for the strength of the perturbation is that, for $|\vc
A|L\Phi_0^{-1}\gg g^{-1/2}$, the dimensionless coupling constant of
the symmetry breaking operator in the action exceeds unity (see
Appendix \ref{app:goldstone}).  Alternatively, one may say that under
these conditions, the mass of the 'Cooperon' greatly exceeds the level
spacing.

\begin{figure}
\begin{center} \epsfig{file=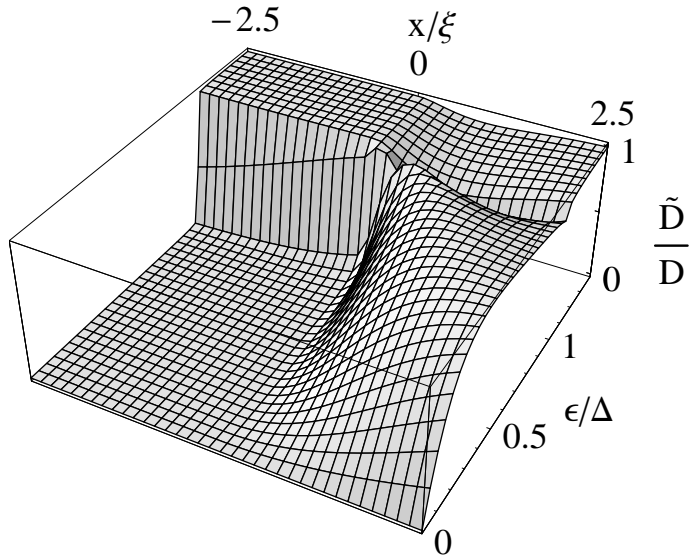,height=8cm}
\end{center}
\caption{The effective diffusion constant, $\tilde{D}$, for the
  Goldstone modes in an SN junction
as a function of both energy and position, for 
$\gamma=0.1$.}
\label{fig:Deff}
\end{figure}

The derivation of the effective action, $S_0$, of the Goldstone modes
is somewhat technical and has been deferred to Appendix
\ref{app:goldstone}. Here we merely state the result,
\begin{equation}
  \label{S0}
  S_0[Q_0] = - c_{\rm sym}\frac{\pi}{4}
  \int \left[\nu_n\tilde D 
   {\, \rm str}_0\left(\partial Q_0\partial Q_0\right)  + 2i
   \omega_+ \nu{\, \rm str}_0\left(Q_0  \sigma_3^{\rm ar}\right)\right],
\end{equation}
where $\tilde D = \frac{D}{2} (1-\vc q_+ \cdot \vc q_-)$ plays the
r\^ole of a space dependent diffusion coefficient. The variation of
$\tilde{D}$ with both position and energy is shown in
fig.~\ref{fig:Deff} for an SN junction with a typical choice of
material parameter.  In addition, $\nu$ is the space dependent, local
DoS, as displayed already in fig.~\ref{fig:3Dsn}. (Notice
that in the bulk of S both the diffusion constant and $\nu$ vanish.
Hence, the support of the action of $Q_0$ is restricted to the N
region.)

Further, in the case of
\begin{itemize}
\item[i)] {\it Orthogonal symmetry:} $c_{\rm sym}=1$, the matrices
  $T_0$ are eight dimensional and obey the time reversal symmetry
  relation (\ref{Tsym}), and of
\item[ii)] {\it Unitary symmetry:} $c_{\rm sym}=2$, the matrices
  $T_0$ are four dimensional (i.e. they do not carry a tr-index
  structure) and Eq.~(\ref{Tsym}) is meaningless.
\end{itemize}
In either case the matrices $T_0$ obey the restricted version of the
symmetry relation (\ref{Usym})\footnote{At first sight it seems like
  we are facing a problem here: A rotation matrix cannot be of the
  form $T=T_0\otimes \openone^{\rm ph}$ and simultaneously obey the
  general relation (\ref{Usym}). The reason that matrices of b) type
  are nonetheless permitted is that the relation (\ref{Usym}) is
  in fact too strict. What matters is the restriction of the symmetry
  to the boson-boson and the fermion-fermion block of the matrices $T$
  (see the corresponding discussion in Ref.~\cite{ZirnMath}.)
  With regard to these bf-diagonal blocks, matrices $T_0$ obeying
  Eq.~(\ref{U0sym}) {\it are} compatible with Eq.~(\ref{Usym}).}
\begin{equation}
\label{U0sym}
  T_0^\dagger = \eta_0 T_0^{-1} \eta_0^{-1},
\end{equation}
where 
\[
\eta_0 = E_{11}^{\rm bf} \otimes \sigma_3^{\rm ar} + E_{22}^{\rm bf}.
\]
In summary, we see that the symmetry of the Goldstone fields is
identical to those of the standard Efetov $Q$-matrix
manifolds\cite{Efetov}.  In other words, by freezing out the
ph-degrees of freedom, the large 16-dimensional $\sigma$-model
manifold collapses to smaller ones of dimensionality 8 (4) which are
symmetrically identical to those encountered in orthogonal (unitary)
applications of the standard $\sigma$-model.

Besides the general symmetry relations, a further condition to be
imposed on the fields is that they obey Neumann boundary conditions
$\partial_\perp T_0(x)=0$ at all SN-interfaces. The derivation of
these boundary conditions is discussed in Appendix
\ref{app:goldstone}.

So far the discussion has been for a general SN geometry. 
In order to actually demonstrate how the b-modes generate
mesoscopic fluctuations we next consider a specific example, namely, the
problem of DoS  fluctuations above the minigap edge in an
SNS-structure.

\subsubsection{Level-Statistics in SNS-structures}
\label{sec:bfluctsns}
It is well known that the single particle spectrum of mesoscopic, {\em
  purely} normal systems is governed by 
various types mesoscopic fluctuations (see
e.g. Ref. \cite{Efetov} for review). The fluctuation behaviour can
be characterized conveniently in terms of correlations $\sim \langle
\nu(\epsilon-\omega/2)\nu(\epsilon+\omega/2)\rangle$ between the DoS's
at different energies. Extensive analyses of correlation functions of
this type have shown that the DoS-correlations become increasingly universal
in character, the lower the energy separation $\omega$ (a fact that follows
heuristically from the interpretation of $\omega$ as an inverse 
time scale). In particular, for energies $\epsilon < E_c$ the
correlations become {\it fully} universal in the sense that they
depend on nothing more than the mean (and constant) level spacing
$\bar d$ and the fundamental symmetries of the system. This is the
regime of Wigner-Dyson statistics. For larger energies, the
Wigner-Dyson behaviour crosses over to other and less universal
types of statistics. Nevertheless, the correlations remain
energetically long-ranged in the sense that they decay algebraically
as a function of $\omega$. 

Here we ask to which extend this behaviour carries over to the
fluctuation behaviour of the SNS DoS in the vicinity of the
minigap. As compared to normal metals, the situation is more intricate
in that already the mean DoS is affected by mechanisms of quantum
coherence. A conceivable situation is that the (strong) modes
of quantum interference, giving rise to the particular structure in the mean
DoS, decouple entirely from the modes responsible for DoS
fluctuations. Another possibility is that one might end up 
with some kind of inseparable conglomerate of modes of
interference, and thereby fundamentally non-universal types of spectral
statistics. Here we demonstrate that the true picture lies  somewhere halfway
between these two extreme options: It is still possible to identify a regime of
universal Wigner-Dyson statistics, albeit superimposed on an
energetically non-uniform DoS. However, its range of validity 
shrinks down to a small energy window beyond which the correlations do
become entirely non-universal.

To obtain specific information
about spectral fluctuations in an
SNS-structure, we apply the action (\ref{S0}) to the analysis of the SNS geometry, shown
in fig.~\ref{fig:sns}.
The set of field configurations obeying Neumann boundary
conditions at the SN-interfaces obviously contains a subset with
trivial spatial dependence, $T_0={\rm const.}$ -- the
'zero mode'. The zero mode action reads:
\begin{eqnarray}
S_0[Q_0={\rm const.}] = -i\frac{\pi}{2}\frac{\omega_+} {\bar{d}
  (\epsilon)}{\rm str}_0 
[Q_0 \sigma_3^{\rm ar}].
\label{zerom}
\end{eqnarray}
Here $\bar{d}(\epsilon) = (\int \nu(\epsilon))^{-1}$ denotes the
average level spacing, which, in contrast to the purely normal case, is
now energy dependent. Note the similarity with the action,
(\ref{ea_C0}), of the 
C-zero mode. The difference is in the pre-factor and in
the physical spaces in which the matrices $Q$ and $Q_C^0$,
respectively, operate.

In order to demonstrate any significance of the isolated zero mode
action, we have to show that it is separated by an energy gap from the
action of all other field configurations. At this point, the boundary
conditions begin to play a crucial role. Expanding the fields in terms
of cosines, that is, a complete set of functions compatible with
Neumann boundary conditions, we see that, next to the zero mode, the
field with least curvature varies as $\sim \rm cos(2\pi x/L)$. Due to
the presence of the gradient term, the action associated with this
configuration is of ${\cal O}(D /(L^2 \bar d)$. Thus, for $\omega_+
\ll E_c$, the zero mode action is energetically gapped against all
fluctuating contributions and plays a dominant role.

We will see in the next section that, due to the presence of the
c-type fluctuations, the range of stability of the zero mode is
actually much smaller than $\omega_+ < E_c$. Yet, restricting
ourselves for a moment to the consideration of the zero mode action,
we can, without any further calculation, draw immediate conclusions
about the level statistics in SNS-systems over small correlation
intervals, $\omega_+$. In fact, actions of the structure (\ref{zerom})
are standard in applications of the $\sigma$-model in N-mesoscopic
physics: they appear a) whenever a model may be subjected to a zero
mode approximation or b) when one is dealing with $\sigma$-model
analyses of a single random matrix ensemble. With regard to spectral
statistics, the existence of the zero mode action (\ref{zerom})
implies that {\it level correlations for small energies are of
  Wigner-Dyson type}.

Furthermore, a comparison of the action (\ref{zerom}) with the
analogous action for N-systems \cite{Efetov} shows the correlations to
depend on an average level spacing that is effectively {\em halved}.
This is due to the strong `hybridisation' of levels at energies $\sim
\epsilon_F \pm \epsilon$ induced by Andreev scattering at the
SN-interface.

A more comprehensive discussion of level
fluctuations, including the differences to the types of spectral
statistics found in N-materials, will be given after the c-type
fluctuations have been incorporated in our analysis.

\subsection{c-Type Fluctuations: Quantum Corrections to Level Statistics}
\label{sec:cfluct}
In the previous section, an effective action for b-type fluctuations
was derived, and the latter were shown to be Goldstone modes of the
theory.  Furthermore, for the SNS geometry at energy scales
$\omega_+\ll E_c$, the effective action was shown to be dominated by a
zero mode which established universality of level statistics within
the ergodic regime. Higher modes give rise to non-universal
corrections on energy scales $\omega_+ \sim E_c$. At the same time,
c-type fluctuations, that is, fluctuations that commute with neither
$\vc{\sigma}^{\rm ph}$ nor $\vc{\sigma}^{\rm ar}$), also become
important.

The aim of this section is to examine the r\^ole played by c-type
fluctuations in limiting the regime over which universal correlations
persist\footnote{It is conceivable that the unfolding procedure of the
  previous section, that allows for a energetically inhomogeneous mean
  DoS, may itself be a source of decorrelation on energy scales
  $\omega_+ \sim E_c$.  Such a mechanism for non-universal corrections
  to the level statistics would be {\em separate} to that described
  here for the c-type fluctuations and so is not contained within our
  present analysis.}. Since, for states above the minigap, c-type
fluctuations incur a mass which is of order $\epsilon/\bar{d}> g\gg
1$, fluctuation corrections to the universal level statistics can be
treated within a perturbative manner. Our approach will be based on
the perturbative treatment introduced by Kravtsov and
Mirlin~\cite{K+Mirlin} in studying similar corrections in normal
disordered conductors.

We begin by employing the general parameterization
\begin{eqnarray}
Q = T_0 T \vc{q}\cdot\vc{\sigma}^{\rm ph} T^{-1} T_0^{-1},\qquad 
T = \exp\left[\sum_{\mu=0}^3 W_\mu\sigma_\mu^{\rm ph}\right],
\end{eqnarray}
where $Q_0=T_0\vc{q}\cdot\vc{\sigma}^{\rm ph}T_0^{-1}$ represents the
spatially homogeneous zero mode. Here we have applied the notation
$\sigma_0 \equiv\openone$, and, separating the zero mode, we impose
the requirement 
that the fluctuations obey the constraint $\int W_0=0$. Here, as
before, $q=\vc{q}\cdot\vc{\sigma}^{\rm ph}$ represents the
saddle-point, or Usadel, solution.

Applying this parameterization, and expanding to quadratic order in the 
fields $W$, the total correction to the zero mode action~(\ref{zerom}) takes 
the form
\begin{eqnarray}
\delta S[Q_0,W]=&&-\frac{\pi\nu_n}{8}\int{\rm str}
\Big\{D\left([\widetilde{\partial}W,q]^2+2[\widetilde{\partial}W,W]q\widetilde
{\partial}q\right)\nonumber\\ &&-2i\epsilon q\left[W,[W,\sigma_3^{\rm ph}]
\right]\nonumber\\
&&+4i\omega_+ U_0\sigma_3^{\rm ph}[W,q]\Big\}
+O(W^3),
\label{totala}
\end{eqnarray}
where $U_0 = T_0^{-1}\sigma_3^{\rm ar}T_0$.  In contrast to a normal
conductor, inhomogeneity of the saddle-point solution $q$ allows a
term {\em linear} in $W$ to survive in the action. The presence of
this term has important consequences on the range over which level
correlations are universal.

To proceed, it is convenient to further separate c-type fluctuations into 
two classes, $W=W^A+W^D$:
\begin{itemize}

\item Modes diagonal in ar-space ($[W_\mu^D,\sigma_3^{\rm ar}] = 0$), but 
off-diagonal in ph-space, are termed $D$ modes; $W^D=W_1^D\otimes
\sigma_1^{\rm ph}+W_2^D \otimes\sigma_2^{\rm ph}$.

\item Modes off-diagonal in ar-space ($[W_\mu^A,\sigma_3^{\rm ar}]_+ =
  0$), but diagonal in ph-space, are termed $A$ modes\footnote{The
    denotation 'A' respectively 'D' modes is again motivated by
    Cartan's classification scheme of symmetric spaces (see
    Ref.\cite{A+Zirn} for a discussion of the scheme focusing on its
    application to the symmetry classification of SN-systems.)};
  $W^A=W_0^A\otimes \openone^{\rm ph}+W_3^A\otimes\sigma_3^{\rm ph}$.

\end{itemize}
On the level of the quadratic action, no mixing between these modes
occurs.  Finally, considerations analogous to those presented in
connection with the a-type fluctuations show the spectrum of these
modes to be discretely spaced, where the typical 'level distance' is
of ${\cal O}(E_c)$. In passing we note that the c-modes, as introduced
above, are in fact not complementary to the a-modes. For example, 
the above $W^D$'s
contain modes $\propto \openone_{\rm ar}$, which, by definition,
belong to type a). However, the present analysis,
regarding the
impact of non-universal fluctuations on the b-type Goldstone mode, 
does not require a separation of the a- and c-modes and it is sufficient
to continue with the present definition of the c-modes.

Once again, to keep our discussion simple, we limit consideration to
pure symmetry classes of either orthogonal or unitary type. This leads
to a simplification of the effective action~(\ref{totala}) allowing an
explicit integration over the fluctuations $W$. Specifically, for pure
symmetry, the action takes the form $\delta S=S_A+S_D$, where
%
\begin{eqnarray}
S_A &=& \pi\nu_n \int {\rm str}\left[W_0^A\left(-\partial
\tilde{D} \partial\right)W_0^A +i\omega_+ W_0^A\sigma_3^{\rm ar}
([q_3]_+-[q_3]_-) U_0 \right], \\
S_D &=& \pi\nu_n \int {\rm str}\left[W_2^D\left(-D\partial^2
-2i\epsilon q_3\right)W_2^D +2\omega_+ W_2^D q_1 U_0
\right]
\end{eqnarray}
%
and $\tilde{D}$ is the space and energy dependent diffusion constant
that has been introduced in section~\ref{sec:bfluct} for the Goldstone
modes.  Note that the $W_3$ fluctuations do not couple linearly to
$U_0$ in the pure symmetry case and so may be dropped.

As can be seen from the general structure of the action, c-type
fluctuations in the vicinity of the minigap are generally 'massive',
that is, governed by an action which is at least of order
$\epsilon/\bar{d} \gtrsim g\gg1$.  It is thus permissible to treat
these fluctuations in a simple Gaussian approximation. Applying the
shift operations 
\begin{eqnarray}
W_0^A &\rightarrow& W_0^A -\frac{i\omega_+}{2}\hat{P}_0 (-\partial \tilde{D}
\partial)^{-1} \sigma_3^{\rm
  ar}([q_3]_+-[q_3]_-)U_0, \\
W_2^D &\rightarrow& W_2^D-\omega_+\hat{P}_2 (-D \partial^2)^{-1} q_1 U_0,
\end{eqnarray}
where $\hat{P}_\mu$ represents a projector onto the field space of
$W_\mu$, then performing the Gaussian integral, we obtain the
renormalized zero mode action
\begin{eqnarray}
S[Q_0]=S_0[Q_0]-\frac{\kappa(\epsilon)}{g}\left(\frac{\omega_+}
{\bar{d}(\epsilon)}\right)^2{\rm str}[\sigma_3^{\rm ar},Q_0]^2,
\label{modzero}
\end{eqnarray}
where $\kappa(\epsilon)\sim O(1)$ represents a constant that depends
on the sample geometry.

Eq.~(\ref{modzero}) has a structure equivalent to that found in the
study of universal parametric correlation functions and explicit
expressions for the two-point correlator of DoS fluctuations for both
orthogonal and unitary ensembles can be deduced from
Ref.~\cite{Simons}\footnote{With reference to the specific correlation
  function $R_2$, we remark that only massive fluctuations in the
  ph-sector contribute to connected correlators of the form
  $\left\langle {\cal G}^A_{\epsilon_1} {\cal
      G}^A_{\epsilon_2}\right\rangle$ allowing such terms to be
  neglected.}. Qualitatively, the additional contribution
in~(\ref{modzero}) counteracts the zero-mode fluctuations for
non-vanishing frequencies $\omega_+$.
 
Furthermore, we find a marked difference in the manner in which level
correlations are suppressed as compared to the purely N case.  Already
for energy separations $\omega_+/\bar{d}(\epsilon)\sim \sqrt{g}$, the
zero-mode integration is largely suppressed which manifests in an
exponential vanishing of the level correlations on these scales. This
represents a qualitatively smaller energy scale than that for the
purely N case, for which Wigner-Dyson statistics prevail all the way
up to frequencies $\omega_+ \simeq E_c$. In addition, the exponential
suppression of correlations in the SN case differs from the purely N
case, for which the Wigner-Dyson statistics are succeeded by other
forms of {\it algebraically} decaying spectral statistics in the high
frequency domain $\omega_+>E_c$\cite{AS}. Note that a similar phenomenon
of zero-mode suppression has recently been observed by Skvortsov et
al.\cite{Feiglman2} in their analysis of the level statistics of
normal core excitations in type II superconductor vortices. 

\section{Discussion}
\label{sec:discuss}
In conclusion a general framework has been developed in which the interplay 
of mesoscopic quantum coherence phenomena and the proximity effect can be 
explored. The connection between the conventional quasiclassical approach
and the field theoretic approach adopted here has been emphasized. In 
applying the effective action we have introduced a classification of 
different modes of fluctuations. 

To keep our analysis simple we have focussed on a regime in which the
contact between the superconductor and normal regions is metallic, and
where $\Delta\gg E_c$. Experimental analyses are often carried out in
the complementary regime where tunnel barriers separate S and N and/or
$\Delta\lesssim E_c$. It is straightforward to modify the theory so as
to accommodate tunnel barriers, small order parameters and, in fact,
altogether different sample layouts. However, in order not to
diversify the present exposition of the formalism even further, we
have restricted ourselves to the analysis of the relatively simple
systems discussed above. Whereas certain of our conclusions (e.g. the
existence of a Wigner-Dyson regime of spectral correlations) carry
over to the case of barrier separated SN systems, others do not. More
specifically, reducing the strength of the order parameter below $E_c$
affects both the behaviour of certain of the fluctuation classes
discussed above and the spatial structure of the solutions of the
Usadel equation\cite{KO}. Rather than attempting to set up a most
general 'phase diagram' of mean field and fluctuation regimes -- given
the diversity of SN-systems with qualitatively different physically
behaviour, certainly a fruitless task -- it is more efficient to treat
different problems individually, i.e. to start out from the most
general form of the action (\ref{ea}) and to restrict the analysis to
those fluctuation modes that encompass the physics particular to the
problem under consideration (see, e.g.  Refs.~\cite{Brouwer98} and
\cite{Zirnbauer98} for recent examples.) Whether or not a certain type
of fluctuation around the Usadel saddle point is 'relevant' or not
can be deduced from the way it couples to the
different contributions to the action.

Finally, we note that, in this paper we have focussed on the influence
of mesoscopic fluctuations on the proximity effect in {\em disordered}
SN-structures. However, mechanisms of quantum interference analogous
to those discussed here also induce mesoscopic fluctuations in
irregular clean or ``quantum chaotic'' structures. Moreover, the
proximity effect is strongly influenced by such coherence phenomena
allowing them to be employed as a potential probe of chaotic
behaviour~\cite{Melsen,Lodder98}. Can the framework developed above
for disordered SN-structures be generalized to account for chaotic or
ballistic SN-structures? To address this question we should begin by
recalling the properties of normal chaotic structures.

In fact the connection between the statistical field theory of normal
disordered conductors and ballistic chaotic structures was motivated
by the quasiclassical approach of Eilenberger discussed previously.
Recognizing that the Usadel equation could be associated with the
equation of motion corresponding to the saddle-point of the action of
the diffusive non-linear $\sigma$-model~\cite{Muz}, Muzykantskii and
Khmel'nitskii proposed that the Eilenberger equation could be
identified with a {\em ballistic} analogue of the non-linear
$\sigma$-model action~\cite{Muz2}. In this case, the diffusive
character of the action was replaced by a kinetic operator. Their work
found support in subsequent investigations based on the study of
energy averaged properties of (again normal) chaotic structures which
led to a microscopic derivation of the ballistic
action~\cite{Andreev96}. Taken together, these studies showed that,
while density relaxation in disordered conductors is diffusive, in
general chaotic structures it is governed by modes of the {\em
  irreversible} classical evolution operator.

The generalization of the ballistic field theory to encompass the
proximity effect follows naturally from the ideas presented in this
paper.  Expanding the field space of $Q$ to accommodate particle/hole
degrees of freedom and, as in the diffusive model, introducing the
inhomogeneous order parameter $\hat{\Delta}$, the effective ballistic
action takes the form
\begin{eqnarray}
S[Q] = i\frac{\pi}{2\bar{d}}\int {\rm str}\left[2i T\sigma_3^{\rm ar}
\otimes\sigma_3^{\rm ph}\{H,T^{-1}\} -\sigma_3^{\rm ph}\otimes
\left(\hat{\Delta}+\epsilon+\frac{\omega_+}{2}\sigma_3^{\rm ar}\right)Q
\right],
\label{ballact}
\end{eqnarray}
where $\{H,\ \}$ represents the Poisson bracket of the {\em classical} 
Hamiltonian, and the supermatrix fields $Q=T\sigma_3^{\rm ar}\otimes
\sigma_3^{\rm ph}T^{-1}$ depend on the $2d-1$ phase space coordinates 
parameterizing the constant energy shell, ${\bf x}_\parallel=
(\vc r,\vc p)_{2d-1}$. (With this definition the integration measure is 
normalized such that $\int\equiv\int d{\bf x}_\parallel=1$.) In the presence 
of a Gaussian distributed $\delta$-correlated impurity potential, the 
ballistic action is supplemented by a further term corresponding to a 
collision integral~\cite{Muz2}
\begin{eqnarray}
S_{\rm coll} = \frac{\pi}{4\bar{d}\tau}\int \frac{d\vc{r}}{L^d}
\frac{d\vc n d \vc n'}{S_d^2}{\rm str}\left[Q(\vc n,\vc r)Q(\vc n',\vc
  r)\right],
\label{ballcol}
\end{eqnarray}
where $\vc n = \vc p/|\vc p|$ and $S_d=\int d\vc n$. Indeed, for a strong 
enough impurity potential, $\bar{d}\tau\ll 1$, a moment expansion of the 
action recovers the diffusive action. Varying the action with respect to $Q$, 
and applying the identification $g(\vc n,\vc r) \leftrightarrow 
Q(\vc n,\vc r)$, the saddle-point equation of motion coincides with the 
Eilenberger equation of transport, Eq.~(\ref{Eilen}).

Although, in principle the ballistic action represents a complete theory of
statistical correlations in chaotic SN-structures, an analytical description
of the modes of the classical evolution operator has proved difficult to 
construct. In particular, the sensitivity of weak localization corrections to
mechanisms of ``quantum diffraction'' and ``irreversibility'' in normal clean 
chaotic structures has proved difficult to quantify~\cite{Aleiner}. In the 
SN-geometry the same mechanisms have a dramatic effect on the single-particle
properties of the device such as the minigap structure in the local 
DoS~\cite{Melsen} (see the discussion in section~\ref{sec:qdiff}). For this reason, we believe 
that SN-structures may provide a versatile arena in which properties quantum 
chaotic systems can be explored.

\begin{acknowledgements}
We are indebted to Boris Altshuler, Anton Andreev,
Dima Khmel'nitskii, Vladimir Fal'ko, Alex Kamenev, Vladimir Kravtsov,
Igor Lerner and Martin Zirnbauer for useful
discussions. One of us (DT-S) acknowledges the financial support of
the EPSRC and Trinity College. 
\end{acknowledgements}


\appendix

\section{Boundary conditions of the Usadel equation}
\label{sec:boundary}
In addition to the transport equations provided by quasiclassics, it
is necessary to specify boundary conditions at the SN-interfaces.  For
the Eilenberger equation, at a planar SN-interface with an arbitrary
transmission coefficient, $T$, these conditions have been derived by
Zaitsev \cite{Zaitsev}. Note that these boundary conditions {\em cannot}
be obtained using the standard quasiclassical Green function
(\ref{quasi}) alone.  Instead, one must go back to a more microscopic
formulation.

Following the general philosophy of this section we shall not review the
(somewhat technical) derivation of the boundary conditions but merely
formulate the main results.  The Eilenberger Green function, $g(\vc
n,\vc{r})$, may be separated into symmetric and antisymmetric parts,
$g = g_s+g_a$, with respect to the operation $\vc{v}_F \rightarrow
-\vc{v}_F$. The antisymmetric part, $g_a$, is continuous across the
interface.  In passing we note that this results in the conservation
of  the supercurrent density, 
\begin{eqnarray}
\vc j = -\frac{p_F^2}{4\pi} \left\langle
\vc n \,{\rm tr} \sigma_3^{\rm ph}
  g^{\rm r}(\vc{n},\vc{r}) \right\rangle_{\vc n},
\label{Idef}
\end{eqnarray}
across the interface. In contrast, the symmetric part experiences a jump
depending on the transmission coefficient. The resulting conditions
are:
\begin{mathletters}
\begin{eqnarray}
&&g_a(+) = g_a(-)\equiv g_a, \label{Zaita}\\
&&g_a\left\{R(1-g_a g_a)+\frac{T}{4}(g_s(+)-g_s(-))^2\right\} =
\frac{T}{4}[g_s(-),g_s(+)],
\label{Zaits}
\end{eqnarray}
\end{mathletters}
where $R=1-T$ is the reflection coefficient, the
r,a superscripts have been dropped, and $g_s(\pm)$ denotes the Green
function infinitesimally to the left respectively right of the junction.

For a perfectly transparent ($T=1$) interface, both parts of $g$ are
continuous.  In the low transparency limit, $T\ll1$, we have $g_a \sim
T$ and (\ref{Zaits}) reduces to
\begin{eqnarray}
g_a= \frac{T}{4R}[g_s(-),g_s(+)].
\label{Zaits2}
\end{eqnarray}
The above boundary conditions simplify further in the dirty limit.  As
shown by Kuprianov and Lukichev \cite{Kuprianov}, the reformulation of
(\ref{Zaita}) and (\ref{Zaits2}) in terms of the Usadel Green function
leads to the pair of conditions,
\begin{mathletters}
\begin{eqnarray}
\sigma(-) g_0 \partial_r g_0(-) &=& \sigma(+) g_0\partial_r g_0(+) \\
&\stackrel{T\ll1 }{=} &\frac{G_T}{2}[g_0(+),g_0(-)],
\end{eqnarray}
\end{mathletters}
where
$\sigma(\pm)$ is the metallic conductance on either side of the interface, and 
\begin{eqnarray}
G_T = \frac{e^2 \nu_n v_F}{2} \int_0^1 d(\cos\alpha)\frac{T}{R}
\cos\alpha,
\label{tunnel_cond}
\end{eqnarray}
is the tunnel conductance of the junction, $\alpha$ is the angle
between $\vc{n}$ and $\vc{r}$, and $\nu_n$ is the bulk, normal
metallic DoS. Note that $T$ may depend here on $\alpha$. 
Note also that the second condition of Eq.~(\ref{Kup2})
applies only in the limit $T\ll1$.  Lambert et al.
\cite{Lambert97} have recently examined this restriction and
how it may be relaxed.  In 
the following, we exclusively consider the opposite case of perfect
transmittance, $T=1$, for which the second condition, Eq.~(\ref{Kup2}),
should be replaced by 
\begin{equation}
  g_0(+)=g_0(-).
\end{equation}

\section{Saddle Points and Analytic Continuation}
This appendix is devoted to a discussion of the question of how the
$Q$-saddle-point configurations that appear in applications with
superconductivity may be accessed from the starting-point of
 the diagonal
saddle-point configuration, $\sigma_3^{\rm ph} \otimes \sigma_3^{\rm ar}$,
characteristic for bulk metallic phases. In order to specify what we
mean by 'access', we first have to summarize some facts
regarding the structure of the field manifold of the
$\sigma$-model. In the polar representation of Efetov~\cite{Efetov},
a general $Q$-matrix is parameterized as 
\begin{equation}
  \label{app_QT}
Q=T (\sigma_3^{\rm ph} \otimes \sigma_3^{\rm ar}) T^{-1},
\end{equation}
where $\sigma_3^{\rm ph} \otimes \sigma_3^{\rm ar}$ is the generalization of
the standard matrix, $\Lambda=\sigma_3^{\rm ar}$, to applications with a
ph-substructure. The rotation matrices obey $T \in {\bf G}/{\bf H}$, where
${\bf G}$ is a group of matrices that fulfill various symmetry
conditions, dictated by both the internal symmetries of the model under
consideration and convergence criteria. The group ${\bf H} \subset
{\bf G}$ is determined by the condition $[{\bf H}, \sigma_3^{\rm ph} \otimes
\sigma_3^{\rm ar}]=0$. The most important constraint for 
the present discussion has to do with convergence and reads,
\begin{equation}
  \label{app_Tsym}
  T^\dagger = \eta T^{-1} \eta^{-1},
\end{equation}
where 
\[
\eta = E_{11}^{\rm bf} \otimes \sigma_3^{\rm ph}
\otimes \sigma_3^{\rm ar} + E_{22}^{\rm bf}.
\]
The functional integration $\int dQ$ extends over the coset space
${\bf G}/{\bf H}$. A key question to address is whether or
not all stationary phase points of the action are accessible within
the integration domain specified by (\ref{app_QT}).

\begin{figure}
\begin{center} \epsfig{file=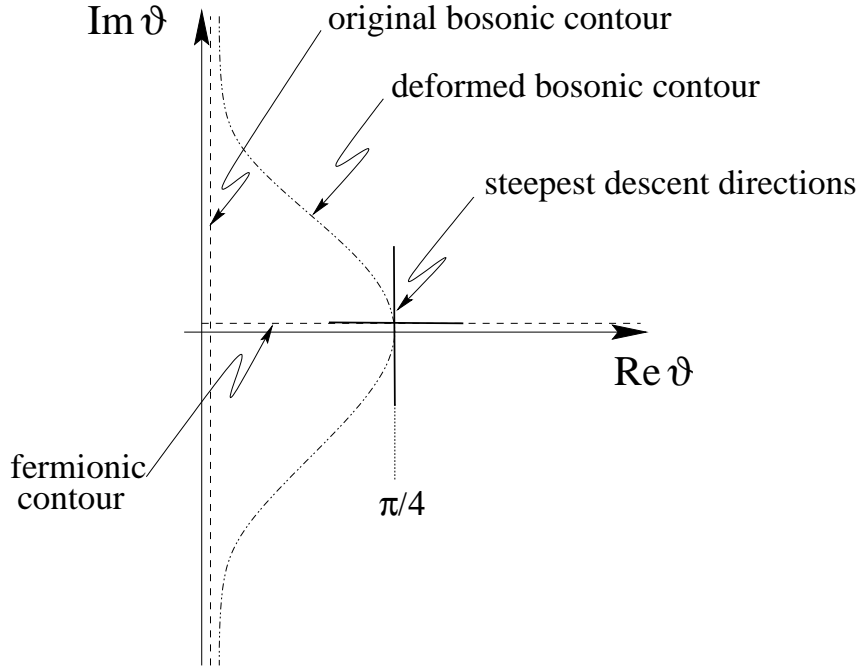,height=9cm}
\end{center}
\caption{Visualization of the deformed integration
  contour.}
\label{fig:analytic}
\end{figure}

To analyse this issue, we first focus on the simple case of a bulk
superconductor in the regime $|\Delta| \gg \epsilon$. In this case, the
saddle point is unique and reads (cf. section \ref{sec:bulk})
\[
\bar Q=\sigma_1^{\rm ph}.
\]
The above question reduces in this case to whether or not there 
exists a solution to the equation,
\begin{eqnarray}
\sigma_1^{\rm ph}\otimes \openone^{\rm ar} \stackrel{!}{=} 
T (\sigma_3^{\rm ph}\otimes \sigma_3^{\rm
  ar}) T^{-1}
\label{rotate}
\end{eqnarray}
Strictly speaking, no such solution exists. To demonstrate this point,
we first note that, since the rhs of Eq.~(\ref{rotate})
is trivial in ar, bf and tr-space, 
it is sufficient to focus on each sector of these spaces separately.
Without loss of generality, we focus on the retarded-retarded (rr) sector,
where Eq.~(\ref{rotate}) takes the form,
\begin{equation}
\label{app_rr_eq}
\sigma_1^{\rm ph} \stackrel{!}{=} T \sigma_3^{\rm ph} T^{-1}
\end{equation}
and we have, for reasons of notational simplicity, denoted the
rr-restricted rotation matrices again by $T$. The analogous equation
for the advanced-advanced block carries an overall minus sign on the
rhs. Specializing the discussion further to the the fermion-fermion (ff)
sector, we find that, in 
this sector, no problem arises in finding a solution to
Eq.~(\ref{app_rr_eq}) with the correct symmetries. When restricted to the
ff-block, the symmetry relation (\ref{app_Tsym}) takes the form
\[
T_{\rm ff}^\dagger = T_{\rm ff}^{-1},
\]
i.e. the ff-matrices are unitary. At the same time, the restriction
of Eq.~(\ref{app_rr_eq}) to the ff-sector reads
\[
\sigma_1^{\rm ph} \stackrel{!}{=} T^{\vphantom{-1}}_{\rm ff}
\sigma_3^{\rm ph}
T_{\rm ff}^{-1},
\]
which may be solved by a unitary rotation matrix. For future
reference, we explicitly write the solution as
\[
T_{\rm ff} = \exp(-i\theta_{\rm ff} \sigma_2^{\rm ph})
\Big|_{{\theta_{\rm ff}}=
  \frac{\pi}{4}}\in {\bf G}/{\bf H}.
\]
In the bb-sector, the situation is more problematic: 
we fail to find a solution to (\ref{app_rr_eq}) with
the correct symmetries. The
restriction of the symmetry criterion (\ref{app_Tsym}) on the retarded/retarded
block reads
\[
T_{\rm bb}^\dagger = \sigma_3^{\rm ph} T_{\rm bb}^{-1} \sigma_3^{\rm ph},
\]
which fails to include any $T_{\rm bb}$ that further fulfill 
\[
\sigma_1^{\rm ph} \stackrel{!}{=} T_{\rm bb}^{\vphantom{-1}}
\sigma_3^{\rm ph}
T_{\rm bb}^{-1}.
\]
The resolution of this problem is provided by analytic continuation. In the
derivation of the $\sigma$-model, the symmetry condition
(\ref{app_Tsym}) is enforced by convergence requirements. As long as
no singularities are encountered, the condition may be relaxed, in the sense
that the integration contours may be analytically continued to regions
where the symmetry criterion is no longer fulfilled. Supposing now we are
integrating over the subset (cf. fig.~\ref{fig:analytic})
\[
\{T=\exp(-i\theta_{\rm bb} \sigma_2^{\rm ph})|\theta_{\rm bb} \in i {\cal R}\}
\subset {\bf G}/{\bf H},
\]
the saddle point we wish to access is reached by distorting the
integration contour so as to cross the point $\theta_{\rm bb} =
\pi/4$. In lifting the integration path off the imaginary axis, no
singularities are encountered. Moreover, a closer analysis shows that, in
accordance with the basic conditions to be imposed on saddle point
integrals, the direction of steepest descent is parallel to the
imaginary axis. Analogous arguments may be applied to the
advanced/advanced sector. Consequently we conclude that $q$ does
represent a proper saddle point of the $Q$-integration.

The question remains as to what happens if we encounter more 
complicated saddle-point configurations, such as those with a 
finite value of $\epsilon/\Delta$ or even with spatial variation.
Although no mathematical proof has been given, we expect that by
analytic continuation such configurations remain accessible.
Independent evidence for the validity of such an assumption is
provided by physical criteria: The disordered mean DoS
may be calculated within the framework of the quasiclassical
approach, while the solution to the quasiclassical equations
coincides with the solution of the $\sigma$-model mean field
equations. Suppose now that the solution were inaccessible in either
block, bb or ff, or both. In this case the functional integral would
exhibit supersymmetry-breaking on the mean field level and it would be
obscure how to reproduce the correct results of quasiclassics. Given such
sources of evidence, we adopt a pragmatic point of view and
take for granted the accessibility of the Usadel-saddle points.

\section{Solutions to the Usadel Equation}
In this section we provide explicit solutions of the Usadel equation
for two simple, quasi-1D geometries: an SN junction and an SNS-junction with
coincident phases of the order parameters of the S regions ($\Delta\varphi=0$). 

\subsection{SN junction}
\label{app:sn}
We begin with the SN geometry. As with the case of
the bulk superconductor, we introduce an angular parameterization for 
$\vc{q}(x)$:
\begin{eqnarray}
\vc{q}(x) = (\sin\theta(x),0,\cos\theta(x)).
\label{angparam}
\end{eqnarray}
Since in this geometry
the phase of the order parameter is spatially constant, the gauge 
transformation of section \ref{sec:saddle} allows us to set the 
second component of $\vc{q}$ to zero. The saddle-point equation for $\vc{q}$,
Eq.~(\ref{Usad2}), becomes a sine-Gordon equation,
\begin{eqnarray}
\frac{D}{2}\partial^2_x \theta + i\epsilon\sin\theta + \Delta(x)\cos\theta = 0,
\label{diff1}
\end{eqnarray}
while the boundary condition at the interface, Eq.~(\ref{Kupbc}), becomes
\begin{eqnarray}
\sigma_s \partial_x \theta(0^-) = \sigma_n \partial_x \theta(0^+).
\label{bcint}
\end{eqnarray}
Also, the symmetry relation, Eq.~(\ref{arsymm}), becomes
\begin{eqnarray}
\theta_-(x) = \pi-(\theta_+)^*.
\label{arsym2}
\end{eqnarray}
In addition, there are further boundary conditions at infinity, at which the 
bulk values of the angle are approached, so that
\begin{eqnarray}
\theta(x) \rightarrow \left\{\begin{array}{lr}
\theta_s, & x\rightarrow-\infty,\\
\theta_n, & x\rightarrow \infty,
\end{array}\right.
\label{bcinf}
\end{eqnarray}
where $\theta_s$ is defined by the equations for the bulk order parameter,
eqns.~(\ref{bulk}), and $(\theta_n)_{+,-} = 0,\pi$. Note that, although 
the conditions (\ref{bcinf})
takes different forms in the ar-sectors, the existence of
the relation (\ref{arsym2}) means that
we need solve only for the retarded component, $\theta_+$, 
and in the following we drop the 
+ subscript.

The solution of the sine-Gordon equation, Eq.~(\ref{diff1}), with boundary
condition (\ref{bcinf}), is of the following (solitonic) form:
\begin{eqnarray}
\theta(x) = \left\{\begin{array}{lr}
\theta_s+4\tan^{-1}\left[\exp\left(-\sqrt{\frac{2\sqrt{R}}{D_s}}x\right)
\tan\frac{\theta(0)-\theta_s}{4}\right], & x<0 \\
4\tan^{-1}\left[\exp\left(-\sqrt{\frac{-2i\epsilon}{D_n}}x\right)
\tan\frac{\theta(0)}{4}\right], & x>0
\end{array}\right.
\label{solsn}
\end{eqnarray}
where $R=|\Delta|^2-\epsilon^2$ as before. 

The integration constant $\theta(0)$ is fixed by
imposing the condition (\ref{bcint}) at the interface, to give
\begin{eqnarray}
\sin\frac{(\theta(0)-\theta_s)}{2}=\gamma \sqrt\frac{\epsilon}{i\sqrt{R}}
\sin\frac{\theta(0)}{2},
\end{eqnarray}
where $\gamma$ is a parameter representing the mismatch between the two 
materials:
\begin{eqnarray}
\gamma = \frac{\sigma_n/\sigma_s}{\sqrt{D_n/D_s}}.
\label{gam}
\end{eqnarray}
In the limit $\gamma\rightarrow0$, the bulk value 
of the angle, $\theta_s$, is imposed asymptotically at the interface, 
$\theta(0)\rightarrow\theta_s$ -- the boundary condition becomes `rigid'.

By Eq. (\ref{DoSexp}), the local DoS is obtained from the relation, 
\begin{eqnarray}
\nu(x) = \nu_n {\rm Re}\cos\theta(x).
\end{eqnarray}

\subsection{SNS Junction with Coincident Phases}
\label{app:sni}
We turn now to the geometry of an SNS-junction, 
of width $L$, and with coincident phases of the order
parameters in the S regions.
Overall symmetry about the origin leads to the
condition, 
\begin{eqnarray}
\partial_x \theta(0) =0.
\label{bcorigin}
\end{eqnarray}
Since an identical condition holds at a normal-insulator interface, 
the solution here also applies
to an SNI junction of width $L/2$. There are further conditions at
infinity,
\begin{eqnarray}
\begin{array}{lr}
\theta(x) \rightarrow \theta_s, & |x|\rightarrow \infty.
\end{array}
\label{bcinf2}
\end{eqnarray}
The solution of Eq.~(\ref{diff1}) for $\theta(x)$, incorporating
eqns.~(\ref{bcorigin}) and (\ref{bcinf2}), is as follows: 
\begin{eqnarray}
\theta(x) =
\cases{\theta_s+4\tan^{-1}\left(\exp\left(-\sqrt{\frac{2\sqrt{R}}
{D_s}}(|x|-L/2)\right)
\tan\frac{\theta(L/2)-\theta_s}{4}\right), & $|x| > L/2$, \cr 
 2 \sin^{-1}\left(\sin\frac{\theta(0)}{2} \, 
{\rm sn}\left(i\left(\frac{-2i\epsilon}{D_n}\right)
^{1/2}x+K\left(\sin\frac{\theta(0)}{2}\right),
\sin\frac{\theta(0)}{2}\right)\right), & $|x| < L/2$. \cr}
\label{SNSsol}
\end{eqnarray}

Here $K$ and sn are the complete elliptic integral of the first kind
and the Jacobi elliptic function, respectively (see \cite{Gradsteyn}).
The two integration constants, $\theta(0)$ and $\theta(L/2)$, are
related by
\begin{eqnarray}
\sin\frac{\theta(L/2)}{2} = 
\sin\frac{\theta(0)}{2} \, {\rm sn}
\left(i\left(\frac{-2i \epsilon}{D_n}\right)^{1/2}\frac{L}{2}+
K\left(\sin\frac{\theta(0)}{2}\right),\sin\frac{\theta(0)}{2}\right), 
\label{C1}
\end{eqnarray}
and the conditions at the interfaces, corresponding to Eq.~(\ref{bcint}),
give the further relation
\begin{eqnarray}
\sin\frac{(\theta(L/2)-\theta_s)}{2} = \gamma
\sqrt{\frac{\epsilon}{i\sqrt{R}}}
\left(\sin^2\frac{\theta(L/2)}{2}-\sin^2\frac{\theta(0)}{2}\right)^{1/2},
\label{bcint2}
\end{eqnarray}
where the parameter $\gamma$ is defined by Eq.~(\ref{gam}) as before. The
integration constants may then be determined by numerical solution
of eqns.~(\ref{C1}) and (\ref{bcint2}).

\section{Effective Action of the Goldstone Mode}
\label{app:goldstone}
The subject of this Appendix is a derivation of the effective action
for the Goldstone modes, represented by rotation matrices $T_0$ such
that $[T_0,{\bf \sigma}^{\rm ph} ]=0$. We consider separately the cases
of orthogonal and unitary symmetry.

\subsection{Time Reversal Invariant Action} 
We begin with the case of orthogonal symmetry.
Substituting the ansatz $Q\equiv T_0 \bar Q T_0^{-1}$ into the
effective action, Eq.~(\ref{ea}), we notice that we obtain two $T_0$
dependent terms: a gradient term, and a term proportional to $\omega_+$.
Note that the two remaining vertices, proportional to $\epsilon$ and $\hat
\Delta$, commute through and do not couple to $T_0$. Focusing on the
gradient term first, we obtain
\begin{eqnarray}
\label{app:gradient}
&&-{\pi D\nu_n\over 8}\int {\rm str}\left(\partial Q\partial
   Q\right) = 
-{\pi D\nu_n\over 8}\int {\rm str}\left((\partial + [T_0^{-1}(\partial
  T_0),\,.\,])\bar Q(\partial + [T_0^{-1}(\partial
  T_0),\,.\,])\bar Q\right)\nonumber\\
&&=-{\pi D\nu_n\over 8}\int {\rm str}\left(
\underbrace{\left([T_0^{-1}(\partial
  T_0),\bar Q]\right)^2}_{{\rm a)}} + 2\underbrace{[T_0^{-1}(\partial
  T_0),\bar Q]\partial \bar Q}_{{\rm b)}} \right) 
+ T_0{\rm -independent}.
\end{eqnarray}
We first observe that the contribution b) vanishes, through the use of
the condition $\bar Q^2 = \openone$.
Turning to the a)-term, we write
\[
{\rm a)} = {\rm str}\left([T_0^{-1}(\partial  T_0),\bar Q]^2\right)=
{\rm str}\left(\check\partial Q\check\partial Q\right),
\]
where $\check \partial$ is defined to be a derivative that acts only
on $T_0$. Making use of the ph-commutativity of the $T_0$'s, we next
introduce
\[
Q= T_0\bar Q T_0^{-1} = \frac{1}{2}\Delta \vc q \cdot \vc \sigma^{\rm ph} Q_0,
\]
where 
\[
Q_0\equiv T_0\sigma_3^{\rm ar} T_0^{-1}
\]
and $\Delta \vc q \equiv \vc q_+ - \vc q_-$. Tracing 
out the ph-indices, we now obtain
\[
{\rm a)} = {\rm str}\left(\check\partial Q\check\partial Q\right) =
\frac{\Delta q \cdot \Delta q}{2}{\rm str}_0\left(\partial Q_0
  \partial Q_0\right).
\]
Substituting this expression into Eq.~(\ref{app:gradient}) 
and using the fact that $\Delta \vc q \cdot \Delta \vc q =
2(1-\vc q_+ \cdot \vc q_-)$, we finally arrive at 
\begin{equation}
  \label{gradient}
   -{\pi D\nu_n\over 8}\int {\rm str}\left(\partial Q\partial
   Q\right)=-\frac{\pi \nu_n}{4} \int \tilde D {\rm
   str}_0\left(\partial Q_0\partial   Q_0\right)  
\end{equation}
as the gradient term of the  Goldstone action. Here, 
\begin{equation}
  \label{diffusion_constant}
  \tilde D \equiv \frac{D}{2} (1-\vc q_+ \cdot \vc q_-)
\end{equation}
plays the role of an effective (and generally space dependent)
diffusion coefficient. Note that the symmetry relation (\ref{arsymm})
implies that this diffusion coefficient is real.

We next turn to the computation of the $\omega_+$-dependent vertex:
\begin{eqnarray}
  \label{omega}
-\frac{i\pi \nu_n \omega_+ }{4} \int {\rm str} \left(Q \sigma_3^{\rm
  ph}
  \otimes \sigma_3^{\rm ar} \right)
&=& -\frac{i\pi \nu_n \omega_+ }{2} \int \Delta q_3{\,
  \rm
  str}_0\left(Q_0  \sigma_3^{\rm ar}\right)\nonumber\\
&=& -\frac{i\pi \omega_+ }{2} \int \nu {\, \rm
  str}_0\left(Q_0  \sigma_3^{\rm ar}\right),
\end{eqnarray}
where $\Delta q_3$ is the 3-component of the vector 
$\Delta \vc q$ and $\nu = \nu_n \Delta q_3$, the space
dependent local DoS (cf. Eq. (\ref{DoSexp})).  Combining
eqns.~(\ref{gradient}) and (\ref{arsymm}), we finally obtain
\begin{equation}
  \label{S0orthogonal}
  S_0[Q_0] = -\int \left[\frac{\pi \nu_n}{4} \tilde D 
   {\rm str}_0\left(\partial Q_0\partial Q_0\right)  +  \frac{i\pi
   \omega_+ 
   \nu}{2}{\rm str}_0\left(Q_0  \sigma_3^{\rm ar}\right)\right]
\end{equation}
as the final expression for the time reversal invariant Goldstone mode
action, as included in Eq.~(\ref{S0}). 
We emphasize that both the diffusion coefficient, $\tilde D$, and
the local DoS, $\nu$, are space dependent.

\subsection{Broken Time Reversal Invariance} 

If the invariance under time reversal is broken by an external
magnetic field and/or significant phase differences between the adjacent
superconductors, rotations operating non-trivially in tr-space are
frozen out. More precisely, if the total flux threading the
system, $\Phi/\Phi_0$, 
exceeds $1/g^{1/2}$, symmetry breaking contributions to
the action with coupling constants greater than unity 
appear\footnote{The fact that
  $1/g^{1/2}$ is the relevant scale follows simply from gauge
  invariance: In the presence of fields, the gradient operator
  appearing in (\ref{gradient}) generalizes to
  the the gauge invariant form $D\partial^2 \rightarrow
  -D(i\partial-e\vc{A}/c)^2$. For a static field of strength $\Phi/{\rm
    Area}$, the vector potential $A\sim \Phi/ {\rm L}$ and the
  'diamagnetic term' is of order $D(eA/c)^2 \sim E_c
  (\Phi/\Phi_0)^2$. Since the basic energy-unit of the action is the
  mean level spacing, $\bar d$, the dimensionless coupling strength is
  of ${\cal O}((\Phi/\Phi_0)^2 E_c/\bar d) = {\cal O}((\Phi/\Phi_0)^2 g)$. For
  $\Phi/\Phi_0>g^{-1/2}$ the coupling strength exceeds unity.}. In the
infrared limit, fluctuations coupling to these vertices become
inessential and only $T$-matrices with tr-block diagonal structure
survive, through their commutativity with the tr-symmetry breaking operators:
\[
T_0= T_1 \otimes E_{11}^{\rm tr} + T_2 \otimes E_{22}^{\rm tr}.
\]
The symmetry relation (\ref{Tsym}) implies that the two blocks are related to
each other by
\begin{equation}
\label{app:Tsym}
T_{1}^T = T_{2}.
\end{equation}
Substituting the block diagonal form into the action of
Eq.~(\ref{ea}), all T-invariance breaking operators drop out. Thus,
similarly to the preceding subsection, we again arrive at
Eq.~(\ref{S0orthogonal}) as an effective Goldstone action. Exploiting
the tr-block structure of the $T$-matrices, the action may be written
as the sum of two contributions, $S=S_1 + S_2$, where the subscript
refers to the tr-index. Due to Eq.~(\ref{app:Tsym}) and the invariance
of the 'str' under matrix transposition, the two contributions are
identical and we obtain
\begin{equation}
  \label{S0unitary}
  S_0[Q_0] = \int \left[\frac{\pi \nu_n}{2} \tilde D {\, \rm str}_0
   \left(\partial Q_0\partial Q_0\right)    -i\pi \omega_+
   \nu {\, \rm str}_0\left(Q_0  \sigma_3^{\rm ar}\right)\right]
\end{equation}
as the final result for the T-non-invariant Goldstone mode action, as
also included in Eq.~(\ref{S0}). In
Eq.~(\ref{S0unitary}), we have used 
\[
Q_0 = T_0 \bar Q T_0^{-1}
\]
where -- for the sake of a homogeneous notation -- we have denoted
$T_1$ again by $T_0$ and the tr-block $\bar Q_{11}$ by $\bar Q$. Note,
however, that the matrix dimension of the fields in (\ref{S0unitary})
is twice as small as the one in (\ref{S0unitary}), as the tr-index
structure is missing, and that there is no symmetry relation such as 
Eq.~(\ref{Tsym}).

\subsection{Boundary Conditions}
\label{app:boundary}

In order to make the gradient terms appearing in eqns.~(\ref{S0orthogonal})
and (\ref{S0unitary}) well defined, boundary conditions at all
interfaces between the N-region and external regions need to be
specified. Whereas the boundary conditions to be imposed at interfaces
to insulators ($\partial T=0$) and idealized leads ($T=\openone$) have
been derived previously\cite{Efetov}, the interfacial behaviour at
SN-boundaries has so far not been analyzed.  Note that the present
analysis applies to the boundary condition of the  Goldstone
  mode as opposed to the behaviour of a single-particle Green
function (the ar-diagonal blocks of the $Q$-matrices) -- The latter has
been summarized already in appendix \ref{sec:boundary}. 

Some insight into the structure of the boundary conditions may be
gained from the fact that {\it two} types of currents across the
SN-boundary may be identified:
\begin{itemize}
\item A potentially non-vanishing {\it electric current}. (Two
  elementary charges flow across the interface whenever Andreev
  scattering takes place.)
\item A {\it quasiparticle current} that vanishes (even in the case of
  a nonzero reflection coefficient at the interface). The quasiparticle
  current democratically counts the flow of electrons and
  holes. Since incoming electrons are either reflected or Andreev
  converted into holes, the normal component of the boundary
  quasiparticle 
current vanishes.
\end{itemize}
Taking into account the fact that the Goldstone mode does not distinguish
between particles and holes, one may anticipate that the boundary
condition reads $\partial_\perp T_0=0$, corresponding to 
zero quasiparticle current flow. We confirm this supposition below. 

First it is necessary to decide on the {\em location} of the NS
boundary. Since superconductive behaviour
penetrates, in the sense of the proximity effect, into the normal region
(and vice versa), the
position of the boundary is to some extent arbitrary and need not
necessarily coincide with the material boundary, at which the jump in
$\Delta$ appears. 
However within the superconductor both the effective diffusion
constant and the DoS vanish on a scale of the order of $\xi=
(D/\Delta)^{1/2}$, which is much smaller than the diffusion length, 
$L_{\epsilon}$
(for $\epsilon\ll \Delta$). With a type of coarse-graining in mind for which
the details of variation over scales of $\xi$ in
the S region becomes irrelevant,  
we make a simple {\it choice} of the physical
SN-interface as the effective one for the Goldstone action. 

To derive boundary conditions for the Goldstone mode, we employ the
method of boundary Ward-identities, as previously used in
Ref.~\cite{Read}. As is usual with Ward identities, the scheme is to
subject the $Q$-degrees of freedom to an infinitesimal gauge
transformation and to exploit the fact that physical expectation
values ought to be invariant, whilst the action need not.

Specifically, we perform the infinitesimal rotation
\begin{equation}
\label{app:Rtrafo}
  Q_0 \rightarrow e^{-R} Q_0 e^R \simeq Q_0 - [R,Q_0],
\end{equation}
where 
\begin{equation}
\label{app:Rstructure}
R(x) = \left(
  \begin{array}{cc}
&R_{+-}(x)\\
R_{-+}(x)&
  \end{array}\right)\otimes \openone_{\rm tr,bf,ph},
\end{equation}
and the matrix structure refers to the ar-indices. A straightforward
calculation then shows that the effective action (\ref{S0orthogonal})
transforms as 
\begin{eqnarray}
  \label{app:ward_action}
  &&S_0[Q_0] \rightarrow S_0[Q_0] + \delta S_{\rm N}[Q_0,R] + \delta S_{\rm
  S/N}[Q_0,R], \nonumber\\
&&\hspace{0.5cm}\delta S_{\rm N}[Q_0,R] = 
\int \left[\pi \nu_n \tilde D {\rm str}_0
   \left(R\partial \left(Q_0\partial Q_0\right)\right)  +  \frac{i\pi
  \omega_+ 
   \nu}{2}{\rm str}_0\left(R[Q_0,\sigma_3^{\rm ar}]\right)
  \right],\nonumber\\ 
&&\hspace{0.5cm}\delta S_{\rm S/N}[Q_0,R]=-\frac{\pi \nu_n}{2}
  \int_{\rm S/N}dS  D_{\rm S/N} {\, \rm str}_0 
\left[   R Q_0\partial_\perp Q_0\right],
\end{eqnarray}
where $\int_{\rm S/N}dS$ denotes a surface integral over the
SN-boundary (induced by an integration by parts necessary to shuffle
all derivatives from $R$ to $Q_0$) and $D_{\rm S/N}$ is the local
diffusion coefficient at the boundary. In order to use
(\ref{app:ward_action}) to construct a boundary condition, we consider
the functional expectation value
\begin{equation}
\label{app:Xdef}
X \equiv \langle F(Q_0(y)) \rangle_{Q_0},
\end{equation}
where $F$ may be an arbitrary function of the matrix $Q_0(y)$ at point
$y\in {\rm N}$. Whereas both the action and $F(Q_0(y))$ need not be
invariant under the transformation (\ref{app:Rtrafo}), the expectation
value, $X$, must be. Expanding the expression (\ref{app:Xdef}) to
first order in $R$ and omitting the matrix arguments in the notation,
we obtain
\[
X \rightarrow X + \langle \delta F + F(\delta S_{\rm N} + \delta
S_{\rm S/N}) \rangle 
\]
and hence
\[
 \langle \delta F + F(\delta S_{\rm N} + \delta
S_{\rm S/N}) \rangle \stackrel{!}{=} 0.
\]
Since the action $\delta S_{\rm S/N}$ is singular at the boundary, its
contribution to the above expression must vanish individually, that is, we
have to demand
\[
\left \langle F(Q_0(y))\int_{\rm S/N}dS D_{\rm S/N} {\, \rm str}_0 \left[ R
    Q_0\partial_\perp Q_0\right] \right \rangle_{Q_0} \stackrel{!}{=} 0
\]
for any function $F$. As $R$ is arbitrary, this can be generally true 
only if
\begin{equation}
\label{app:BC1}
\left \langle F(Q_0(y)){\, \rm str}_0 \left[R
    Q_0(x)\partial_\perp Q_0(x)\right] \right \rangle = 0,\hspace{0.5cm}
\forall y\in {\rm N\,}, x\in {\rm S/N},
\end{equation}
where $R$ may be any matrix of the structure (\ref{app:Rstructure}).
In order to transform Eq.~(\ref{app:BC1}) into a more practical form,
by
which we mean an effective condition to be imposed on the differential
operator governing the action, we subject both the action and the
expectation value in Eq.~(\ref{app:BC1}) to a perturbative
expansion. Introducing
\begin{equation}
\label{app:Qrep}
Q_0=e^W \sigma_3^{\rm ar} e^{-W}, \hspace{0.5cm}W = \left(
  \begin{array}{cc}
&B\\
\bar B&
  \end{array}\right),
\end{equation}
and expanding the action to lowest order in $B$, we obtain
\begin{eqnarray*}
  &&S_0[Q_0] \rightarrow \int {\, \rm str}_0 \left( \bar B \Pi^{-1} B
  \right) + \dots,
\end{eqnarray*}
where $\Pi$ is a shorthand for the diffusion type operator governing
the quadratic action. Note that for the present discussion the 
detailed structure of $\Pi$ is of no concern.
Further, choosing
\[
F(Q_0(y)) = {\rm str}_0 \left( Q_0(y) E_{21}^{\rm ar} \otimes
  \sigma_3^{\rm bf} \right),
\]
Eq.~(\ref{app:BC1}) takes the form
\begin{eqnarray*}
 \left \langle F(Q_0(y)){\, \rm str}_0 \left[R Q_0(x)\partial_\perp
      Q_0(x)\right] \right \rangle &\rightarrow&
  \left \langle {\, \rm str}_0(B(y)
    \sigma_3^{\rm bf}) {\, \rm str}_0(\partial_\perp \bar B(x))\right
  \rangle_{B,\bar B} \\
&=&0
\hspace{0.5cm} \forall y\in {\rm N\,}, x\in {\rm
    S/N},
\end{eqnarray*}
where $\langle \dots \rangle_{B,\bar B}$ stands for a functional
average with respect to the above quadratic action.  Computing the
expectation value by means of Wick's theorem, we obtain finally
\[
\partial_\perp \Pi(x,y)=0,\hspace{0.5cm} \forall y\in {\rm N\,}, x\in
{\rm S/N}.
\]
This is the required boundary condition. It implies that
the eigenfunctions of the operator $\Pi$ must be drawn from the set of
functions obeying Neumann boundary conditions at the S/N interface. In
order to install this condition generally, we restrict the functional
integration to the set of field configurations, $B$, that obey the
same boundary condition, $\partial_\perp B=0$. Since $T=\exp(W)$,
where $W$ is given in Eq.~(\ref{app:Qrep}), an alternative formulation
reads $\partial_\perp T=0$.

\section{Renormalization of the Minigap Edge}
\label{app:renorm}

In section~\ref{sec:a_sns_pert} the r\^ole of a-type fluctuations on
the single-particle properties was investigated within the framework
of a perturbative expansion around the inhomogeneous saddle-point
solution of the Usadel equation. There it was shown that, in the SNS
geometry, the minigap induced by the proximity effect is not destroyed
by quantum fluctuations.  However, this calculation failed to account
for the {\em shift} of the minigap edge resulting from the quantum
renormalization of the diffusion constant due to mechanisms of weak
localization. This phenomenon is described below.

To take into account weak localization corrections, we apply a
conventional momentum shell renormalization group procedure to the
effective action as it is detailed e.g. in \cite{Efetov}.  Beginning
with the parameterization defined in Eq.~(\ref{parQ}), we factorise
rotations $T=T_>T_<$ into fast $T_>$ and slow $T_<$ degrees of
freedom. Here rotations $T_>$ ($T_<$) involve spatial fluctuations on
scales shorter (longer) than $b/\Lambda$, where $\Lambda$ represents
an ultraviolet cut-off and $0<b<1$. Applying the parameterization
\begin{eqnarray}
Q= U Q_> U^{-1},\qquad Q_>=T_>\sigma_3^{\rm ph}T_>^{-1},\qquad
U=R T_<,
\label{paramrg}
\end{eqnarray}
where the rotation $R$ defines the saddle-point~(\ref{parQ}), 
we obtain
\begin{eqnarray}
S=-\frac{\pi\nu_n}{8}\int {\rm str}\left[D\left(
(\partial Q_>)^2+4Q_>\partial Q_> 
\cdot \vc{\Phi} +[\vc{\Phi},Q_>]^2\right)
+4i\epsilon Q_> U^{-1} \sigma_3^{\rm ph}U\right],
\label{sf1}
\end{eqnarray}
where $\vc{\Phi}=U^{-}\partial U$. Setting $T_>=e^W$, and expanding
$Q_>$ to quadratic order in in the generators of rotations $W$, the
action separates into three contributions:
\begin{eqnarray}
S=S_{\rm S} +S_{\rm SF}+S_{\rm F},
\end{eqnarray}
where, defining $Q_<=U\sigma_3^{\rm ph}U^{-1}$ and $\breve{Q}_< =
U^{-1} \sigma_3^{\rm ph}$,
%
\begin{eqnarray}
S_{\rm S}&=&-\frac{\pi\nu_n}{8}\int {\rm str}
\left[D\left(\partial Q_<\right)^2+4i\epsilon
\sigma_3^{\rm ph}Q_<\right], \\
S_{\rm SF}&=&-\pi\nu_n\int {\rm str}\left[
D\left\{[W,\partial W]\cdot
\vc{\Phi}+\left(\vc{\Phi}\sigma_3^{\rm ph}W\right)^2+\left(\vc{\Phi}
\sigma_3^{\rm ph}\right)^2W^2
-\partial W\cdot\vc{\Phi} +W(\vc{\Phi}\sigma_3^{\rm ph})^2\right\}
\right.\nonumber\\
&&\left.\mbox{}+i\epsilon\left(
W\sigma_3^{\rm ph}\breve{Q}_< +W^2\sigma_3^{\rm ph}
\breve{Q}_<\right)\right], 
\label{SF2}\\
S_{\rm F}&=&\frac{\pi\nu_n}{2}\int D{\rm str}\left[(\partial W)^2+{\Lambda^2
\over b^2}W^2\right].
\end{eqnarray}


We now integrate over the fast fluctuations, at the one-loop level, 
to obtain a new effective
action, of the form $S=S_{\rm S}+ \langle S_{\rm SF}\rangle_{\rm
  F}$. 
Taking the energy cut-off $\Lambda/b$ to be far above $E_c$, it is
sufficient to neglect the linear terms in $W$
and to integrate over the whole range of
energies $[\Lambda/b,\infty)$ with a {\em constant} $R$. We find
\begin{eqnarray}
\langle S_{\rm SF}\rangle_{\rm F} =\frac{\pi\nu_n}{4} \int d\vc{r}\ 
D\Pi(\vc{r},\vc{r}) {\rm str}(\partial Q_<)^2,
\label{SFav}
\end{eqnarray}
where $\Pi(\vc{r},\vc{r}^\prime)$ represents the diffusion propagator
\begin{eqnarray}
2\pi\nu_n D\left(-\partial^2+{\Lambda^2\over b^2}\right)
\Pi(\vc{r},\vc{r}^\prime)=\delta(\vc{r}-\vc{r}^\prime).
\end{eqnarray}
Altogether, applying the rescaling, at one-loop we obtain the renormalized 
action 
\begin{eqnarray}
S^\prime=-\frac{\pi\nu_n}{8} \int {\rm str}\left[D_{\rm eff} (\partial Q)^2
+4i\epsilon\sigma_3^{\rm ph}Q\right],
\end{eqnarray}
where $D_{\rm eff}=D[1-2\Pi(0,0)]$
denotes the renormalized diffusion 
constant. From this result, we see that the bare diffusion constant is 
subject to a standard weak localization correction~\cite{Abrahams}, albeit 
derived from purely within the particle/hole sector.

As a result of the renormalization procedure, no new terms are generated in the
effective action. Instead, we obtain the usual kinetic term but with a 
renormalized diffusion coefficient. However, as a consequence of the 
renormalization, $\vc{q}\cdot\sigma_3^{\rm ph}=R^{-1}\sigma_3^{\rm ph}R$ no 
longer represents the saddle-point of the theory. Accordingly, it is necessary 
to recalculate the saddle-point solution in the presence of the renormalized 
diffusion constant. The result is a corresponding renormalization of the 
minigap edge discussed in the text. 


\begin{references}
\bibitem{Beenakker97} Beenakker, C. W. J., 1997, {\em Rev. Mod. Phys.}, 
{\bf 69}, 3, 731. 
\bibitem{Spivak} Spivak, B. Z., and Khmel'nitskii, D. E., 1982,
{\em JETP Lett.}, {\bf 35}, 8, 412.
\bibitem{A+Spivak} Al'tshuler, B. L., and Spivak, B. Z., 1987,
 {\em Sov. Phys. JETP}, {\bf 65}, 2, 343.
\bibitem{S+Nazarov} Stoof, T. H.,  and Nazarov, Y. V., 1996, {\em
Phys. Rev. B}, {\bf 53}, 21, 14496.
\bibitem{Marmorkos93} Marmorkos, I. K, Beenakker, C. W. J., and
Jalabert, R. A., 1993, {\em Phys. Rev. B}, {\bf 48}, 2811.
\bibitem{Eilenberger} Eilenberger, G., 1968, {\em Z. Phys. B}, 
{\bf 214}, 2, 195.
\bibitem{Usadel} Usadel, K. D., 1970, {\em Phys. Rev. Lett.},  
{\bf 25}, 8, 507.
\bibitem{Hartog} Den Hartog, S. G., Kapteyn, C. M. A., van Wees,
B. J., Klapwijk, T. M, van der Graaf, W., and Borghs, G., 
1996, {\em Phys. Rev. Lett.}, {\bf
76}, 4592; Den Hartog, S. G., Kapteyn, C. M. A., van Wees,
B. J., Klapwijk, T. M., and Borghs, G., 1996, {\em Phys. Rev. Lett.}, 
{\bf 77}, 24, 4954.
\bibitem{Hecker} Hecker, H., Hegger, H., Altland, A.,  and Fiegle, K.,
1997, {\em  Phys. Rev. Lett.}, {\bf 79}, 8, 1547.
\bibitem{A+Zirn} Altland, A., and Zirnbauer, M. R., 1996,  
{\em Phys. Rev. Lett.}, {\bf 76}, 18, 3420; 1997, {\em Phys. Rev. B}, 
{\bf 55}, 2, 1142. 
\bibitem{Takane} Takane, Y., and Ebisawa, H., 1991, {\em
J. Phys. Soc. Jpn.}, {\bf
60}, 3130; 1992, {\em J. Phys. Soc. Jpn.}, {\bf 61}, 5, 1685; 1992, 
{\em J. Phys. Soc. Jpn.}, {\bf 61}, 2858;
1993, {\em J. Phys. Soc. Jpn.}, {\bf 62}, 1844.
\bibitem{AST} Altland, A., Simons, B. D., and Taras-Semchuk, D., 1997, 
{\em Pis'ma v. ZhETF}, {\bf 67}, 1, 21
[1998, {\em JETP Lett.}, {\bf 67}, 1, 22].
\bibitem{Beenakker93} Beenakker, C. W. J., 1993, {\em Phys. Rev. B}, 
{\bf 47}, 15763.
\bibitem{Hui} Hui, V. C., and Lambert, C. J., 1993, {\em
Europhys. Lett.}, {\bf 23}, 203.
\bibitem{Lambert93} Lambert, C. J., 1993, {\em J. Phys. Cond. Mat.}, 
{\bf 5}, 707. 
\bibitem{Brouwer95} Brouwer, P. W., and Beenakker, C. W. J., 1995, {\em
Phys. Rev. B}, {\bf 52}, 16772.
\bibitem{Brouwer95a} Brouwer, P. W., and Beenakker, C. W. J., 1995,
{\em Phys. Rev. B}, {\bf 52}, 3868.
\bibitem{Brouwer96} Brouwer, P. W., and Beenakker, C. W. J., 1996,
{\em Phys. Rev. B}, {\bf 54}, 12705.
\bibitem{Frahm} Frahm, K. M., Brouwer, P. W., Melsen, J. A., and
Beenakker, C. W. J., 1996, {\em Phys. Rev. Lett.}, {\bf 76}, 16, 2981. 
\bibitem{Beenakker91} Beenakker, C. W. J., 1991, {\em
Phys. Rev. Lett.}, {\bf 67}, 3836.
\bibitem{BTK} Blonder, G. E., Tinkham, M., and Klapwijk, T. M., 1982,
{\em Phys. Rev. B}, {\bf 25}, 7, 4515.
\bibitem{Lambert91} Lambert, C. J., 1991, {\em J. Phys. Cond. Mat.}, 
{\bf 3}, 6579.
\bibitem{Beenakker92} Beenakker, C. W. J., 1992, {\em Phys. Rev. B}, 
{\bf 46}, 12841. 
\bibitem{Muz2} Muzykantskii, B. A., and Khmelnitskii, D. E., 1995, {\em JETP
  Lett.}, {\bf 62}, 76. 
\bibitem{Mehta} Mehta, M. L., 1991, {\it Random Matrices} 
(Academic, New York).
\bibitem{Andreev} Andreev, A. F., 1964, {\em Zh. Eksp. Teor. Fiz.}, 
{\bf 46}, 1823; 1965, {\bf 49}, 655; and 1966, {\bf 51}, 1510
[1964, {\em Sov. Phys. JETP}, {\bf 19}, 1228;
1966, {\bf 22}, 455; and 1967, {\bf 24}, 1019].
\bibitem{Kulik} Kulik, I. O., 1969, {\em Zh. Eksp. Teor. Fiz.}, 
{\bf 57}, 1745 (1969)
[1970, {\em Sov. Phys. JETP} {\bf 30}, 944].
\bibitem{Tinkham} Tinkham, M., 1996, {\em Introduction to Superconductivity}
(R. E. Krieger, Malabar, FL; 2nd. edition, McGraw-Hill, New York).
\bibitem{Imry} Imry, Y., 1997, {\em Introduction to Mesoscopic Physics} 
(Oxford University Press, Oxford).
\bibitem{Likharev} Likharev, K. K., 1979, {\em Rev. Mod. Phys.}, 
{\bf 51}, 1, 101. 
\bibitem{Zaikin} Zaikin, A. D., 1988, see
{\em Nonequilibrium Superconductivity}, 
ed. V. L. Ginzburg (Nova Science Publications).
\bibitem{Likharev76} Likharev, K. K., 1976, {\em
Sov. Tech. Phys. Lett.}, {\bf 2}, 12.
\bibitem{Z+Zarkov} Zaikin, A. D., and Zarkov, G. F., 1981, {\em Fiz. 
Nizk. Temp.}, {\bf 7}, 3, 375.
\bibitem{Petrashov} Petrashov, V. T., Antonov, V. N., Delsing, P., and
 Claeson, T., 1995, {\em Phys. Rev. Lett.}, {\bf 74}, 26, 5268.
\bibitem{Zhou} Zhou, F., Spivak, B., and Zyuzin, A., 1995, {\em 
Phys. Rev. B}, {\bf 52}, 6, 4467.
\bibitem{KO} Kulik, I. O., and Omelyanchuk, A. N., 1975, {\em 
Sov. Phys. JETP Lett.}, {\bf 21}, 96.
\bibitem{Melsen} Melsen, J. A., Brouwer, P. W., Frahm, K. M., and
 Beenakker, C. W. J., 1996, {\em Europhys. Lett.}, {\bf 35}, 1, 7.
\bibitem{Gutzwiller} Gutzwiller, M. C., 1990, {\em Chaos in Classical and
  Quantum Mechanics} (Springer-Verlag, New York).
\bibitem{Argaman} Argaman, N., Smilansky, U., and Imry, Y., 1993,
{\em Phys. Rev. B}, {\bf 47}, 4440.
\bibitem{Aleiner} Aleiner, I. L., and Larkin, A. I., 1996, {\em 
Phys. Rev. B}, {\bf 54}, 20, 14423.
\bibitem{Brouwer96b} Brouwer, P. W., and Beenakker, C. W. J., 
1996, {\em J. Math. Phys.}, {\bf 37}, 4904.
\bibitem{Gorkov} Gorkov, L. P., 1958, {\em Sov. Phys. JETP}, 
{\bf 7}, 505; 1959, {\bf 36}, 1918.
\bibitem{Larkin} Larkin, A. I., and Ovchinnikov, Y. N., 1968, {\em
Sov. Phys. JETP}, {\bf 28}, 6, 1200.
\bibitem{Rammer} Rammer, J., and Smith, H., 1986, {\em
Rev. Mod. Phys.}, {\bf 58}, 2, 323.
\bibitem{Volkov} Volkov, A. F., and Pavlovskii, V. V., 1997, Report presented
at the {\em Euroschool on Physics of Mesoscopic Systems} (Sienna,
Italy); preprint cond-mat/9711251.
\bibitem{G+Landau} Ginzburg, V. L., and Landau, L. D., 1950,
{\em Zh. Eksp. Teor. Fiz.}, {\bf 20}, 1064.
\bibitem{Zaitsev} Zaitsev, A. V., 1994, {\em Sov. Phys. JETP}, 
{\bf 59}, 5, 1015.
\bibitem{Shelankov} Shelankov, A. L., 1985, {\em Jnl. Low Temp. Phys.}, {\bf
  60}, 1/2, 29.
\bibitem{Lambert98} Lambert, C. J., and Raimondi, R.,
1998, {\em J. Phys. Cond. Mat.}, {\bf 10}, 5, 901.
\bibitem{Lambert97} Lambert, C. J., Raimondi, R., Sweeney, V., and
  Volkov, A. F., 1997, {\em Phys. Rev. B}, {\bf 55}, 6015.
\bibitem{Kuprianov} Kuprianov, M. Y.,  and Lukichev, V. F., 
1988, {\em Sov. Phys. JETP}, {\bf 67}, 1163.
\bibitem{Nazarov} Nazarov, Y. N., 1994, {\em Phys. Rev. Lett.}, 
{\bf 72}, 10, 1420.
\bibitem{Vegvar} de Vegvar, P. G. N., Fulton, T. A., Mallison, W. H., and
Miller, R. E., 1994, {\em Phys. Rev. Lett.}, {\bf 73}, 1416.
\bibitem{Courtois} Courtois, H., Gandit, P., Mailly, D., and
Pannetier, B., 1996, {\em Phys. Rev. Lett.}, {\bf 76}, 1, 130.
\bibitem{Charlat} Charlat, P., Courtois, H., Gandit, P., Mailly, D.,
Volkov, A. F., and Pannetier, B., 1996, {\em Phys. Rev. Lett.}, 
{\bf 77}, 24, 4950.
\bibitem{Gueron} Gueron, S., Pothier, H., Birge, N. O., Esteve, D.,
and Devoret, M. H., 1996, {\em Phys. Rev. Lett.}, {\bf 77}, 14, 3025.
\bibitem{Kastalsky} Kastalsky, A., Kleinsasser, A. W., Greene, L. H., 
Bhat, R., Milliken, F. P., Harbison, J. P., 1991, {\em
Phys. Rev. Lett.}, {\bf 67}, 21, 3026; 
Kleinsasser, A. W., and Kastalsky, A., 1993, {\em Phys. Rev. B}, 
{\bf 47}, 8361.
\bibitem{Keldysh} Keldysh, L. V., 1964, {\em Zh. Eksp. Teor. Fiz.}, 
{\bf 47}, 1515 [1965, {\em Sov. Phys.-JETP}, {\bf 20}, 1018].
\bibitem{Beenakker93b} Beenakker, C. W. J., 1993, {\em
Phys. Rev. Lett.}, {\bf 70}, 1155; Beenakker, C. W. J., and Rejaei,
B., 1993, {\em Phys. Rev. Lett.}, {\bf 71}, 3689; 
1994, {\em Phys. Rev. B}, {\bf 49}, 7499.
\bibitem{Caroli} Caroli, C.,  de Gennes, P. G., and Matricon, J.,
1964, {\em Phys. Lett.}, {\bf 9}, 307.
\bibitem{Feiglman1} Feigl'man, M. V.,  and  Skvortsov, M. A., 
1997, {\em Phys. Rev. Lett.}, {\bf 78}, 2640.
\bibitem{Feiglman2} Skvortsov, M. A., Kravtsov, V. E., and Feigl'man,
M. V., preprint cond-mat/9805296.
\bibitem{Zirnbauer98} Bundschuh, R., Casanello, C., Serban, D., and
Zirnbauer, M. R., preprint cond-mat/9806172. 
\bibitem{Efetov} Efetov, K. B., 1983, {\em Ann. Phys.}, {\bf 32}, 53;
Efetov, K. B., 1997, {\em Supersymmetry in Disorder and Chaos} (Cambridge
University Press: Cambridge).
\bibitem{B+Frahm} Brouwer, P. W., and Frahm, K., 1996, {\em 
Phys. Rev. B}, {\bf 53}, 3, 1490.
\bibitem{Oppermann} Oppermann, R., 1987, {\em Nuclear Phys. B}, 
{\bf 280}, 753.
\bibitem{Kravtsov} Kravtsov, V. E., and Oppermann, R., 1991, {\em 
Phys. Rev. B}, {\bf 43}, 13, 10865.
\bibitem{Bohigas} Bohigas, O., 1984, {\em Springer Lecture Notes in Physics},
{\bf 209}, 1.
\bibitem{Kuprianov82} Kuprianov, M. Y., and Lukichev, V. F., 1982, 
{\em Fiz. Nizh. Temp.}, {\bf 8}, 10, 1045 [1982, {\em Sov. J. Low
Temp. Phys.}, {\bf 8}, 10, 526].
\bibitem{Bruder} Belzig, W., Bruder, C., and Schon, G., 
1996, {\em Phys. Rev. B}, {\bf 54}, 13, 9443.
\bibitem{Zhou98} Zhou, F., Charlat, P., Spivak, B., and Pannetier, B.,
1998, {\em Jnl. Low Temp. Phys.}, {\bf 110}, 3-4, 841.
\bibitem{Wegner} Wegner, F. J., 1970, {\em Z. Phys. B}, {\bf 35}, 207.
\bibitem{ZirnMath} Zirnbauer, M. R., 1996, {\em J. Math. Phys.}, {\bf
37}, 10, 4986.
\bibitem{Abrahams} Abrahams, E., Anderson, P. W., Licciardello, D. C., and
  Ramakrishnan, T. V., 1979, {\em Phys. Rev. Lett.}, {\bf 42}, 673.
\bibitem{Morse} Morse, P. M., and Feschbach, H., 1953, {\em Methods of
Theoretical Physics} (McGraw-Hill: New York).
\bibitem{Magnus} Magnus, W., and Winkler, S., 1966, {\em Hill's Equation}
(Dover Publications, Inc.: New York).
\bibitem{Smith} Smith, R. A., Reizer, M. Y., and Wilkins, J. W., 1995, 
{\em Phys. Rev. B}, {\bf 51}, 10, 6470.
\bibitem{Gradsteyn} Gradshteyn, N. S., and Ryzhik, I. M., 1994, {\em Table of
    Integrals, Series and Products} (London: Academic). 
\bibitem{K+Mirlin} Kravtsov, V. E., and Mirlin, A. D., 1994, {\em JETP
Lett.}, {\bf 60}, 65.
\bibitem{Muz} Muzykantskii, B. A., and Khmelnitskii, D. E., 1995, {\em
Phys. Rev. B}, {\bf 51}, 5480.
\bibitem{IWZ} Iida, S., Weidenm\"uller, H. A., and Zuk, J. A.,
1990, {\em Ann.Phys.}, {\bf 200}, 219.  
\bibitem{Falko} Fal'ko, V. I., and Efetov, K. B., 1995, {\em
Europhys. Lett.}, {\bf 32}, 627.
\bibitem{Mirlin} Mirlin, A. D., 1996, {\em Phys. Rev. B}, 
{\bf 53}, 1186.
\bibitem{Simons} Simons, B. D., and Altshuler, B. L., 1993, {\em 
Phys. Rev. Lett.}, {\bf 70}, 4063.
\bibitem{AS} Altshuler, B. L., and Shklovskii, B. I., 1986, {\em 
Sov. Phys. JETP}, {\bf 64}, 127.
\bibitem{Finkelstein} Finkelstein, A. M., 1984, {\em
Zh. Eksp. Teor. Fiz.}, {\bf 84}, 168 [1984, {\em Sov. Phys. JETP},
{\bf 57}, 97].
\bibitem{Andreev96} Andreev, A. V., Agam, O., Simons, B. D., and
  Altshuler, B. L., 1996, {\em Phys. Rev. Lett.}, {\bf 76}, 21, 3947. 
\bibitem{Weidenmueller} Guhr, T. , M\"uller-Groeling, A., and 
Weidenm\"uller, H. A., 1998, {\em Phys. Rep.}, {\bf 299}, 4, 190.
\bibitem{Brouwer98} 
Brouwer, P., Oreg, Y., Simons, B. D.,  and Altland, A. (unpublished).
\bibitem{Lodder98} Lodder, A., and Nazarov, Y. V., preprint cond-mat/9801310.
\bibitem{Read} Xiong, S. H., Read, N., and Stone, A. D., 1997,
{\em Phys. Rev. B}, {\bf 56}, 3982.
\end{references}
\end{document}